\documentstyle{article}
 
\newcommand{\x}{\chi}
\renewcommand{\O}{{\Omega}}

\renewcommand{\L}{\Lambda}
\renewcommand{\l}{\lambda}

\renewcommand{\t}{\tau}

\newcommand{\var}{\varepsilon}
\newcommand{\e}{\epsilon}

\newcommand{\h}{\eta}

\newcommand{\r}{\rho}
\renewcommand{\P}{\Pi}
\newcommand{\s}{\sigma}
\renewcommand{\S}{\Sigma}
\newcommand{\th}{\theta}

\newcommand{\D}{\Delta}
\renewcommand{\d}{\delta}
\newcommand{\g}{\gamma}
\newcommand{\G}{\Gamma}
\newcommand{\m}{\mu}

\newcommand{\n}{\nu}
\renewcommand{\a}{\alpha}

\renewcommand{\b}{\beta}
\newcommand{\nn}{\nonumber}
\newcommand{\ft}[2]{{\textstyle\frac{#1}{#2}}}

\begin {document}

\begin{titlepage}
\begin{flushright} {\em \bf U.N-PHY 26-I-01}\\
January - 2001\\
\end{flushright}
\vfill
\begin{center}
{\LARGE\bf $OSp(N|4)$ group and their contractions to $P(3,1)\times Gauge$} 
\vskip 2 cm  
{\large \bf Mauricio Ayala}$^1$
\vskip 4 mm
{\large \bf Tutor: Richard Haase}$^2$ 
\vskip 2 cm  
{\large \em Departamento de F\'{\i}sica}\\
{\large \em Universidad Nacional, Bogot\'a - Colombia}
\end{center}
\vfill
\begin{center}
{\bf Abstract}
\end{center}
\begin{quote}
Starting from $SO(n,m)$ groups, we are in search of groups that:\\
{\bf 1.)} in a simple way, include $N$ supersymmetric generators. (see \cite{Nahm})\\
{\bf 2.)} contain as subgroup: the de Sitter group $SO(4,1)$ or the Anti-de Sitter group $SO(3,2)$\\
{\bf 3.)} permit nontrivial gauge symmetry groups.\\
The smallest groups satisfyng above conditions are the $OSp(N|4)$ groups, which contain $Sp(4)\times SO(N)$ ($Sp(4) \sim SO(3,2)$) or $OSp(1|4)\times SO(N-1)$. Because of this, it is possible to generate $P(3,1)\times G$ using groups contraction mechanism, which may be: 
$$SO(3,2)\to P(3,1) \ \ \ \ \ o \ \ \ \ \ OSp(N|4)\to \ ^SP(3,1|N)$$
where $P(3,1)$ is the Poincar\'e group and $G$ is a gauge group, say $SO(N)$ or $SO(N-1)$. This group contraction mechanism and its consequences upon different groups representations including $SO(3,2)$ or $SO(4,1)$, is clarified and extended to $OSp(N|4)$ representations (see \cite{Nicolai}), contracted to its $N$-extensi\'on SuperPoincar\'e group $\ ^SP(3,1|N)$.$^\diamondsuit$
\vskip 1 cm  

{\large \bf Keywords:} Wigner-In"n Contraction, Poincar\'e and SuperPoincar\'e Groups, (Anti)de Sitter Groups, Super-AdS Groups.
\vfill      
\hrule width 14.cm
\vskip 2.mm
{\noindent $^\diamondsuit$ This Document is a physics degree tittle requeriment}\\
\hrule width 7.cm
{\small \noindent $^1$ mao@estudiantes.fisica.unal.edu.co}\\
{\small \noindent $^2$ rhaase@ciencias.ciencias.unal.edu.co}\\
\end{quote}
\end{titlepage}

\tableofcontents{}

\section {Introducci\'on}
Se puede decir con seguridad que: la Teor\'{\i}a General de la Relatividad y la Teor\'{\i}a Cu\'antica de Campos, son los dos marcos te\'oricos que se imponen en la actualidad al describir las $1+3$ interacciones fundamentales conocidas. Teniendo en cuenta que la gravedad no es considerada una interacci\'on en el contexto de la Relatividad General (Forma cl\'asica de ver la naturaleza: teor\'{\i}a macrosc\'opica) y al buscar su modelo cu\'antico muestra problemas tales como: la no-renormalizaci\'on (no se pueden absorver las divergencias en las ecuaciones de campo). \\ 
\\ 
Por otro lado la Teor\'{\i}a Cu\'antica de Campos (t. microsc\'opica) tiene \'exito en el Modelo Standard, al considerar los tres grupos gauge de simetr\'{\i}a: $U(1)_Y\times SU(2)_L\times SU(3)_C$. Los cuales son relevantes para la construcci\'on de una teor\'{\i}a que incluye las interacciones: Electromagn\'etica (EM), D\'ebil (D) y Fuerte (F). Con los respectivos generadores $Y,T_i,G_a$ ($i=1,2,3$)($a=1,2..,8$) correspondientes a simetr\'{\i}as de: hipercarga, quilaridad y color en su orden. Estos se introducen en las ecuaciones de campo por medio de la derivada covariante como:(ver \cite{kaku})
$$D_\mu = I \partial_\mu +gYA_{\mu}(x)+g'T_iW_\mu^{i}(x)+g''G_aV_\mu^a(x)
\ \ \ \ \ \ \ \ \ \ 
A_{\mu}(x),\ W_\mu^{i}(x),\ V_\mu^a(x)\ :\ \ \mbox {Campos de interacci\'on.}
$$
Las constantes $\ g,\ g',\ g''$ son las constantes de acoplamiento, donde cada una caracteriza la magnitud de la correspondiente interacci\'on. En esta ecuaci\'on se puede notar que hay tantos campos bos\'onicos o de interacci\'on como generadores de simetr\'{\i}a tiene el grupo asociado, los cuales son etiquetados como sigue: 
$$
\mbox {(EM)}\ \mapsto\ \mbox{ 1 fot\'on: }A_{\m}(x),\ \ \ \ \ \ \ \mbox {(D)}\ \mapsto\ \mbox{ 3 bosones d\'ebiles: }W_\m^{i}(x),\ \ \ \ \ \ \ \mbox {(F)}\ \mapsto\ \mbox{ 8 gluones: }V_\m^a(x)
$$
La Teor\'{\i}a Cuantica de Campos (T.C.C.) surgi\'o como una uni\'on de la Teor\'{\i}a de Grupos (en particular el grupo Poincar\'e: $P(3,1)$ o el grupo de Lorentz: $SO(3,1)\subset P(3,1)$) y la Mec\'anica Cu\'antica. Donde la estructura de grupo fija de una forma \'unica la matriz $S$ bajo ciertos par\'ametros que especifican las interacciones. De una forma m\'as precisa se puede exijir las siguientes condiciones para construir una T.C.C., la cual se impone aqu\'{\i}.
\begin{itemize}
\item Cualquier campo debe transformarse como una representaci\'on irreducible (irrep) del grupo Poincar\'e y alg\'un grupo gauge o grupo de simetr\'{\i}a interna.
\item La teor\'{\i}a debe ser unitaria, su acci\'on causal, renormalizable e invariante bajo estos grupos
\end{itemize}
Estos postulados exijen fuertes condiciones sobre la teor\'{\i}a, ya que los campos solo pueden ser masivos/no-masivos con esp\'{\i}n $0,1/2,1,..$etc. (Fermiones $\Rightarrow$ part\'{\i}culas de esp\'{\i}n semientero) (Bosones $\Rightarrow$ part\'{\i}culas de esp\'{\i}n entero). Sin embargo esta condici\'on no determina la acci\'on: dado que es posible encontrar teor\'{\i}as que son invariantes, no-causales y no-unitarias. Por ejemplo las teor\'{\i}as con derivadas de tercer orden o superior satisfacen esta condici\'on, sin embargo poseen part\'{\i}culas con norma negativa, las cuales violan la unitaridad. Por lo tanto la segunda condici\'on nos ayuda a fijar la acci\'on bajo ciertas representaciones y las constantes de acoplamiento de las interacciones. En resumen: a este marco te\'orico se le impone dos tipos de simetr\'{\i}as continuas:
\begin{itemize}
\item {\em Simetr\'{\i}as del espacio-tiempo:} Estas incluyen el grupo Poincar\'e, con $P(3,1)\equiv T_{3,1}\wedge SO(3,1)$ y son simetr\'{\i}as no-compactas. Es decir el rango de sus par\'ametros es dado por un intervalo abierto sobre los reales (no-compacto) y no confinado (no es posible identificar sus puntos extremos). Por ejemplo: la velocidad de una part\'{\i}cula masiva puede tomar valores entre $v=0$ y $v=c$, pero no puede tomar el valor $v=c$. Adem\'as estos valores extremos no pueden ser identificados ($v=0\neq c$). 
\item {\em Simetr\'{\i}as internas:} Por definici\'on $G$ es un grupo gauge si $G$ conmuta  con $P(3,1)\ \Rightarrow [G,P(3,1)]=0$, las cuales son simetr\'{\i}as que `mezclan' part\'{\i}culas. Por ejemplo el grupo $SU(3)$ de simetr\'{\i}as sirve para recombinar el color de los $3$ quarks $\{u,d,s\}$. Esta simetr\'{\i}a interna rota los campos y las part\'{\i}culas en un espacio abstracto: el `espacio isot\'opico'. Estos grupos son compactos y el rango de los par\'ametros es finito. Por ejemplo el grupo de las rotaciones se puede parametrizar usando \'angulos con un rango $[0,2\pi]$ identificando $0$ con $2\pi$. Estas simetr\'{\i}as pueden ser globales (independientes del espacio-tiempo) o locales (pueden variar en cada punto del espacio-tiempo) en una teor\'{\i}a gauge. 
\end{itemize}
Una b\'usqueda fundamental ser\'{\i}a: encontrar una teor\'{\i}a unificada de campos que incluya las cuatro interacciones, donde la Relatividad General se pueda considerar como un l\'{\i}mite a bajas energias en una teor\'{\i}a cu\'antica de la gravedad. Pero surge una inhabilidad para buscar un grupo gauge que combine el espectro de part\'{\i}culas con la gravedad cu\'antica, dado el trabajo de Coleman y Mandula \cite{ColemanMandula}. El cual se resume en el siguiente teorema:\\
\\
{\em Teorema de `Coleman-Mandula'}: {\it Sea la matriz $S$  no trivial, donde la amplitud de dispersi\'on es una funci\'on anal\'{\i}tica de: el cuadrado de la transferencia de momento y el cuadrado de la energ\'{\i}a de centro de masa. Donde:\\
$\diamondsuit$- El espectro de estados de masa de una part\'{\i}cula, es un conjunto aislado de valores positivos (posiblemente infinito) con un n\'umero finito de tipos de part\'{\i}culas.\\
$\diamondsuit$- Sea $F$ una simetr\'{\i}a de la matriz $S\ \Rightarrow[F,S]=0$, con $P(3,1)\subset F$. Donde los generadores del grupo Poincar\'e pueden ser construidos al menos localmente por operadores integrales en el espacio de momento.\\
\\
Entonces si $F $ contiene simetr\'{\i}as internas, este debe ser localmente isomorfo al producto directo:} 
$$P(3,1)\times G \ \ \ \ \ \ \ G\Rightarrow \mbox {\it grupo de simetr\'{\i}a interna}$$
Este teorema nos indica que no es posible encontrar una extensi\'on no trivial para las simetr\'{\i}as del espacio-tiempo (grupo Poincar\'e) en la matriz $S$. Esto imposibilita mezclar simetr\'{\i}as del espacio-tiempo y simetr\'{\i}as gauge en un mismo grupo de simetr\'{\i}a bos\'onica (los generadores cumplen relaciones de conmutaci\'on). De una forma m\'as comprensible se tiene el siguiente resultado:\\
\\
{\em Teorema `No-go'}: {\it Dado un grupo de Lie no-compacto, no es posible encontrar una representaci\'on unitaria de dimensi\'on finita para este.}(ver \cite{kaku})\\
\\
Entonces: cualquier uni\'on no trivial entre grupos compactos y no-compactos con una representaci\'on unitaria de dimensi\'on finita falla. Si se insiste en construir una representaci\'on unitaria de un grupo no-compacto esta debe ser necesariamente de dimensi\'on infinita. La cual puede tener propiedades no f\'{\i}sicas tales como: un n\'umero infinito de part\'{\i}culas en una irrep o un espectro continuo de masas en cada irrep.\\
\\
Por lo tanto combinar gravedad con los grupos gauge de interacci\'on no resulta nada sencillo. Es aqu\'{\i} donde surge una soluci\'on: considerar unos nuevos par\'ametros que anticonmuten dentro del grupo de simetr\'{\i}as del espacio-tiempo (variables de Grassmann). Donde los nuevos generadores de grupo cumplen relaciones de anticonmutaci\'on (generadores de una nueva simetr\'{\i}a: supersimetr\'{\i}a), los cuales no son considerados en la derivaci\'on original del {\it teorema Coleman-Mandula}. Supersimetr\'{\i}a evade este teorema. Por lo tanto alguna gente impone esta simetr\'{\i}a adicional (ver el trabajo de Berenstein \cite{Berenstein} como un complemento). En resumen: 
\begin{itemize}
\item {\em Supersimetr\'{\i}a:} es una simetr\'{\i}a relativista que combina simetr\'{\i}as bos\'onicas y fermi\'onicas de una manera no trivial, donde los generadores de simetr\'{\i}a son de tipo fermi\'onico (obedece relaciones de anticonmutaci\'on). Supersimetr\'{\i}a nace al considerar todas las extensiones del grupo Poincar\'e compatibles con la mec\'anica cu\'antica. Actualmente este es el \'unico camino disponible para unificar simetr\'{\i}as internas y simetr\'{\i}as del espacio-tiempo de la matriz $S$ en una teor\'{\i}a relativista de part\'{\i}culas.
\end{itemize}
Al construir una teor\'{\i}a supersim\'etrica en un espacio de Minkowski (m\'etrica plana), se permiten part\'{\i}culas en un mismo multiplete con esp\'{\i}n diferente y los dem\'as n\'umeros cu\'anticos iguales. Entonces se puede mezclar fermiones y bosones en el mismo multiplete, salvo que tienen un mismo valor de masa.\\ 
\\
En la d\'ecada de los setenta: Ferrara \cite{Ferrara-PVN}, Freedman \cite{Freedman} y sus colaboradores mutuos construyeron una nueva clase de teor\'{\i}as de supergravedad con una $N$-extensi\'on de supersimetr\'{\i}a no-local. Teor\'{\i}a descrita en un espacio de Sitter que permite campos gauge de esp\'{\i}n-$\ft 32$, los cuales pertenecen a una representaci\'on vectorial de $O(N)$. Estas teor\'{\i}as tienen una particularidad: combinan simetr\'{\i}as internas con simetr\'{\i}as del espacio-tiempo de una forma no trivial y pueden ser candidatos a unificar la gravedad con las otras interacciones.\\
\\
En resumen al tomar el grupo de simetr\'{\i}as de un espacio-tiempo Anti-de Sitter $SO(3,2)$ isomorfo a $Sp(4)$ en su \'algebra y al acoplar $N$ generadores de supersimetr\'{\i}a, surge un nuevo grupo  $OSp(N|4)$, el cual es la extensi\'on graduada de $Sp(4)\times SO(N)$. Aparece de una forma natural un nuevo grupo $SO(N)$ de simetr\'{\i}a interna. Con la ventaja que supersimetr\'{\i}a Anti-de Sitter no implica que las part\'{\i}culas fermi\'onicas y bos\'onicas en un mismo multiplete tengan masas iguales.\\
\\
Aqu\'{\i} se muestra la posibilidad de deducir el grupo $P(3,1)\times G$ apartir del supergrupo $OSp(N|4)$ usando mecanismos de contracci\'on de grupo. M\'etodo que da una nueva posibilidad de obtener las simetr\'{\i}as del espacio-tiempo usando otros grupos m\'as grandes, los cuales pueden conllevar a nuevas relaciones o propiedades f\'{\i}sicas no antes vistas localmente. Por ejemplo una simetr\'{\i}a $SO(3)$ sobre la superficie de la tierra localmente puede verse como una simetr\'{\i}a $E_2\equiv T_2\wedge SO(2)$. Entonces existe una contracci\'on $SO(3)\to T_2\wedge SO(2)$.\\ 
\\
Con este prop\'osito, se da una exposici\'on aumentando el nivel de dificultad: empezando con el m\'etodo de contracciones de grupo en la secci\'on (\ref{contraccion}) para familiarizar al lector con esta t\'ecnica y sus resultados, ya que al reescalar los generadores, las representaciones tambi\'en lo son. Ah\'{\i} se muestran las contracciones: $SO(3) \to E_2\equiv T_2\wedge SO(2)$ \ $SO(3,1)\to G_3\equiv K_3\wedge SO(3)$, util en la secci\'on (\ref{anti/de-sitter}) al exponer el grupo de Sitter, sus representaciones y su contracci\'on al grupo Poincar\'e. Luego en la secci\'on (\ref{superpoincare}) se expone el esquema de supersimetr\'{\i}a acoplada al grupo Poincar\'e y su clasificaci\'on de irreps, con el prop\'osito de construir sus extensiones a grupos $SO(n,m)$ teniendo en cuenta la clasificaci\'on de Nahm\cite{Nahm} (Sec. \ref{susy-ads}). Siendo $OSp(N|4)$ el grupo supersim\'etrico minimal que contiene $SO(3,2)$, luego se muestran sus irreps unitarias y su contracci\'on a SuperPoincar\'e. Por \'ultimo, suponiendo un grupo $OSp(8|4)$ se clasifica su espectro (ver Nicolai \cite{Nicolai}) y se usa la contenencia: 
$$OSp(8|4)\supset OSp(1|4)\times SO(7) \ \ \ \ \ o \ \ \ \ \ OSp(8|4)\supset Sp(4)\times SO(8)$$
para indicar las posibles contracciones a $P(3,1)\times G$, con $G=SO(7)$ o $SO(8)$, lo que completa este trabajo.

\section {Mecanismo de Contracci\'on de In\"{o}n\"{u}-Wigner}\label{contraccion}

Este m\'etodo surge al estudiar el l\'{\i}mite de la mec\'anica relativista a la mec\'anica cl\'asica. Ya que al tomar el l\'{\i}mite de la velocidad de la luz a infinito $c\ \rightarrow \infty$ (o l\'{\i}mite a bajas velocidades comparado con la velocidad de la luz $v\ \rightarrow 0$) el grupo de Poincar\'e se transforma en el grupo de Galileo. Otro ejemplo es la transici\'on de la mec\'anica cu\'antica a la mec\'anica cl\'asica considerando el l\'{\i}mite a cero de la constante de Planck ($\hbar \rightarrow 0$) en algunos ejemplos f\'{\i}sicos. Esta idea fue propuesta y estudiada inicialmente por In\"on\"u y Wigner en 1953 \cite{Inonu-Wigner}, la cual act\'ua como un proceso l\'{\i}mite sobre los generadores de grupo y considerando su \'algebra bajo cambio de par\'ametros de escala conduce a la creaci\'on de un nuevo grupo. Tambi\'en ha indicado algunas relaciones importantes en el comportamiento asint\'otico de las funciones especiales de la f\'{\i}sica matem\'atica \cite {Izmest}.

\subsection {Formalismo}
Sea $L$ un \'algebra de Lie asociada a  $G$. $X_a(a=1,2,..n)$ una base para $L$ como espacio vectorial, donde:

\begin{equation}
[X_a,X_b]=\sum_{c=1}^n D_{ab}^c X_c\ \ \ \ \ \ \ (1 \leq \{a,b\} \leq n)
\end{equation}
$(D_{ab}^c)$ es llamada la constante de estructura del \'algebra de Lie con respecto a la base dada, donde la identidad de Jacobi impone ciertas condiciones sobre estas constantes:

\begin{equation}
\sum_{c=1}^n (D_{bd}^c D_{ac}^e + D_{da}^c D_{bc}^e + D_{ab}^c D_{dc}^e)=0
\label{form2}
\end {equation}  
Suponiendo que:

\begin {itemize}
\item Es conocida una sucesi\'on infinita ${[X_a]^u}$ de bases de $L$ con $u=1,2,...$ y sus correspondientes constantes de estructura ${[D_{ab}^c]^u}$.

\item El $\lim_{u \rightarrow \infty}\ [D_{ab}^c]^u=\ [D_{ab}^c]^{\infty}$ existe para todo  ${a,b,c}$.

\item La ecuaci\'on (\ref {form2}) se preserva bajo el l\'{\i}mite.
\end {itemize}
Se tiene que ${[D_{ab}^c]^{\infty}}$ genera una nueva \'algebra $L'$ y se dice que {\bf $L'$ es generado bajo la contracci\'on de $L$}. Formalizando esto de otra manera obtenemos:\\
\\
$\Diamond$ {\bf Definici\'on: } Dado $g \in G$ se define el {\bf Automorfismo Interior $Ad \ g$} como:

\begin {equation}
Ad\ g(g') \equiv g\ g'\ g^{-1}\ \ \ \ \ \ \ g' \in G
\end {equation}
$\Diamond$ {\bf Definici\'on: } Sea $F$ y $F'$ subgrupos de un grupo de Lie $G$. Se dice que {\bf $F'$ es el l\'{\i}mite de $F$ en $G$} si existe una sucesi\'on $g_1,g_2,...$ de elementos en $G$ tal que dada una sucesi\'on $f_1, f_2,f_3,...$ de elementos en $F$, la sucesi\'on $Ad\ g_1(f_1), Ad\ g_2(f_2),...$ converge a un elemento de $F'$.\\
\\
Dado que es m\'as conveniente trabajar con \'algebras de Lie que con grupos de Lie, se presenta una versi\'on infinitesimal de esta definici\'on. (ver \cite{Hermann})\\ 
\\
$\Diamond$ {\bf Definici\'on: } Sea $G$ un grupo de Lie, $L$ y $L'$ sub\'algebras de $G$. Se dice que {\bf $L'$ es el l\'{\i}mite de $L$ en $G$} si existe una sequencia $g_1,g_2,...$ de elementos en $G$ tal que dada una sucesi\'on $X_1, X_2,X_3,...$ de elementos en $L$, la sucesi\'on $Ad\ g_1(X_1), Ad\ g_2(X_2),...$ converge a un elemento de $L'$.\\
\\
Si $L$ y $L'$ est\'an relacionados por este camino se escribe:
\begin {equation}
L'=\lim_{n \rightarrow \infty} Ad\ g_n (L)
\end {equation}
En lo que sigue se muestran algunos ejemplos de contracci\'on de grupo, \'utiles en las otras secciones.  

\subsection {Contracci\'on de $SO(3) \to E_2\equiv T_2\wedge SO(2)$}
Dado $SO(3)$, el grupo de las rotaciones\footnote {Aqu\'{\i} $SO(3)$ puede significar la isotrop\'{\i}a del espacio 3$-Dim$.} en $R^3$, el cual posee 3 generadores de rotaci\'on infinitesimal $\{ J_1,J_2,J_3 \}$ y un \'algebra correspondiente:
\begin{equation}
\left[ J_r,J_s \right] = C_{rs}^t J_t \ \ \ \ \ \ \ \ \ \ C_{rs}^t \equiv i\epsilon_{rs}^t \ \ \ (r,s,t)=1,2,3
\label{alg-so3}
\end {equation}
$\epsilon_{rs}^t$ representa un tensor totalmente antisim\'etrico $(\epsilon_{12}^3=1)$. Para generar la contracci\'on, se construye la siguiente sucesi\'on de elementos del \'algebra de $SO(3)$ dependientes de $R$:
\begin{equation}
J_3 \Rightarrow J_0 \equiv J_3, \ \ \ \ \ J_r \Rightarrow \P_r \equiv {J_r \over R} \ \ \ o \ \ \ J_r = R\ \P_r \ \ \ \mbox{ con }\ r=1,2
\label{susti-so3}
\end {equation}
Con un \'algebra que mantiene la misma forma dada por (\ref{alg-so3}), salvo las constantes de estructura asociadas: 
\begin{equation}
\left[ \P_1,\P_2 \right] = {i \over R^2} J_0, \ \ 
\left[ \P_2,J_0 \right] = i \P_1, \ \ 
\left[ J_0,\P_1 \right] = i \P_2, 
\label{alg-so}
\end {equation}
Por otro lado se define la siguiente relaci\'on, que dar\'a paso a la contracci\'on:
\begin {equation}
\lim_{R \to \infty} \P_r \equiv K_r \ \ \ \ \ \ \ \mbox{ con }\ r=1,2
\label{k}
\end {equation}
$K_r$ denotar\'a el generador de traslaci\'on en $E_2$.\\
\\
Por lo tanto al tomar el l\'{\i}mite de contracci\'on en la ecuaci\'on (\ref{alg-so}) ($\lim_{R \to \infty}$) y usando la ecuaci\'on (\ref{k}), el grupo de rotaci\'ones en $R^3$ se transforma en el grupo de rotaci\'on-traslaci\'on en el plano $E_2=T_2 \wedge SO(2)$. Con generadores de traslaci\'on ${K_1,K_2}$ y el correspondiente generador de rotaci\'on $J_0$. El \'algebra que satisface $E_2$ toma la forma:
\begin {equation}
\left[ K_1,K_2 \right] = 0, \ \ 
\left[ K_2,J_0 \right] = iK_1, \ \ 
\left[ J_0,K_1 \right] = iK_2, 
\end {equation}
Estas relaciones definen localmente el grupo Euclidiano de Traslaci\'on-Rotaci\'on, donde la rotaci\'on generada por $J_3$ se conserva. Las rotaci\'ones generadas por $J_1, J_2$ se transforman bajo el l\'{\i}mite como generadores de traslaci\'on en las mismas componentes, con la particularidad que $SO(3) \supset SO(2)$ y al contraer se obtiene $SO(3) \to T_2 \wedge SO(2)$.\\
\\
Tomando el Casimir de $SO(3)$:
\begin{equation}
J^r J_r =J.J=J_1^2+J_2^2+J_3^2
\end {equation}
Aplicando (\ref {susti-so3}), se obtiene:  
\begin{equation}
J.J=R^2(\P_1^2+\P_2^2)+J_0^2
\end {equation}
Construyendo el Casimir para $E_2$, al tomar el l\'{\i}mite en la contracci\'on se obtiene:
\begin{equation}
C_{E_2}\equiv \lim_{R \to \infty}{J.J \over R^2}=\lim_{R \to \infty}(\P_1^2+\P_2^2)+J_0^2/R^2=K_1^2+K_2^2=K.K
\end {equation}
El anterior m\'etodo de contracci\'on se puede ver de otro modo: dado $S_2$ (la esfera 2-Dim), que representa la variedad de simetr\'{\i}a del grupo $SO(3)$ y $R^2$ (el plano) la variedad de simetr\'{\i}a generada por $E_2$. Se obtiene entonces que la contracci\'on $SO(3)\to E_2$ genera una descompactificaci\'on de la esfera 2-Dim al tomar el l\'{\i}mite de curvatura a 0 ($R \to \infty$). En forma expl\'{\i}cita tomando una representaci\'on en coordenadas esf\'ericas $(u_1,u_2,u_3)$, se tiene:
\begin {equation}
u_0=R\ Cos(\th_1), \ \ \ \ \ u_1=R\ Sen(\th_1)\ Cos(\th_2) \ \ \ \ \ u_2=R\ Sen(\th_1)\ Sen(\th_2)
\end {equation}  
Donde $(0 \leq \th_1 < \pi)$, $(0 \leq \th_2 < \pi)$. Tomando el l\'{\i}mite apropiado ($\lim_{R \rightarrow \infty}$) y ($\th_1 \simeq r/R$), obtenemos el nuevo sistema coordenado $(x_1,x_2)$ en $E_2$:
\begin {equation}
y_1=R\ Tan(\th_1)\ Cos(\th_2)\ \rightarrow x_1=r\ Cos(\th_2), \ \ \ \ \ \ y_2=R\ Tan(\th_1)\ Sen(\th_2)\ \rightarrow x_2= r\ Sen(\th_2) 
\end {equation}  
Aqu\'{\i} la circunferencia del ecuador ($\th_1=\pi/2$) `se env\'{\i}a' al infinito. (ver \cite{Izmest})

\subsubsection {Representaciones Asociadas a $SO(3)$ y su Esquema de Contracci\'on}\label {so3} 
En lo que sigue por completitud se presenta un an\'alisis standard de los grupos: $SO(3)$ y $E_2$, util en las siguientes secciones (ver \cite{simetries}). Este se puede obviar hasta la ecuaci\'on (\ref{sol-am_bm}) si lo desea.\\
\\
Definiendo el \'algebra {$J_+,J_-,J_3$}, para $SO(3)$ como:
\begin {equation}
J_+ =J_1+iJ_2 \ \ \ \ \ , \ \ \ \ \ J_-=J_1-iJ_2 \ \ \ \ \ \ \ \ \ \ (J_{\pm})^\dagger=J_{\mp}
\end{equation}
Estos cumplen el \'algebra:
\begin {equation}
\left[ J_3,J_+\right] = J_+, \ \ 
\left[ J_+,J_- \right] = 2J_3, \ \ 
\left[ J_3,J_- \right] = -J_-, 
\label{new-alg-so-c}
\end {equation}
Por lo tanto:
\begin{equation}
J.J={1 \over 2}(J_+J_-+J_-J_+)+J_3^2    
\label {150}
\end {equation}  
Podemos ahora encontrar los vectores propios comunes para $J.J$ y para $J_3$ ($SO(3)$) con sus correspondientes valores propios $j(j+1),\ m$. 
\begin {equation}
J.J |j,m\rangle=j(j+1) |j,m\rangle \ \ \ \ \ , \ \ \ \ \ J_3 |j,m\rangle=m |j,m\rangle
\label {new-val}
\end {equation}
De (\ref{new-alg-so-c}) y (\ref{new-val}) se tiene que:
\begin {eqnarray}
J_3J_+|j,m\rangle&=&J_+(J_3+1)| j,m\rangle=(m+1)J_+| j,m\rangle\label {new-val20}\nn\\
\Rightarrow J_+ | j,m\rangle &\equiv& a_m | j,m+1\rangle
\label {new-val10}
\end {eqnarray}
y
\begin {eqnarray}
J_3J_-| j,m\rangle&=&J_+(J_3-1)| j,m\rangle=(m-1)J_-| j,m\rangle\label {new-val200}\nn\\
\Rightarrow J_- | j,m\rangle &\equiv& b_m | j,m-1\rangle
\label {new-val100}
\end {eqnarray}
Para conocer $a_m,b_m$ usando la ecuaci\'on (\ref{150}), se obtiene: 
\begin {eqnarray}
J_+J_-| j,m\rangle=J.J-J_3(J_3-1)| j,m\rangle=[j(j+1)-m(m-1)]| j,m\rangle\label {160}\\
J_-J_+| j,m\rangle=J.J-J_3(J_3+1)| j,m\rangle=[j(j+1)-m(m+1)]| j,m\rangle\label {170}
\end {eqnarray}
Tomando:
\begin {eqnarray}
<j,m | J_-J_+ | j,m\rangle=[<jm | J_+]^+[J_+ | j,m\rangle]\geq 0 \label {180}\\
<j,m | J_+J_- | j,m\rangle=[<jm | J_-]^+[J_- | j,m\rangle] \geq 0 \label {190}
\end {eqnarray}
Por lo tanto:
\begin {eqnarray}
j(j+1)-m(m+1)=(j-m)(j+m+1) \geq 0 \label {200}\\
j(j+1)-m(m-1)=(j+m)(j-m+1) \geq 0 \label {210}
\end {eqnarray}
De (\ref{180}) se tiene que:
\begin{equation}
\|a_m\|^2=j(j+1)-m(m+1)
\end{equation}
De (\ref{190}):
\begin{equation}
\|b_m\|^2=j(j+1)-m(m-1)
\end{equation}
Teniendo en cuenta (\ref{200},\ref{210})
\begin{equation}
-j \leq m \leq j
\label {230}
\end{equation}
La posible construcci\'on apartir de $|j,m\rangle$, est\'a dada como:
\begin{equation}
|j,m\rangle,\ J_+ |j,m\rangle,\ ...,\ J_+^p|j,m\rangle \ \ \ \ \ \mbox { con}\ \ \ (p>0) \in Z 
\end{equation}
Con los correspondientes valores propios  de $J_z$
\begin{equation}
m,\ m+1,\ ...,\ m+p=j  
\end{equation}
Y tomando:
\begin{equation}
|j,m\rangle,\ J_-|j,m\rangle,\ ...,\ J_-^q|j,m\rangle \ \ \ \ \ \mbox { con}\ \ \ (q>0) \in Z 
\end{equation}
Con los correspondientes valores propios para $J_z$
\begin{equation}
m,\ m-1,\ ...,\ m-q=-j
\end{equation}
Dada la dependencia de $|j,m\rangle$ y $|j,-m\rangle$ con ($p>0$) y ($q>0$) se tiene que:
\begin{eqnarray}
p+q&=&2j  \\
\Rightarrow j&=&0,\ \ft 12,\ 1,\ \ft 32,...
\end{eqnarray}
Del \'algebra (\ref{new-alg-so-c}), se obtiene para ($m$) que:
\begin{equation}
m=0,\ \pm \ft 12,\ \pm 1,\ \pm \ft 32,\ ...,\ \pm j
\end{equation}
Con la siguiente condici\'on para $m=\pm j$:
\begin{eqnarray}
J_+|j,m\rangle&=&0 \ \ \mbox { para }\ m=j\\
J_-|j,m\rangle&=&0 \ \ \mbox { para }\ m=-j
\end{eqnarray}
Al hacer contracci\'on a $E_2$ tenemos el \'algebra {$J_0,K_1,K_2$}, que se transforma en el \'algebra {$J_0,K_+,K_-$} como:  
\begin{equation}
K_+=K_1+iK_2 \ \ \ \ \ K_-=K_1-iK_2
\end{equation}
Con conmutadores:
\begin {equation}
\left[K_+,K_-\right] = 0, \ \ 
\left[K_-,J_0 \right] = K_-, \ \ 
\left[J_0,K_+ \right] = K_+, 
\label{250}
\end {equation}
Donde el Casimir asociado toma la forma:
\begin {equation}
K.K =K_+K_-={1 \over 2}\left \{ K_+,K_- \right \} =K_-K_+
\label {260}
\end {equation}
Los valores propios y vectores propios asociados est\'an dados como:
\begin {equation}
K.K |k,m\rangle=k^2 |k,m\rangle \ \ \ \ \ , \ \ \ \ \ J_0|k,m\rangle=m |k,m\rangle
\label {270}
\end {equation}
De (\ref{270}) y (\ref{250}) se tiene que:
\begin {eqnarray}
J_0 K_+|k,m\rangle&=&K_+(J_0+1)| k,m\rangle=K_+(m+1)|k,m\rangle=(m+1)K_+|k,m\rangle\label {280}\nn\\
\Rightarrow K_+ |k,m\rangle &\equiv& \bar a_m | k,m+1\rangle \label {290}
\end {eqnarray}
De igual forma para $K_-$, usando las ecuaciones (\ref{270}) y (\ref{250}) tenemos que:
\begin {eqnarray}
J_0 K_-|k,m\rangle&=&K_-(J_0-1)| k,m\rangle=K_-(m-1)|k,m\rangle=(m-1)K_-|k,m\rangle\label {290-1}\nn\\
\Rightarrow K_- |k,m\rangle &\equiv& \bar b_m | k,m-1\rangle \label {300}
\end {eqnarray}
Para conocer $\bar a_m, \bar b_m$. De (\ref{260},\ref{290}) y (\ref{300}) se obtiene: 
\begin {eqnarray}
K.K|k,m\rangle=K_+K_-|k,m\rangle=\bar b_m K_+|k,m-1\rangle=\bar b_m \bar a_m |k,m\rangle=k^2|k,m\rangle
\end {eqnarray}
Por lo tanto:
\begin {equation}
\bar b_m \bar a_m =k^2
\end {equation}
Ademas:
\begin {equation}
\langle k,m | K_-K_+ |k,m\rangle=\langle k,m|K_+]^\dagger [K_+ |k,m\rangle=[\langle k,m|\bar a_m]^\dagger [\bar a_m|k,m\rangle] \geq 0 \label {300-1}
\end {equation}
Dado que $(\langle k,m|k,m\rangle) > 0$ $\Rightarrow$ $k > 0$. De igual forma se obtiene que:
\begin {eqnarray}
\bar a_m=\bar b_m = k
\label{sol-am_bm}
\end {eqnarray}
Los valores propios $(m)$ y vectores propios $|k,m\rangle$ para $J_0$ en $E_2$ conservan la misma estructura asociada a $J_3$ en $SO(3)$, adem\'as el valor propio $(m)$ toma un valor libre e independiente de $k$ al hacer contracci\'on. Esto se puede ver usando la ecuaci\'on (\ref{230}), debido a:
\begin{equation}
j \to j(R)\equiv j_R \ \ \ \mbox{ con }\ \ \ \lim_{R \to \infty} j_R=\infty
\end{equation}
Donde:
\begin {equation}
\lim_{R \to \infty}{J_{\pm}\over R}=K_{\pm} \ \ \ \ \ \mbox{y} \ \ \ \ \ \lim_{R \to \infty}|j_R,m\rangle =|k,m\rangle
\end {equation}
Esto permite definir una nueva base indexada como:
\begin{equation}
\lim_{R \to \infty}{j_R \over R}= k
\label {360}
\end {equation}
Se puede notar de la ecuaci\'on (\ref{360}) y de (\ref{susti-so3}) que:
\begin{equation} 
j\sim R\ k \ \ \ \Rightarrow \ \ \ \hbar\ j\sim R\ (\hbar k)=R\ p
\end {equation}
An\'alogo a la norma del momento angular cl\'asico $L=R\times P$. En el l\'{\i}mite de contracci\'on se tiene que:
\begin {eqnarray}
K.K|k,m\rangle= K_+K_-|k,m\rangle=k^2|k,m\rangle&=&\lim_{R\to\infty}{J_+J_-\over R^2}\lim_{R \to \infty}|j_R,m\rangle\nn\\
&=& \lim_{R \to \infty}{j_R (j_R+1)-m(m-1) \over R^2} |j_R,m\rangle
\label {350}
\end {eqnarray}
Usando (\ref{360}) en (\ref{350}):
\begin {equation}
\lim_{R \to \infty} \left({j_R \over R}\right)^2 |j_R,m\rangle =k^2|k,m\rangle 
\label {370}
\end {equation}

\subsection {Contracci\'on de $SO(3,1) \to {\bf G}_3 \equiv K_3 \wedge SO(3)$}
Considerando el grupo de Lorentz homogeneo, este tiene 6 generadores $J_r,B_r$ $(r=1,2,3)$. Estos generadores obedecen el siguiente \'algebra:   
\begin {equation}
\left[B_r,B_s \right] =-i\e_{rs}^t J_t \ , \ \ \ 
\left[J_r,B_s \right]=i\e_{rs}^t B_t \ ,\ \ \ 
\left[J_r,J_s \right]=i\e_{rs}^t J_t \ \ \ \ \ (r,s,t)=1,2,3
\label{xyz}
\end {equation}
$B_r$ representa los boosts de Lorentz, $J_r$ los generadores de rotaci\'on. Las rotaciones espaciales se muestran aqu\'{\i} expl\'{\i}citamente, las cuales forman un subgrupo. Se puede escribir (\ref{xyz}) de una forma compacta como:
\begin{equation}
\left[ J_{\m \n} ,J_{\r \s} \right] =i(\h_{\n\r}J_{\m\s}-\h_{\m \r} J_{\n \s}+\h_{\m\s}J_{\n \r}-\h_{\n\s}J_{\m\r})\ \ \ \ \ \h_{\m\n}\equiv diag(+++-)\ \ \ (\m,\n=1,2,3,4)
\label{compacta}
\end{equation}
$J_{\m\n}$ se define como:
\begin {equation}
J_{\m\n}\ \Rightarrow \ \ \  
\left\{
\begin {array}{cll} 
J_{\m\n}&\equiv \e_{\m\n}^{\s} J_\s &\mbox {si }\ \ \ \m,\n,\s=1,2,3\\
J_{4\n}&\equiv B_\n & \mbox {si }\ \ \ \m=4 \ \ \mbox{ y }\ \ \n=1,2,3
\end{array}
\right\}\ \ \ \ \ \mbox{ donde }\ 
J_{\m\n}=-J_{\n\m} 
\end {equation}
Los Casimir del grupo de Lorentz propio est\'an dados como:
\begin{equation}
C_2^{1,3}\equiv{1\over 2}J^{\m \n}J_{\m\n}=J^rJ_r-B^rB_r,\ \ \ \ \ \ \ 
C_4^{1,3}\equiv{1\over 2}J^{\m \n}\tilde J_{\m\n}=J^rB_r+B^rJ_r=2(J^rB_r) 
\label{cas-lor}\\
\end{equation}
El dual de $J_{\m\n}$ es definido como:
\begin{equation}
\tilde J^{\m \n}\equiv {1\over 2}\e^{\m\n\r\s}J_{\r\s}
\end{equation}
$\e^{\m\n\r\s}$ representa un tensor completamente antisim\'etrico ($\e^{1234}=-1$).\\
\\ 
Para hacer la contracci\'on $(c \rightarrow \infty)$ en (\ref{xyz}), se define:
\begin {eqnarray}
J_r&=& J_r \label{u}\\
\O_r&\equiv& {1 \over ic} B_r \ \ \ \ \ \ \ \ G_r \equiv \lim_{c \to \infty} \O_r\ \ \ \ \ \ \ \ r=1,2,3  
\label{v}
\end {eqnarray}
Al tomar el l\'{\i}mite cuando $(c \rightarrow \infty)$, las relaciones dadas por (\ref{xyz}) toman la forma:
\begin {eqnarray}
\left[ J_r ,J_s \right] &=& i\e_{rs}^t J_t\nn\\
\left[ J_r ,G_s \right] &=& i\e_{rs}^t G_t\nn\\
\left[ G_r ,G_s \right] &=& 0 \label{n}
\end {eqnarray}
Teni\'endose que estas relaciones definen localmente el grupo de Galileo homogeneo ${\bf G}_3$, el cual contiene las transformaciones de cambio de sistemas de referencia en movimiento $G_r$ y las rotaciones espaciales $J_r$. Por supuesto este grupo es isomorfo al grupo Euclidiano en 3-Dim $E_3=T_3\wedge SO(3)$. Al usar la sustituci\'on dada por (\ref{v}) en los Casimir de Lorentz dados por (\ref{cas-lor}), redefiniendo los Casimir y tomando el l\'{\i}mite, se encuentra:     
\begin{equation}
C_2^3\equiv \lim_{c\to \infty}{C_2^{1,3}\over c^2}=\lim_{c\to \infty}{1\over 2c^2}J^{\m \n}J_{\m\n}=\lim_{c\to \infty}(J^rJ_r/c^2+\O^r\O_r)=G^rG_r=G.G
\label{cas-gal}
\end{equation}
y
\begin{equation}
C_4^3\equiv \lim_{c\to \infty}{C_4^{1,3}\over ic}=\lim_{c\to \infty}{1\over ic}J^{\m \n}\tilde {J}_{\m\n}=\lim_{c\to \infty}2(J^r \O_r)=2(J^rG_r)=2(J.G)
\label{cas-gal-1}
\end{equation}
De manera similar a $T_2$ en la secci\'on (\ref{so3}), los n\'umeros $g$ y $m$ son independientes y no hay una restricci\'on para $m$ dependiente de $g$. 

\subsubsection {Representaciones Finitas Asociadas y su Esquema de Contracci\'on} 
Se define a continuaci\'on el siguiente conjunto de operadores en el grupo de Lorentz propio, como:
\begin {equation}
{\cal A}_r^- \equiv {1\over2}(J_r-iB_r) \ \ \ \ \ , \ \ \ \ \ {\cal A}_r^+ \equiv {1\over2}(J_r+iB_r)
\end{equation}
$({\cal A}_r^\pm)^{\dagger}={\cal A}_r^{\pm}$, siendo $J^\dagger=J$ y $B^\dagger=-B$, donde:
\begin {equation}
\left[ {\cal A}_r^-,{\cal A}_s^-\right] = i\e_{rs}^t {\cal A}_t^-, \ \ 
\left[ {\cal A}_r^+,{\cal A}_s^-\right] = 0, \ \ 
\left[ {\cal A}_r^+,{\cal A}_s^+\right] = i\e_{rs}^t {\cal A}_t^+, \ \ 
\label{lo-new-alg-so-c}
\end {equation}
De (\ref{lo-new-alg-so-c}) puede notarse que esta nueva \'algebra conlleva a un isomorfismo entre el \'algebra de $SO(3,1)$ y el \'algebra de $SO(3)\times \bar{SO}(3)$. Este conjunto de generadores se puede escribir como ${\cal A}_r^-\equiv \bar {\cal A}_r \otimes I$ y ${\cal A}_r^+ \equiv I \otimes {\cal A}_r$ actuando sobre el espacio producto directo $\n_{j_1} \otimes \n_{j_2}$, los cuales corresponden a dos espacios vectoriales independientes bajo transformaciones del grupo de Lorentz. La representaci\'on del grupo de Lorentz puede ser etiquetada como $[j_1 \otimes j_2]$, donde la dimensi\'on de esta irrep es $(2j_1+1)(2j_2+1)$. Usando resultados del algebra de $SO(3)$, los Casimirs de estas \'algebras que conmutan, est\'an dados como:
\begin{equation}
{\cal A}^-.{\cal A}^-={\cal A}^{-\ r}{\cal A}_r^-,\ \ \ \ \   
{\cal A}^+.{\cal A}^+={\cal A}^{+\ r}{\cal A}_r^+, 
\label {lo1-150}
\end {equation}
Con sus correspondientes valores propios $j_1(j_1+1)$ y $j_2(j_2+1)$ respectivamente. Donde los Casimirs del grupo de Lorentz pueden ser expresados en t\'erminos de estos como:
\begin{equation}
C_2^{1,3}=2({\cal A}^- .{\cal A}^-+{\cal A}^+.{\cal A}^+),\ \ \ \ \ C_4^{1,3}=2i({\cal A}^-.{\cal A}^--{\cal A}^+.{\cal A}^+)    
\label {lo-150}
\end {equation}
Con los siguientes valores propios asociados actuando sobre $|j_1m_1,j_2m_2\rangle$:
\begin{eqnarray}
C_2^{1,3} &\Rightarrow& 2[j_1(j_1+1)+j_2(j_2+1)]=2(j_1+j_2+1)(j_1+j_2)-4j_1j_2 \label {loA-150}\\
C_4^{1,3} &\Rightarrow& 2i[j_1(j_1+1)-j_2(j_2+1)]=2i(j_1+j_2+1)(j_1-j_2) \label {loB-150}
\end {eqnarray}
Construyendo los vectores base acoplados $|j_1m_1,j_2m_2\rangle$ como un producto directo de vectores base de $SO(3)$. Donde $\{{\cal A}_r^-\}$ y $\{{\cal A}_r^+\}$ forman 2 conjuntos de generadores independientes de $SO(3)$ (ver \cite{Haase}). Por lo tanto se puede definir los operadores vectoriales esf\'ericos $A_\m$ y $\bar A_\m$ con $\m=(+,0,-)$ para cada conjunto de generadores, tal como en la secci\'on (\ref{so3}):
\begin {eqnarray}
A_+\equiv {\cal A}_1^+ + i {\cal A}_2^+ \ \ \ \ \ \ \ \ \ A_-\equiv{\cal A}_1^+ - i {\cal A}_2^+ \ \ \ \ \ \ \ \ \ A_0\equiv{\cal A}_3^+ 
\end {eqnarray}
La base para $A_\m$ es dada como en la secci\'on (\ref{so3}), etiquetada con los n\'umeros $\{j_2,m_2\}$: 
\begin {eqnarray}
A_{\pm}|j_2m_2\rangle&=&N_{\pm}(j_2m_2)|j_2m_2\pm 1\rangle\ \ \ \ \ \mbox {con} \ \ N_{\pm}(j_2m_2)=\sqrt{j_2(j_2+1)-m_2(m_2\pm 1)}\nn\\
A_0|j_2m_2\rangle&=&m_2|j_2m_2\rangle
\end {eqnarray}
De manera similar se definen los generadores $\bar A_\m$ para el conjunto de generadores $\{{\cal A}_r^-\}$, los cuales est\'an etiquetados con los n\'umeros $\{j_1,m_1\}$.
Por otro lado para conocer la acci\'on de $J_r$ y $B_r$ sobre $|j_1m_1,j_2m_2\rangle$ siguiendo el desarrollo para el \'algebra de $SO(3)$ como en la secci\'on (\ref{so3}), tenemos que:
\begin {equation}
J_r={\cal A}_r^-+{\cal A}_r^+ \ \ \ \ \ \ \ B_r=i({\cal A}_r^--{\cal A}_r^+)
\end {equation}
Por lo tanto:
\begin {eqnarray}
J_r|j_1m_1,j_2m_2\rangle&=&{\cal A}_r^-|j_1m_1,j_2m_2\rangle+{\cal A}_r^+ |j_1m_1,j_2m_2\rangle\nn\\  
B_r|j_1m_1,j_2m_2\rangle&=&i({\cal A}_r^-|j_1m_1,j_2m_2\rangle-{\cal A}_r^+ |j_1m_1,j_2m_2\rangle)
\end {eqnarray}
En particular, se tiene para $J_3$ y $B_3$.
\begin {eqnarray}
J_3|j_1m_1,j_2m_2\rangle&=&{\cal A}_3^-+{\cal A}_3^+|j_1m_1,j_2m_2\rangle
=(m_1+m_2)|j_1m_1,j_2m_2\rangle \nn\\  
B_3|j_1m_1,j_2m_2\rangle&=&i({\cal A}_3^--{\cal A}_3^+)|j_1m_1,j_2m_2\rangle=i(m_1-m_2)|j_1m_1,j_2m_2\rangle
\label{j3b3}
\end {eqnarray}
Todo el an\'alisis para $SO(3,1)$ puede ser seguido de los resultados ya obtenidos para $SO(3)$ teniendo en cuenta que $SO(3,1)\sim SO(3)\times \bar{SO}(3)$ en su \'algebra. Al hacer contracci\'on a $G_3\sim E_3=T_3 \wedge SO(3)$ tenemos el \'algebra asociada {$J_r,G_r$} con $r=1,2,3$. Donde el Casimir est\'a dado por:
\begin {equation}
C_2^3=G.G=G^rG_r=G_1^2+G_2^2+G_3^2
\label {lo-260}
\end {equation}
Los valores y vectores propios asociados est\'an dados como:
\begin {equation}
G.G |g,m,s\rangle=g^2|g,m,s\rangle \ \ \ \ \ \ \ J_3|g,m,s\rangle=m|g,m,s\rangle \ \ \ \ \ \ \ G_3|g,m,s\rangle=s|g,m,s\rangle
\label {lo-270}
\end {equation}
Los generadores $\{J_r\}\in G_3$ conservan la misma estructura de $\{J_r\}\in SO(3,1)$. Donde $m_a\ \ (a=1,2)$ toma un valor libre independiente de $g$ al hacer contracci\'on, teniendo en cuenta la ecuaci\'on (\ref{230}) en $SO(3)$, debido a:
\begin{equation}
j_a \to j_a(c)= j_{a,c} \ \ \ \mbox{ con }\ \ \ \lim_{c \to \infty} j_{a,c}=\infty \ \ \ \ \ \ \ a=1,2
\end{equation}
Del Casimir cuadr\'atico al hacer contracci\'on, usando las ecuaciones (\ref{v}) y (\ref{cas-gal},\ref{loA-150},\ref{lo-270}), se obtiene que:
\begin {equation}
{g^2\over 2} = \lim_{c\to\infty} \left [{j_{1,c}^2+j_{2,c}^2+j_{1,c}+j_{2,c}\over c^2}\right ]
\end {equation}
Pero al tener en cuenta el Casimir cu\'artico, de (\ref{cas-gal-1}) y (\ref{loB-150}) se tiene que: 
\begin{equation}
{2i(j_{1,c}+j_{2,c}+1)(j_{1,c}-j_{2,c})\over ic}=2c \left ({j_{1,c}^2+j_{2,c}^2+j_{1,c}+j_{2,c}\over c^2}-2{j_{2,c}^2+j_{2,c}\over c^2}\right )
\label{c-c4-1}
\end{equation}
Y al tomar el l\'{\i}mite de contracci\'on:
\begin{equation}
\lim_{c\to\infty}{2i(j_{1,c}+j_{2,c}+1)(j_{1,c}-j_{2,c})\over ic}=(g^2-2\b)\lim_{c\to\infty}c= \left\{\begin {array}{cl} (g^2-2\b<\infty)\infty &\to \infty\\ (g^2-2\b=0)\infty &\to 0 \end{array}\right\}
\ \ \ \b=\lim_{c\to\infty}{2(j_{2,c}^2+j_{2,c})\over c^2}
\label{c-c4-2}
\end{equation}
Por lo tanto bajo la contracci\'on, se presentan 2 posibilidades para la representaci\'on del Casimir cu\'artico:
\begin{equation}
2(J.G)=\lim_{c\to\infty}2(J.\O) \to \left\{\begin {array}{cc}\infty&\ J \parallel G \\ 0&\ J\perp G \end{array}\right\}
\end{equation}
El caso de interes f\'{\i}sico se presenta cuando $2(J.G)=0$ (operador nulo). Esto tiene soluci\'on si $j_{1,c}=j_{2,c}$ (ver ecuaci\'on (\ref{c-c4-2})). Adicionalmente se tiene que $2(J.G)$ ya no es un Casimir dado que $[J.G,G]\neq 0$, entonces se define una nueva base indexada como: 
\begin {equation}
\lim_{c \to \infty}{j_{1,c}\over c}=\lim_{c \to \infty}{j_{2,c}\over c}={g \over 2}
\label {lo-400}
\end {equation}
De las ecuaciones (\ref{v},\ref{lo-270}) y (\ref{j3b3}):
\begin{equation}
\begin{array}{c}m_1\to m_1(c)=m_{1,c} \\ m_2\to m_2(c)=m_{2,c}\end{array} \ \ \ \ \ \ \ 
\lim_{c \to \infty} {m_{1,c}+m_{2,c}}= m \ \ \ \ \ \ \ \lim_{c \to \infty} {(m_{1,c}-m_{2,c}) \over c}=s 
\end{equation}
Una interpretaci\'on lineal a esto se puede dar como:
\begin{equation}
\begin{array}{c}m_{1,c}=a_1c+b_1\\m_{2,c}=a_2c+b_2\end{array}\ \ \ \ \ \ \ \mbox{con }\ \ \ a_1=-a_2={s\over 2}\ \ \ y \ \ \ b_1+b_2=m\end{equation}
Entonces
\begin{equation} 
\lim_{c \to \infty} {m_{1,c}+m_{2,c}}=\lim_{c \to \infty} {(a_1+a_2)c+(b_1+b_2)}= m \ \ \ \ \ \ \ \lim_{c \to \infty} {(m_{1,c}-m_{2,c}) \over c}=\lim_{c \to \infty} {(a_1-a_2)c+(b_1-b_2)\over c}=s 
\end{equation}
Adem\'as
\begin {equation}
\lim_{c \to \infty}|j_{1,c}m_{1,c},j_{2,c}m_{2,c}\rangle=|g,m,s\rangle
\end {equation}
Por lo tanto en el l\'{\i}mite de contracci\'on se tiene que:
\begin {eqnarray}
G.G|g,m,s\rangle&=&\lim_{c\to\infty}{C_2^{1,3}\over c^2}\lim_{c\to\infty}|j_{1,c}m_{1,c},j_{2,c}m_{2,c}\rangle\nn\\
&=& \lim_{c \to \infty}{2(j_{1,c}+j_{2,c}+1)(j_{1,c}+j_{2,c})-4j_{1,c}j_{2,c} \over c^2} |j_{1,c}m_{1,c},j_{2,c}m_{2,c}\rangle
\label {lo-350}
\end {eqnarray}
Y usando (\ref{lo-400}) en (\ref{lo-350}), se obtiene:
\begin {equation}
\lim_{c \to \infty} {2(j_{1,c}^2+j_{2,c}^2) \over c^2}|j_{1,c}m_{1,c},j_{2,c}m_{2,c}\rangle=g^2|g,m,s\rangle 
\label {lo-370}
\end {equation}
Para $B_3$ en el l\'{\i}mite de contracci\'on se tiene que:
\begin {eqnarray}
G_3|g,m,s\rangle&=& \lim_{c \to \infty}{B_3\over ic}|j_{1,c}m_{1,c},j_{2,c}m_{2,c}\rangle\nn\\
&=& \lim_{c \to \infty}{(m_{1,c}-m_{2,c})\over c} |j_{1,c}m_{1,c},j_{2,c}m_{2,c}\rangle=s|g,m,s\rangle 
\end {eqnarray}
De aqu\'{\i} se puede obtener la contracci\'on $P(3,1)\equiv T_{3,1}\wedge SO(3,1)\to G(3,1)\equiv T_3\wedge G_3$ con $G_3\equiv K_3\wedge SO(3)$. Dado que esta solo afecta a los generadores de cambio de sistema de referencia o `boosts', la contracci\'on se restringe a $SO(3,1)\to G_3$, lo cual se acaba de mostrar.

\section {El Grupo de Sitter/Anti-de Sitter}\label{anti/de-sitter}
Los grupos (Anti) de Sitter, no es algo nuevo, ya que en la d\'ecada de los treinta, Dirac consider\'o ecuaciones de onda que eran invariantes bajo grupos Anti-de Sitter\cite{Dirac1}. Tiempo despues Fronsdal y sus colaboradores discutieron sus representaciones asociadas\cite{Fronsdal}.\\
\\
En modelos cosmol\'ogicos son los \'unicos que pueden representar un universo curvo y uniforme en el espacio-tiempo (espacios Riemannianos con constante de curvatura no nula). Estos se clasifican segun su curvatura: 
\begin {itemize}
\item Si la curvatura es positiva, corresponde al espacio (4,1) de Sitter ($dS$), representado como una superficie 4-Dim de una hiperesfera en un espacio plano 5-Dim con m\'etrica $\h_{ij}=diag(++++-)$    
\item El caso de curvatura negativa (3,2) o espacio Anti-de Sitter ($AdS$) corresponde a una hiperesfera en un espacio con m\'etrica $\h_{ij}=diag(+++--)$   
\end {itemize}
Ambos espacios describen un universo en expansi\'on, donde las lineas de universo corresponden a las geod\'esicas de movimiento con velocidad radial, proporcional a la distancia radial desde cualquier punto del espacio.\\
\\
Es posible generar el grupo de Poincar\'e $P(3,1)$ como contracci\'on (en el l\'{\i}mite de curvatura a cero) del grupo $SO(5-h,h)$ (con $h=1$ o $2$) de una manera no trivial \cite{Gradechi}, en la misma forma como es contraido el grupo de Lorentz al grupo de Galileo \cite{Haase}. Donde el \'algebra de $SO(5-h,h)$ genera el \'algebra de Poincar\'e  tomando el l\'{\i}mite $R\rightarrow \infty$ y manteniendo $K_\m$ constante en la relaci\'on $J_{z \m}=R_z K_\m=-J_{\m z}$ con $z=1,5$.
\begin {itemize}
\item Si $z=1$, $SO(4,1)$: se obtiene una contracci\'on `espacial' al tomar $R\rightarrow \infty$ en la relaci\'on $J_{1 \m}=R_1 K_\m$.
\item Si $z=5$, $SO(3,2)$: se obtiene una contracci\'on `temporal' al tomar $R\rightarrow \infty$ en la relaci\'on $J_{5 \m}=R_5 K_\m$. 
\end {itemize}
Aqu\'{\i} se presenta el grupo de Sitter/Anti-de Sitter, sus propiedades b\'asicas y dado que este es un grupo no-compacto (como se sabe no es posible encontrar una representaci\'on de dimensi\'on finita unitaria) se dan las condiciones de unitaridad, \'utiles para construir las representaciones supersim\'etricas. Por otro lado se muestra su contracci\'on al grupo Poincar\'e, donde el par\'ametro de contracci\'on puede ser el radio de curvatura $R$ (o la constante de curvatura $a$) del espacio-tiempo $AdS$, con $R\to \infty$ ($a\to 0$).\\

\subsection {El Grupo $SO(5-h,h)$ y su Contracci\'on a $ P(3,1)\equiv T_{3,1}\wedge SO(3,1)$}\label{sittertopoincare}
En general el grupo $SO(5-h,h)$ ($h=1$ o $2$) corresponde al grupo de simetr\'{\i}as maximal en un espacio $dS$/$AdS$ que se puede  describir como una hipersuperficie embuida en un espacio $5$-dimensional. Usando las coordenadas $y^i$ con $i= 1,2,3,4,5$, la hipersuperficie es definida por: 
\begin{equation}
(y^1)^2+(y^2)^2 +(y^3)^2 \pm (y^4)^2 - (y^5)^2 =\h_{ij} \,y^i y^j=-a^{-2}\ \ \ \ \ \mbox{ con} \ \ \ a\equiv {1\over R}
\end{equation}
Donde $y^5$ representa la coordenada extra. Es claro que la hipersuperficie es invariante bajo transformaciones lineales que preserven la  m\'etrica $\h_{ij}  =diag(+++\pm-)$. Estas transformaciones constituyen el grupo de Sitter/Anti-de Sitter $SO(5-h,h)$ con ${1\over 2}5(5-1)=10$ generadores denotados por $J_{ij}$ que representan los operadores de rotaci\'on en $E_5$. Estos satisfacen el \'algebra:
\begin{equation}
\left[ J_{ij} ,J_{kl} \right] = i(J_{ik}  \h_ {jl}-J_{il}  \h_ {jk}+J_{jl}  \h_ {ik}-J_{jk}  \h_ {il}) \ \ \ \ \ \ \ \hbox{con }\ \ \ \{i,j,k,l,m\}=1,2,3,4,5
\label{sitter}
\end {equation}
Los dos Casimirs de Sitter/Anti-de Sitter son: 
\begin {equation}
C_2 \equiv {1\over 2}J_{ij} J^{ij} \ \ \ \ \ \ \ \ \ \ C_4 \equiv W^iW_i
\label{cas-sitter}
\end {equation}
Donde $W_i$ representa un vector 5-Dim definido como: 
\begin {equation}
W_i \equiv {1\over 2}\var_{ijklm} J^{jk}J^{lm}
\label{pauli}
\end {equation}
$\var$ es un tensor totalmente antisim\'etrico con 5 \'{\i}ndices, siendo $J^{ij}=J_{kl}\h^{ik}\h^{jl}$. Para la contracci\'on se define:
\begin {equation}  
\P_{\m}\equiv {1\over R_z} J_{z\m} \ \ \ \ \mbox {con}\ z=1\ \ \mbox {o}\ \ 5 
\label{7}
\end {equation}
Reescribiendo (\ref {sitter}) en la forma 4-Dim usando (\ref {7}), se obtiene:
\begin {equation}
\left[ J_{\m \n} ,J_{\r \s} \right] = i(J_{\m \r}  \h_ {\n \s}-J_{\m \s}  \h_ {\n \r}+J_{\n \s}  \h_ {\m \r}-J_{\n \r}  \h_ {\m \s}) 
\label{a}
\end {equation}
\begin {equation}
\left[ \P_{\m} ,J_{\r \s} \right] = i(\P_{\r}  \h_ {\m \s}-\P_{\s}  \h_ {\m \r})\ \ \ \ \ \ \ \ \ \ \ \ \ \ \ \left[ \P_{\m} ,\P_{\n} \right] = {i \over R^2_z} J_{\m \n} 
\label{b}
\end {equation}
Con $\{\m,\n,\r,\s,\l\} = 1,2,3,4$. En el l\'{\i}mite de contracci\'on se define:
\begin {equation}
\lim_{R_z \rightarrow \infty} \P_{\m}\equiv K_{\m}
\label{8}
\end {equation}
donde $K_{\m}$ denota el operador de translaci\'on en el espacio-tiempo plano. Aqu\'{\i} se puede ver que una rotaci\'on en el plano $(x_1,x_{\m})$ o $(x_{\m},x_5)$ se transforma en una traslacci\'on espacio-temporal en el l\'{\i}mite de curvatura cero. Obteni\'endose los generadores del grupo Poincar\'e $K_{\m}$, $J_{\n \s}$ bajo la contracci\'on. Donde el \'algebra que este satisface, coinciden con (\ref {a}) y al tomar $\lim_{R_z \rightarrow \infty}$ en (\ref {b}) se recuperan las siguientes relaciones:
\begin {equation}
\left[ J_{\m \n} ,J_{\r \s} \right] = i(J_{\m \r}  \h_ {\n \s}-J_{\m \s}  \h_ {\n \r}+J_{\n \s}  \h_ {\m \r}-J_{\n \r}  \h_ {\m \s}) \label{a-1}
\end {equation}
\begin {equation}
\left[ K_{\m} ,J_{\r \s} \right] = i(K_{\r}  \h_ {\m \s}-K_{\s}  \h_ {\m \r})\ \ \ \ \ \ \ \ \ \ \ \ \ \ \ \ \ \left[ K_{\m} ,K_{\n} \right] = 0
\label{b-1}
\end {equation}
Reescribiendo los invariantes `de Sitter' en (\ref{cas-sitter}) usando (\ref{7}), se tiene para el primer invariante:
\begin {equation}
C_2 ={1\over 2}J_{ij} J^{ij}= R_z^2 \P_{\m}\P^{\m}+{1\over 2}J_{\m \n} J^{\m \n}
\label{10}
\end {equation}
Por lo tanto al tomar el l\'{\i}mite de curvatura cero ($\lim_{R_z \rightarrow \infty}$). Se obtiene el primer invariante de Poincar\'e:   
\begin {equation}
C^{3,1}_2 \equiv \lim_{R_z \rightarrow \infty} {C_2 \over R_z^2}= K_{\m}K^{\m}
\label{11}
\end {equation}
Tomando un procedimiento similar usando (\ref{7}) en (\ref{pauli}), el segundo invariante `de Sitter' toma la forma:
\begin {eqnarray}
C_4 = W_i W^i &=& {1\over 4}R_z^2\var_{\l z \r\m \n} \P^{\r}J^{\m \n} \var^{\l z \r' \m' \n'} \P_{\r'}J_{\m' \n'} 
+ 
{1 \over 8}\var_{z \m\n\r\s}J^{\m\n} J^{\r\s}
\var_{z \m'\n'\r'\s'}J^{\m'\n'} J^{\r'\s'}\nn\\
&=&{1\over 4}R_z^2\var_{\l z \r\m \n} \P^{\r}J^{\m \n} \var^{\l z \r' \m' \n'} \P_{\r'}J_{\m' \n'} 
+ 
{1 \over 4} J_{\m \n} J^{\m \n} J_{\m' \n'} J^{\m' \n'} 
\label{13}
\end {eqnarray}
Al desarrollar el primer t\'ermino del lado derecho usando (\ref{b}) obtenemos:
\begin {eqnarray}
{1\over 4}R_z^2\var_{\l z\r\m \n} \P^{\r}J^{\m \n} \var^{\l z \r' \m' \n'} \P_{\r'}J_{\m' \n'}&=&R_z^2\var_{\r\m \n} \P^{\r}J^{\m \n} \var^{ \r' \m' \n'} \P_{\r'}J_{\m' \n'}\nn\\ 
&=& R_z^2 \P^{\r}J^{\m \n} \P_{\r}J_{\m \n} - \P^\r J^{\m\n} \P_{\m}J_{\n\r} \nn\\ 
&=& R_z^2 J^{\m\n}J_{\m\n} \P^\r \P_\r - J^{\n\r}\P_{\n} J_{\m\r}\P^{\m}
\end {eqnarray}
El segundo t\'ermino del lado derecho en la ecuacion (\ref{13}) desaparece al tomar el l\'{\i}mite de curvatura cero ($\lim_{R_z \rightarrow \infty}$), por lo tanto el segundo invariante de Poincar\'e toma la siguiente forma:
\begin {equation}
C^{3,1}_4 \equiv \lim_{R_z \rightarrow \infty} {C_4\over R_z^2} = 
J^{\m\n}J_{\m\n}K^\r K_\r - J^{\n\r}K_{\n}J_{\m\r}K^{\m}=(J:J)(P.P)-(J.P).(J.P)
\label{cas4-sitter}
\end {equation}
Este invariante puede ser llevado a la siguiente forma:
\begin {equation}
C^{3,1}_4 =W^\m W_\m \ \ \ \ \ \ \ \mbox{ con } \ \ W_\m = {1 \over 2 } \var_{\m\n\r\s} K^\n J^{\r\s}
\label{c4}
\end {equation} 
Aqu\'{\i} el vector $W_\m$ representa el vector de Pauli-Lubanski. El cual se puede escribir en t\'erminos de los generadores de momento angular y los `boosts': 
\begin {equation}
W_4= J.K \ \ \ \ \ \ \ W=-J K_4 - B \times K
\end {equation} 
El cual cumple el siguiente \'algebra:
\begin {equation}
\left [ W_\m,J_{\r\s} \right ]= i(\h_{\m\r}W_\s-\h_{\m\s}W_\r)\ \ \ \ \ \ \ 
\left [ W_\m,W_{\n} \right ]= i W^\r K^\s \var_{\r\s\m\n}\ \ \ \ \ \ \ 
\left [ W_\m,K_\n \right ]=0 \ \ \ \ \ \ \ W.K=0
\end {equation} 
De aqu\'{\i} que $W_\m$ conmuta con $K_\n$ y es ortogonal a este mismo. Dado el prop\'osito de este trabajo y teniendo en cuenta los resultados en la secci\'on (\ref{superadS}), se enfocar\'a nuestro inter\'es en las representaciones unitarias del \'algebra Anti-de Sitter, util en la \'ultima secci\'on. En lo que sigue: como complemento ver \cite{Heidenreich,GunaNieuWarn,Nicolai,dewitt}

\subsection {Representaciones Unitarias del \'Algebra Anti-de Sitter}\label{representacion-unitaria-so32} 
Al tomar los generadores $J_{ij}$ de $SO(3,2)$, estos permiten una representaci\'on en forma espinorial dada por las matrices gamma $(\langle ..|J_{ij}|..\rangle=\G_{ij})$, las cuales son una representaci\'on de dimensi\'on finita: 
\begin{equation}
J_{ij} \Rightarrow \ \ \ \left\{ 
\begin{array}{lll}
\ft 12 \G_{ij}&\mbox{para} & i,j=1,2,3,4\\[3mm]
\ft 12\G_j  &\mbox{para} & i=5\,, j= 1,2,3,4 \end{array}\right\}
\ \ \ \ \ \mbox{ donde }\ \ J_{ij} =-J_{ji} 
\end{equation}
Estas satisfacen la propiedad de Clifford $\{\G^\m\,,\, \G^\n\} = 2\, \h^{\m\n}\, {\bf 1}$, con $\h^{\m\n} = {\rm diag}\,(+,+,+,-)$. Por lo tanto forman una representaci\'on del grupo de cubertura $Sp(4)$ de $SO(3,2)$. Donde $Sp(n)$ se define como el conjunto de operadores unitarios ($J^\dagger J=I$) que cumplen $M^\dagger JM=J$ con $M^\dagger M=I$ y $J=\s\otimes I$, donde $J^2=-I$. Este posee $\ft 12 (n+1)n$ generadores. En esta representaci\'on no todos los generadores son unitarios, por ejemplo: 
\begin {eqnarray}
J_{ij}&=&J_{ij}^\dagger \ \ \ \ \hbox{ para }\ J_{rs},\ J_{45} \ \ \ \hbox{ con }\ r,s= 1,2,3 \nn\\
J_{ij}&=&-J_{ij}^\dagger \ \ \ \hbox{para }\ J_{4r},\ J_{r5}  
\label{nohermitico}
\end {eqnarray}
La anterior relaci\'on es consecuencia de no poder obtener una representaci\'on unitaria de dimensi\'on finita para un grupo no-compacto. Para obtener la hermiticidad de la representaci\'on en los generadores $J_{ij}=J_{ij}^\dagger$, se requiere una representaci\'on de dimensi\'on infinita la cual si tiene relevancia f\'{\i}sica. De la ecuaci\'on (\ref{nohermitico}) se distingue los generadores compactos $J_{45}$ y $J_{rs}$ de los no-compactos $J_{r5},J_{4r}$. Esta clasificaci\'on es v\'alida tambi\'en en una representaci\'on de dimensi\'on infinita. Los operadores $J_{rs}$ y $J_{45}$ generan un sub\'algebra compacta maximal asociada al grupo $SO(3)\times SO(2)$. Los generadores $J_{rs}$ satisfacen el \'algebra usual de momento angular.
\begin {equation}
J_{rs}=\e_{rs}^t J_{t} \ \ \ \hbox{ con }\ r,s,t= 1,2,3 
\label {mo-angular}
\end {equation}
Se define ahora los siguientes operadores en $SO(3,2)$ como:\footnote{se asume desde ahora que $J_{ij}=J_{ij}^\dagger$ en su representaci\'on matricial} 
\begin {equation}
M_{r}^+= i J_{4r}+ J_{5r} \ \ \ \ \ 
M_{r}^-= i J_{4r}- J_{5r} \ \ \ \ \ \ \ \ \ \ \ \ 
M_{r}^-=-(M_{r}^+)^\dagger  \ \ \ \ \ r=1,2,3 
\label{boost}
\end {equation}
Con un \'algebra asociada:
\begin {equation}
\left [M_{r}^+,M_{s}^- \right ]= 2(\d_{rs} J_{45} + iJ_{rs}) \ \ \ \ \ \ \ 
\left [M_{r}^+,M_{s}^+ \right ]= \left [M_{r}^-,M_{s}^- \right ]= 0
\label{conm-pm}
\end {equation}
Tambi\'en se tiene que:
\begin {equation}
\left [J_{45},M_{r}^+ \right ]= M_{r}^+ \ \ \ \ \ \left [ J_{45},M_{r}^- \right ]=-M_{r}^-  
\label{conm-pm-j45}
\end {equation}
Donde $M_r^+$ Y $M_r^-$ sube y baja respectivamente en una unidad los valores propios de energ\'{\i}a de los estados sobre el cual son aplicados. M\'as adelante se usar\'an los operadores $M_{1+2i}^\pm$ y $M_{1-2i}^\pm$ definidos como:
\begin {equation}
M_{1+2i}^\pm={1\over \sqrt {2}}(M_{1}^\pm+iM_{2}^\pm)\ \ \ \ \ \ \  
M_{1-2i}^\pm={1\over \sqrt {2}}(M_{1}^\pm-iM_{2}^\pm)
\label{deboost}
\end {equation}
Estos suben y bajan respectivamente la componente z de esp\'{\i}n por una unidad. Estos cumplen un \'algebra: 
\begin {equation}
\left[M_{1+2i}^\pm,M_{1-2i}^\pm\right]=0\ \ \ \ \ \ \ 
\left[M_{1+2i}^\pm,M_{1-2i}^\mp\right]=2(J_{45}\pm J_3)\ \ \ \ \ \ \
\left[M_{1\pm 2i}^+,M_{1\pm 2i}^-\right]=0 
\label{rel-mpp-1}
\end {equation}
Adicionalmente para $J_{45}$ y ($J_3=J_{12}$):
\begin {equation}
\left[J_{3},M_{1\pm 2i}^+ \right]=\pm M_{1\pm 2i}^+\ \ \ \ \ \left[ J_{3},M_{3}^+ \right] =0  \ \ \ \ \ \ \ 
\left[J_{45},M_{1\pm 2i}^+ \right]= M_{1\pm 2i}^+\ \ \ \ \ \left[ J_{45},M_{1\pm 2i}^-,J_{45} \right] =-M_{1\pm 2i}^- 
\label{rel-mpp-2}
\end {equation}
Asumiendo que existe una representaci\'on espacial sobre los operadores $J_{ij}$. Es conveniente etiquetar los estados por los valores propios de $J_{45}$, $J^rJ_r$ y $J_3$, los cuales son de la forma:
\begin {eqnarray}
J^{r}J_{r}\ |(...)E_0\ s\ m\rangle&=&s(s+1)\ |(...)E_0,\ s,\ m\rangle \nn\\
J_{45}\ |(...)E_0\ s\ m\rangle&=&E_0\ |(...)E_0,\ s,\ m\rangle\nn\\
J_{3}\ |(...)E_0\ s\ m\rangle&=&m\ |(...)E_0,\ s,\ m\rangle 
\label{t-ecua-propia}
\end {eqnarray}
Donde $(...)$ denota un conjunto no especificado de etiquetas. El operador Casimir $C_2$ puede ser escrito como:
\begin {equation}
C_2=(J_{45})^{2} + {1\over 2}J^{rs}J_{rs}+J^{4r}J_{4r} + J^{5r}J_{5r} 
   =(J_{45})^{2} + J^{r}J_{r} + {3\over 2} \{M_{r}^+,M_{r}^- \}  
\label{newc2}
\end {equation}
Note que: \ $\ft 12 J^{rs}J_{rs}=J^rJ_r$ \ , \ $J^{4r}J_{4r}=-3(J_{4r})^2$ \ , \ $J^{5r}J_{5r}=-3(J_{5r})^2$, donde $\{M_{r}^+,M_{r}^- \}=-2(J_{4r}^2+J_{5r}^2)$. Las posibles representaciones se separan en dos clases: 
\begin{itemize}
\item No existen estados que puedan ser aniquilados por los operadores de bajada $M_{r}^-$. Ya que el operador de energ\'{\i}a $J_{45}$ no es acotado hacia abajo y cualquier estado de energ\'{\i}a es obtenido por aplicaci\'on de $M_r^-$. Este caso no es de inter\'es f\'{\i}sico.\footnote{Cabe anotar que la ecuaci\'on (\ref{valor-propio-c2}) no es v\'alida para esta representaci\'on `continua'}
\item Existe un conjunto de estados que son aniquilados por $M_r^-$. Donde el espectro de energ\'{\i}a es acotado hacia abajo. Si se denota los valores propios de energ\'{\i}a por $E_0$ y los valores momento angular por $s$, el vacio consiste de $(2s+1)$ estados $|(E_0,s)E_0,\ s,\ m\rangle$, $m=-s,-s+1,...,s-1,s$. Aunque se podr\'{\i}a usar los valores propios de los Casimir $C_2$ y $C_4$ para etiquetar las representaciones, es conveniente usar $E_0$ y $s$. 
\end{itemize}
Para evaluar $C_2$ sobre el estado de vacio $|E_0,s\rangle$, se usa:
\begin {equation}
M_{r}^-\ |(E_0,s)E_0,\ s,\ m\rangle=0
\label{ecua-propia}
\end {equation}
Al reemplazar $\{M_r^+,M_r^-\}$ en (\ref{newc2}) por $-[M_r^+,M_r^-]$, teniendo en cuenta (\ref{ecua-propia}) junto con (\ref{t-ecua-propia}) y (\ref{conm-pm}):
\begin {equation}
C_2|(E_0,s)E_0,\ s,\ m\rangle=E_0(E_0-3)+s(s+1)|(E_0,s)E_0,\ s,\ m\rangle \label{valor-propio-c2}
\end {equation}
Al considerar una T.C.C. en un espacio de Minkowski, la representaci\'on del espacio de Fock se construye al actuar los operadores de creaci\'on ${\cal A}_i^+$ sobre el estado de vacio $|(E_0,s)E_0,\ s, \ m\rangle$. De esta forma al hacer la extensi\'on a espacios $AdS$, una representaci\'on del espacio $\aleph$ es escrita como:
\begin {equation}
\aleph = \oplus_{n=0}^\infty \ (\aleph_n)\ \ \ \ \ \ \ \ \ \ \ \ \ \ \ \mbox{con} \ \ \ \ \ 
J_{45}\aleph_n = (E_0+n)\aleph_n \ \ \ \ \ dim \ \aleph_n < \infty
\end {equation}
Donde $\aleph_n$ es generado por todos los vectores de la forma:
\begin {equation}
\sum_{n_1+n_2+n_3=n} \ \ \ C_{n_1 n_2 n_3}(M_1^+)^{n_1}(M_2^+)^{n_2}(M_3^+)^{n_3}|(E_0,s)E_0,\ s,\ m\rangle \ \ \ \ \ \  C_{n_1 n_2 n_3} \in C
\end {equation}
Como tal $\aleph$ no es un espacio de Hilbert, aunque posee un producto interior. La estructura de espacio de Hilbert puede ser impuesta sobre $\aleph$ con un producto escalar positivo.
\begin {equation}
||\psi_n ||^2 = (\psi_n,\psi_n)>0 \ \ \ \ \ \hbox{para } 0\neq \psi_n  \in \aleph_n
\end {equation}
Esto garantiza que las representaciones sean unitarias. $\aleph$ consiste de todos los vectores de la forma.
\begin {equation}
\psi = \S_{n=0}^\infty \ \psi_n\ \ \ \ \ ||\psi ||^2= \S_{n=0}^\infty \ ||\psi_n||^2 < \infty
\end {equation}
El estado $|(E_0,s)E_0,\ s,\ m\rangle$ se asume ortonormal. La norma de los otros estados son calculados a continuaci\'on, incrementando la energ\'{\i}a, al mover los operadores boost $M_r^+$ a la izquierda usando la condici\'on (\ref{ecua-propia}) y las relaciones (\ref{conm-pm}) hasta (\ref{rel-mpp-2}) (ver \cite{Nicolai,dewitt}). En ciertos casos l\'{\i}mites la norma de algunos estados es cero, por lo tanto el espacio de Hilbert f\'{\i}sico se obtiene al excluir estos estados mediante la factorizaci\'on: 
$$
\aleph_{Fis.}=\aleph\ /\ \aleph^{(0)}\ \ \ \ \ \ \ \aleph^{(0)}=\{\psi \in \aleph,\ \ ||\psi||=0\}
$$

\subsection {Condiciones de Unitaridad para $SO(3,2)$}\label{condicion-unitaria}
Asumiendo que el espectro de energ\'{\i}a est\'a acotado por abajo ($E\geq E_0$) y considerando las irreps unitarias de bajo peso, donde los estados son etiquetados como $|E_0,s,m\rangle$. Tal que $E_0$ denota el valor propio de energ\'{\i}a y $s$ el valor del momento angular total. Existen m\'as n\'umeros cu\'anticos asociados con el operador de momento angular sobre alg\'un eje pero no se consideran y se suprimen aqu\'{\i}. Dado que los estados con $E<E_0$ pueden aparecer, se establece la condici\'on (\ref{ecua-propia}) sobre los estados base. Por lo tanto las representaci\'ones  pueden ser construidas al actuar los operadores de subida sobre el estado de vacio $\vert E_0,s,m\rangle$. Es decir todos los estados de energ\'{\i}a $E=E_0+ n$ son construidos por $n$-productos de operadores de creaci\'on $M^+_r$, de esta forma se obtiene los estados de valores propios $E$ superiores con esp\'{\i}n superior. Cada estado es $(2s+1)$-veces degenerado. Para obtener las restricciones sobre los n\'umeros cu\'anticos $E_0$ y $s$ del vacio que determinan la unitaridad, se deja actuar los operadores de energ\'{\i}a y los boost de esp\'{\i}n $M_{1+2i}^+$ sobre el estado de vacio, obteniendose:
\begin {eqnarray} 
M_{1+2i}^+|(E_0,s)E_0,\ s,\ m\rangle &=& R_+ \langle sm11|s+1,m+1\rangle|(E_0,s)E_0+1,\ s+1,\ m+1\rangle+\nn\\
&&R_0 \langle sm11|s,m+1\rangle|(E_0,s)E_0+1,\ s,\ m+1\rangle+ \label{m1+2-base} \\
&&R_- \langle sm11|s-1,m+1\rangle|(E_0,s)E_0+1,\ s-1,\ m+1\rangle\nn
\end {eqnarray}
Donde el operador $M_{1+2i}^+$ al ser aplicado sobre un estado de momento angular $s$, obtiene 3 estados de momento angular total $s+1,\ s,\ s-1$. Los coeficientes de Clebsch-Gordan en (\ref{m1+2-base}) son dados por:
\begin {equation}
\begin{array}{cll} 
\langle sm11|s+1,m+1\rangle&=&\left ({(s+m+1)(s+m+2)\over (2s+1)(2s+2)}\right )^{1\over 2}\\
\langle sm11|s,m+1\rangle&=&-\left ({(s+m+1)(s-m)\over 2s(s+1)}\right )^{1\over 2}\\
\langle sm11|s-1,m+1\rangle&=&\left ({(s-m)(s-m+1)\over 2s(2s+1)}\right )^{1\over 2}
\end{array}
\end {equation}
Al restringir (\ref{m1+2-base}) para cada uno de los casos con momento angular $s$ $(s=0,\ft 12,1,...)$, se obtiene:
\begin{itemize}
\item Para $s=0$ los \'ultimos dos coeficientes de Clebsch-Gordan son indeterminados y los \'ultimos dos t\'erminos del lado derecho de (\ref{m1+2-base}) desaparecen.
\item Para $s=1/2$ el \'ultimo t\'ermino del lado derecho de (\ref{m1+2-base}) desaparece.
\item Para $s\geq 1$ todos los t\'erminos del lado derecho de (\ref{m1+2-base}) est\'an presentes.
\end{itemize}
\setlength{\unitlength}{0.5mm}
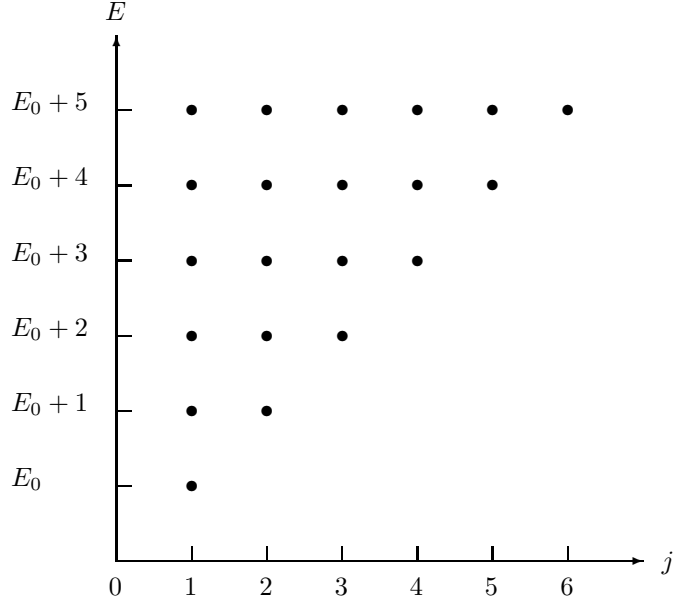
\begin{figure}[t]
\begin{picture}(260,160)(-50,0)
\put(30,10){\vector(1,0){140}}
\put(175,8){$j$}
\put(30,10){\vector(0,1){140}}
\put(27,154){$E$}
\put(28,1){$0$}
\put(48,1){$1$}
\put(68,1){$2$}
\put(88,1){$3$}
\put(108,1){$4$}
\put(128,1){$5$}
\put(148,1){$6$}
\put(50,10){\line(0,1){4}}
\put(70,10){\line(0,1){4}}
\put(90,10){\line(0,1){4}}
\put(110,10){\line(0,1){4}}
\put(130,10){\line(0,1){4}}
\put(150,10){\line(0,1){4}}
\put(30,30){\line(1,0){4}}
\put(30,50){\line(1,0){4}}
\put(30,70){\line(1,0){4}}
\put(30,90){\line(1,0){4}}
\put(30,110){\line(1,0){4}}
\put(30,130){\line(1,0){4}}
\put(2,30){$E_0$}
\put(50,30){\circle*{3}}
\put(2,50){$E_0+1$}
\put(50,50){\circle*{3}}
\put(70,50){\circle*{3}}
\put(2,70){$E_0+2$}
\put(50,70){\circle*{3}}
\put(70,70){\circle*{3}}
\put(90,70){\circle*{3}}
\put(2,90){$E_0+3$}
\put(50,90){\circle*{3}}
\put(70,90){\circle*{3}}
\put(90,90){\circle*{3}}
\put(110,90){\circle*{3}}
\put(2,110){$E_0+4$}
\put(50,110){\circle*{3}}
\put(70,110){\circle*{3}}
\put(90,110){\circle*{3}}
\put(110,110){\circle*{3}}
\put(130,110){\circle*{3}}
\put(2,130){$E_0+5$}
\put(50,130){\circle*{3}}
\put(70,130){\circle*{3}}
\put(90,130){\circle*{3}}
\put(110,130){\circle*{3}}
\put(130,130){\circle*{3}}
\put(150,130){\circle*{3}}
\end{picture}
\caption{Estados de la representaci\'on con esp\'{\i}n $s=1$ en t\'erminos de los valores propios de energ\'{\i}a $E$ y momento angular $j$.} 
\end{figure}
\setlength{\unitlength}{0.5mm}
\begin{figure}[t]
\begin{picture}(260,160)(-50,0)
\put(30,10){\vector(1,0){140}}
\put(175,8){$j$}
\put(30,10){\line(0,1){19}}
\put(30,31){\line(0,1){38}}
\put(30,71){\line(0,1){38}}
\put(30,111){\vector(0,1){40}}
\put(27,154){$E$}
\put(28,1){$0$}
\put(48,1){$1$}
\put(68,1){$2$}
\put(88,1){$3$}
\put(108,1){$4$}
\put(128,1){$5$}
\put(148,1){$6$}
\put(50,10){\line(0,1){4}}
\put(70,10){\line(0,1){4}}
\put(90,10){\line(0,1){4}}
\put(110,10){\line(0,1){4}}
\put(130,10){\line(0,1){4}}
\put(150,10){\line(0,1){4}}
\put(31,30){\line(1,0){3}}
\put(30,50){\line(1,0){4}}
\put(31,70){\line(1,0){3}}
\put(30,90){\line(1,0){4}}
\put(31,110){\line(1,0){3}}
\put(30,130){\line(1,0){4}}
\put(2,30){$E_0$}
\put(30,30){\circle{2}}
\put(40,30){\circle*{3}}
\put(2,50){$E_0+1$}
\put(50,50){\circle{2}}
\put(40,50){\circle*{3}}
\put(60,50){\circle*{3}}
\put(2,70){$E_0+2$}
\put(30,70){\circle{2}}
\put(40,70){\circle*{3}}
\put(70,70){\circle{2}}
\put(60,70){\circle*{3}}
\put(80,70){\circle*{3}}
\put(2,90){$E_0+3$}
\put(50,90){\circle{2}}
\put(40,90){\circle*{3}}
\put(60,90){\circle*{3}}
\put(90,90){\circle{2}}
\put(80,90){\circle*{3}}
\put(100,90){\circle*{3}}
\put(2,110){$E_0+4$}
\put(30,110){\circle{2}}
\put(40,110){\circle*{3}}
\put(70,110){\circle{2}}
\put(60,110){\circle*{3}}
\put(80,110){\circle*{3}}
\put(110,110){\circle{2}}
\put(100,110){\circle*{3}}
\put(120,110){\circle*{3}}
\put(2,130){$E_0+5$}
\put(50,130){\circle{2}}
\put(40,130){\circle*{3}}
\put(60,130){\circle*{3}}
\put(90,130){\circle{2}}
\put(80,130){\circle*{3}}
\put(100,130){\circle*{3}}
\put(130,130){\circle{2}}
\put(120,130){\circle*{3}}
\put(140,130){\circle*{3}}
\end{picture}
\caption{Estados de la representaci\'on $s=\ft12$. El c\'{\i}rculo peque\~no denota el multiplete original $s=0$ desde el cual se ha construido el multiplete de esp\'{\i}n- $\ft12$, tomando el producto directo.} 
\end{figure}
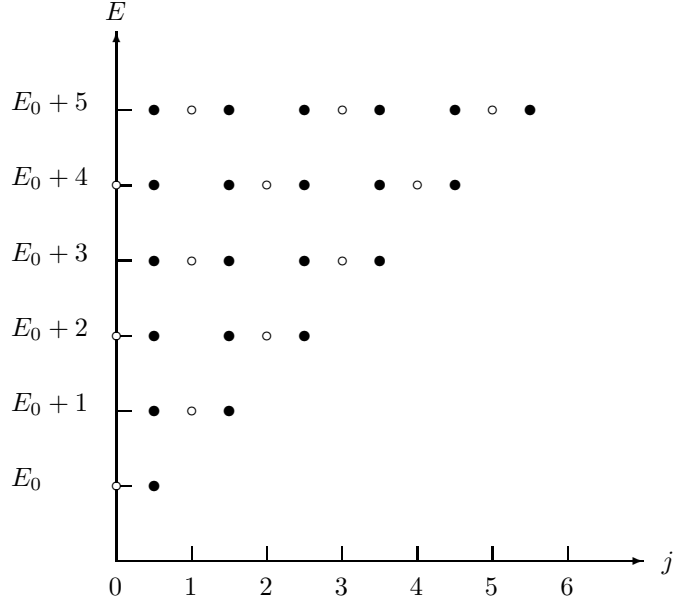
\setlength{\unitlength}{0.5mm}
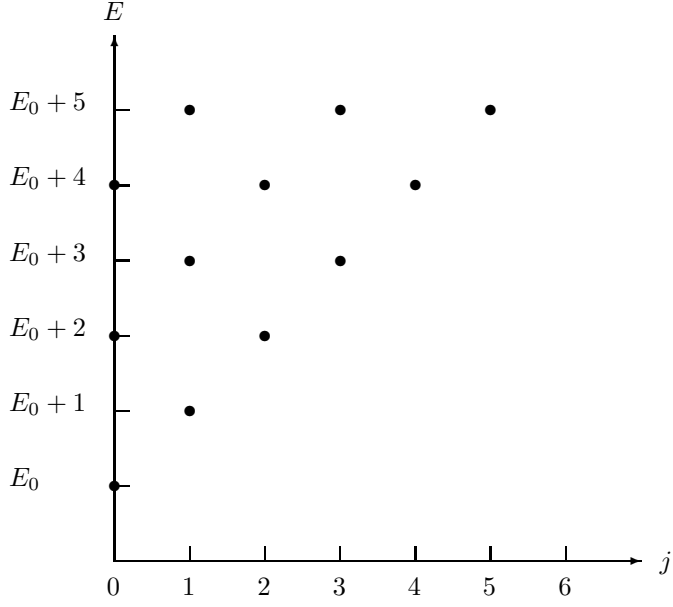
\begin{figure}[t]
\begin{picture}(260,160)(-50,0)
\put(30,10){\vector(1,0){140}}
\put(175,8){$j$}
\put(30,10){\vector(0,1){140}}
\put(27,154){$E$}
\put(28,1){$0$}
\put(48,1){$1$}
\put(68,1){$2$}
\put(88,1){$3$}
\put(108,1){$4$}
\put(128,1){$5$}
\put(148,1){$6$}
\put(50,10){\line(0,1){4}}
\put(70,10){\line(0,1){4}}
\put(90,10){\line(0,1){4}}
\put(110,10){\line(0,1){4}}
\put(130,10){\line(0,1){4}}
\put(150,10){\line(0,1){4}}
\put(30,30){\line(1,0){4}}
\put(30,50){\line(1,0){4}}
\put(30,70){\line(1,0){4}}
\put(30,90){\line(1,0){4}}
\put(30,110){\line(1,0){4}}
\put(30,130){\line(1,0){4}}
\put(2,30){$E_0$}
\put(30,30){\circle*{3}}
\put(2,50){$E_0+1$}
\put(50,50){\circle*{3}}
\put(2,70){$E_0+2$}
\put(30,70){\circle*{3}}
\put(70,70){\circle*{3}}
\put(2,90){$E_0+3$}
\put(50,90){\circle*{3}}
\put(90,90){\circle*{3}}
\put(2,110){$E_0+4$}
\put(30,110){\circle*{3}}
\put(70,110){\circle*{3}}
\put(110,110){\circle*{3}}
\put(2,130){$E_0+5$}
\put(50,130){\circle*{3}}
\put(90,130){\circle*{3}}
\put(130,130){\circle*{3}}
\end{picture}
\caption{Estados de la representaci\'on $s=0$ en t\'erminos de ($E,j$). Cada punto es $(2j+1)$-veces degenerado.} 
\end{figure}
Para calcular los t\'erminos $\{R_+,R_0,R_-\}$ se escoje $(m=s)$: solo el caso con momento angular $s+1$ contribuye en (\ref{m1+2-base}). Se calcula la norma en ambos lados de la ecuaci\'on, teniendo en cuenta que $(M_{1+2i}^+)^\dagger=-M_{1-2i}^-$ aniquila el estado de vacio y usando (\ref{conm-pm}) se sustituye el producto $(M_{1+2i}^+)^\dagger M_{1+2i}^+$ en t\'erminos del conmutador:
\begin {equation}
-\left [M_{1-2i}^-,M_{1+2i}^+ \right ] = 2(J_{45} + J_{3}) 
\label{conm-pm-1}
\end {equation}
Se obtiene que:
\begin {equation}
|R_+|^2= 2(E_0+s) 
\label{matriz-red0}
\end {equation}
Usando el mismo m\'etodo para $R_0$ y $R_-$ con $m=s-1$ y $m=s-2$, se obtiene las ecuaciones:
\begin {eqnarray}
2(E_0+s-1)&=&{2(E_0+s)(2s+1)s\over (2s+1)(s+1)}+{|R_0|^2 (s)\over s(s+1)}+{|R_-|^2\over s(2s+1)}\nn\\
2(E_0+s-2)&=&{2(E_0+s)(2s-1)s\over (2s+1)(s+1)}+{|R_0|^2(2s-1)\over s(s+1)}+{3|R_-|^2\over s(2s+1)}
\end {eqnarray}
Con soluci\'on:
\begin {eqnarray}
|R_0|^2 &=& 2(E_0-1)  \label {matriz-red1}\\
|R_-|^2 &=& 2(E_0-s-1) \label{matriz-red2}
\end {eqnarray}
De las ecuaci\'on (\ref{matriz-red0},\ref{matriz-red1},\ref{matriz-red2}) se puede concluir:
\begin{itemize}
\item Para $s\geq 1$, el requerimiento unitario es dado como:
\begin {equation}
E_0 \geq s+1 \ \ \ \ \ \hbox{para } s\geq 1 \ \ \ \hbox{ con }\ s=1,\ft32,2,...
\label{condicion-1}
\end {equation}
Si $E_0=s+1$, los valores propios de $C_2$ son negativos si $s>1$ y cero si $s=1$, usando (\ref{valor-propio-c2}) se obtiene:   
\begin {equation}
C_2|(E_0=s+1,s)(E_0=s+1),\ s,\ m\rangle=2(s^2-1)|(E_0=s+1,s)(E_0=s+1),\ s,\ m\rangle 
\end {equation}
\item Para $s=1/2$ de (\ref{matriz-red1}) se puede inferir que:
\begin {equation}
E_0 \geq 1 \ \ \ \ \ \hbox{para } s= 1/2 
\label{condicion-2}
\end {equation}
Para el caso particular $E_0=1$ se tiene la representaci\'on singleton de esp\'{\i}n-$\ft12$ (encontrada por Dirac \cite{Dirac1}). Esta deja solo un estado para cualquier valor de esp\'{\i}n. El Casimir tiene un valor propio negativo, donde: 
\begin{equation}
C_2 |(E_0=1,s)(E_0=1),\ s,\ m\rangle= -\ft 54 |(E_0=1,s)(E_0=1),\ s,\ m\rangle \end{equation}
\item Para $s=0$ se procede como sigue:
\begin {equation} 
M_{1+2i}^+ M_{1+2i}^+|(E_0,0)E_0,0,0\rangle = \sqrt{2E_0} M_{1+2i}^+|(E_0,0)E_0+1,1,1\rangle =R^{'}\sqrt{2E_0}|(E_0,0)E_0+2,2,2\rangle
\end {equation}
Como se mostro anteriormente se calcula la norma a ambos lados de la ecuaci\'on, obteniendose:
\begin {equation}
|R^{'}|^2= 4(E_0+1) 
\end {equation}
Y considerando:
\begin {eqnarray} 
M_{3}^+M_{1+2i}^+|(E_0,0)E_0,0,0\rangle &=&\sqrt{2E_0}(2\sqrt{E_0+1}\langle 1110|21\rangle |(E_0,0)E_0+2,2,1\rangle+\nn\\
&&R^{''} \sqrt{2E_0} \langle 1110|11\rangle|(E_0,0)E_0+2,1,1\rangle)
\label {m3-m1+2i}
\end {eqnarray}
Usando \ $\langle 1110|21\rangle\ =\ \langle 1110|11\rangle\ =\ 1/\sqrt{2}$ y las relaciones dadas por (\ref{rel-mpp-1},\ref{rel-mpp-2}) al tomar la norma de (\ref{m3-m1+2i}), teniendo en cuenta la condici\'on (\ref{ecua-propia}) al aplicar sobre un estado base, se tiene que:
\begin {equation} 
(M_{3}^+M_{1+2i}^+)^\dagger (M_{3}^+M_{1+2i}^+)|(E_0,0)E_0,\ 0,\ 0\rangle=4(J_{45}+1)(J_{45}+J_3)|(E_0,0)E_0,\ 0,\ 0\rangle
\end {equation}
Entonces:
\begin {equation} 
4(E_0+1)(E_0+(s=0))=2E_0(2(E_0+1)+E_0|R^{''}|^2)
\end {equation}
Por lo tanto $R^{''}=0$, de aqu\'{\i} se obtiene que el estado con $j=1$ est\'a ausente. Finalmente:
\begin {eqnarray} 
M_{1-2i}^+M_{1+2i}^+|(E_0,0)E_0,\ 0,\ 0\rangle&=&\sqrt{2E_0}\{2\sqrt{E_0+1}\langle 111-1|20\rangle |(E_0,0)E_0+2,\ 2,\ 0\rangle+\nn\\
&&R^{'''} \langle 111-1|00\rangle|(E_0,0)E_0+2,\ 0,\ 0\rangle\}
\end {eqnarray}
Sustitutendo los coeficientes de Clebsch-Gordan y tomando la norma, se obtiene:
\begin {equation}
|R^{'''}|^2= 4(E_0-1/2) 
\end {equation}
Donde:
\begin {equation}
E_0 \geq 1/2 \ \ \ \ \ \hbox{para }\ s=0 
\label{condicion-3}
\end {equation}
\end{itemize} 
En resumen: las condiciones dadas por (\ref{condicion-1}),(\ref{condicion-2}) y (\ref{condicion-3}) son suficientes para garantizar la positividad de la norma en todos los sectores de altas energ\'{\i}as $\aleph_n$ con $n\geq 2$. Donde todas las representaciones son univocamente caracterizadas por los n\'umeros cu\'anticos $E_0$ y $s$ y se denotan como $D(E_0,s)$, con el siguiente an\'alisis: 
\begin {itemize}
\item El primer caso denota el estado de vacio con esp\'{\i}n cero ($s=0$) y para un valor propio $E$, El estado de esp\'{\i}n superior es dado por los productos sim\'etricos con traza cero de operadores $M^+_a$ sobre el estado base $E=E_0$. Estos estados se muestran en la figura Fig. (3.). Se puede notar que el diagrama para $s=0$ difiere del diagrama para espines superiores, dado el resultado $R^{''}=0$ en la ecuaci\'on (\ref{m3-m1+2i}). 
\item Para obtener el caso de esp\'{\i}n-$\ft12$ $(s=\ft 12)$ se toma los productos directos de estados con esp\'{\i}n-$\ft12$. Esto implica que cualquier punto con esp\'{\i}n $j$ en la Fig. (3) genera dos puntos con esp\'{\i}n $j\pm\ft12$, con la excepci\'on de los puntos asociados a $j=0$, los cuales simplemente se mueven a $j=\ft12$. Ver Fig. (2.)
\item Continuando se puede tomar los productos directos con estados de esp\'{\i}n-1 $(s=1)$, pero la situaci\'on es m\'as complicada ya que el multiplete resultante no siempre es irreducible. En principio, cada punto con esp\'{\i}n $j$ genera ahora tres puntos, asociados con $j$ y $j\pm 1$, Con la excepci\'on de los puntos $j=0$, los cuales simplemente se mueven a $j=1$. El resultado de este procedimiento se muestra en Fig. (1.)
\end{itemize}
Al hacer contracci\'on teniendo en cuenta (\ref{10}) y (\ref{11}), se obtiene el primer invariante de Poincar\'e:
\begin {equation}
C^{3,1}_2 \equiv \lim_{R_5 \rightarrow \infty} {C_2 \over R_5^2}
  =\lim_{R_5 \rightarrow \infty}(\Pi_{4})^{2} + \left (\lim_{R_5 \rightarrow \infty}{1\over R_5^2}J^{r}J_{r}\to 0\right )+\lim_{R_5 \rightarrow \infty}{3\over 2R_5^2} \{M_{r}^+,M_{r}^- \}=K_{\m}K^{\m}  
\label{contra-newc2}
\end {equation}
Al tomar el limite de curvatura a cero en la representaci\'on $(E_0,s)$ con: $E_0\to E_{0,R}\equiv E_0(R)$ se obtiene que:
\begin {eqnarray}
K_\m K^\m |k,\lambda\rangle=\lim_{R_5\rightarrow\infty}{C_2\over R_5^2}|(E_{0,R},s)E_{0,R},\ s,\ m\rangle &=&\lim_{R_5\rightarrow\infty}{E_{0,R}(E_{0,R}-3)+s(s+1)\over R_5^2}|(E_{0,R},s)E_{0,R},\ s,\ m\rangle \nn\\
&=&\lim_{R_5\rightarrow\infty}{E_{0,R}(E_{0,R}-3)\over R_5^2}|(E_{0,R},s)E_{0,R},\ s,\ m\rangle \label{propio-contra-newc2}
\end {eqnarray}
Usando $E_{0,R}=k\ R_5\ $ en (\ref{propio-contra-newc2}), siendo $k=mc/\hbar$ ($m\Rightarrow$ masa en reposo de la part\'{\i}cula), la contracci\'on es dada como: 
\begin {equation}
K_\m K^\m |k,\lambda\rangle =\lim_{R_5\rightarrow\infty}{E_{0,R}^2\over R_5^2}|(E_{0,R},s)E_{0,R},\ s,\ m\rangle\ =\ k^2|k,\lambda\rangle 
\end {equation}

\section  {Supersimetr\'{\i}a}\label{superpoincare}
Hist\'oricamente SUSY (supersimetr\'{\i}a) fue introducido en la Teor\'{\i}a de la Matriz $S$. Donde una simetr\'{\i}a de $S$ tiene el efecto de reordenar los estados asint\'oticos singletes y multipletes. Una simetr\'{\i}a de $S$ en f\'{\i}sica de part\'{\i}culas puede significar:
\begin {itemize}
\item Invariancia de Poincar\'e: el producto semidirecto de traslaciones y rotaciones de Lorentz, con los respectivos generadores de grupo $K_\m$ y $J_{\m\n}$
\item Simetr\'{\i}as discretas: paridad $P$, inversi\'on temporal $T$, y conjugaci\'on de la carga $C$. 
\item Simetr\'{\i}as internas globales, relacionadas a la conservaci\'on de n\'umeros cu\'anticos, tales como: carga el\'ectrica e isosp\'{\i}n. Los generadores de simetr\'{\i}a $B_i$ asociados son escalares de Lorentz y generan un \'algebra de Lie, siendo $U(1)_Y\times SU(2)_L \times SU(3)_C$ el grupo de simetr\'{\i}a interna asociado al Modelo Standard, con un \'algebra de Lie:
$$
[B_i,B_j]=ic_{ij}^kB_k
$$
$c_{ij}^k$ representa las constantes de estructura.
\end {itemize}
En 1967 Coleman y Mandula proporcionaron un argumento riguroso, el cual clasifica las posibles simetr\'{\i}as de la matriz $S$ asumiendo que el \'algebra de una simetr\'{\i}a de $S$ solo involucra conmutadores \cite{ColemanMandula}. En 1971 Gol'fond y Likhtman mostraron que debilitando esta condici\'on e involucrando generadores de simetr\'{\i}a que anticonmutan era posible construir una simetr\'{\i}a m\'as grande, una supersimetr\'{\i}a que incluye Poincar\'e y grupos de simetr\'{\i}a interna de una forma no trivial. Con la particularidad que estos generadores de simetr\'{\i}a que anticonmutan transforman como representaciones espinoriales del grupo de Lorentz y no como representaciones tensoriales, por esto SUSY no es una simetr\'{\i}a interna. Luego se defini\'o SUSY como una extensi\'on del \'algebra de $P(3,1)$ con generadores espinoriales que anticonmutan, y en 1975, Haag, Lopusz\'anski y Sohnius probaron que SUSY era la \'unica simetr\'{\i}a adicional de la Matrix $S$ \cite{HLS}. De aqu\'{\i} que est\'a sea la \'unica extensi\'on posible de las simetr\'{\i}as del espacio-tiempo de la f\'{\i}sica de part\'{\i}culas: que al extender en la teor\'{\i}a de campos se obtiene ($Q_a \Psi_{\m}^a( X , \th )$):
\begin {itemize}
\item El `espacio' es extendido adicionando unos nuevos par\'ametros  de Grassmann, $(X) \rightarrow (X,\th )$.
\item El campo es reemplazado por un supercampo. 
\end {itemize}
Donde los generadores de supersimetr\'{\i}a est\'an definidos de tal forma que transforma estados fermi\'onicos  en bos\'onicos y estados bos\'onicos en fermi\'onicos, es decir:
$$
Q | f \rangle = | b \rangle \ \ \ \ \ \ \ Q | b \rangle = | f^{'} \rangle
$$
La importancia de SUSY radica en que soluciona problemas en la f\'{\i}sica m\'as all\'a del Modelo Standard:
\begin {itemize}
\item Permite una unificaci\'on entre la Relatividad General y la Mec\'anica C\'uantica.
\item Predice Gravedad $\Rightarrow$ posibilita la existencia de part\'{\i}culas sin masa con sp\'{\i}n 2.
\item Es una teor\'{\i}a finita $\Rightarrow$ surgen cancelaciones entre las contribuciones fermi\'onicas y bos\'onicas. 
\end {itemize}
Partiendo de un espacio-tiempo plano $4-Dim.$ con un grupo de simetr\'{\i}a asociado $P(3,1)=T_{3,1}\wedge SO(3,1)$ y una m\'etrica $(+++-)$, se muestra una $N$-extensi\'on supersim\'etrica denotada como $\ ^SP(N|3,1)$ y la clasificaci\'on de irreps asociadas. En particular $\ ^SP(N|3,1)$ no permite generadores de esp\'{\i}n $\ft 32$. Por otro lado supersimetr\'{\i}a es consistente con una teor\'{\i}a de esp\'{\i}n-2 si:
$$N \leq 8$$ 
Un an\'alisis detallado de teor\'{\i}as de supergravedad se puede encontrar en: Zanelli \cite{Zanelli}, Tanii \cite{Tanii} y Berenstein \cite{Berenstein}. En esta secci\'on se muestra parte del trabajo de R. Haase en sus lecturas \cite{Haase}, el cual se ha ampliado donde es necesario, ver tambi\'en \cite{Srivastava,Annecy,van}.

\subsection  {El \'Algebra de SuperPoincar\'e 4-Dim $\ ^SP(N|3,1)$}
Denotando los generadores de SUSY por $Qs$ y sus correspondientes par\'ametros pos $\th s$ (variables de grassmann), con $\th^i\th^j=-\th^j\th^i$. 
De acuerdo a la notaci\'on espinorial 4-Dim de Weyl, las N parejas de generadores de SUSY son escritos como: $Q_{\a a}$ y $\bar Q_{\dot\b  b}=(Q_{\b b})^\dagger$ con ($a,b=1,..N$) y los \'{\i}ndices espinoriales: ($\a,\b=1,2$) y  ($\dot \a ,\dot \b=\dot 1,\dot 2$). Los cuales se transforman bajo la acci\'on del grupo de Lorentz, como:
\footnote{
Dado que la teor\'{\i}a de representaci\'on de $SO(3)$ es conocida en la aplicaci\'on a la teor\'{\i}a del momento angular y al grupo de isosp\'{\i}n. Este es util para establecer  un isomorfismo del \'algebra de $SO(3,1)$ con $SO(3) \times SO(3)$ para construir las propiedades de los irreps del grupo de Lorentz. Se obtiene que las irreps de dimensi\'on finita del grupo de Lorentz es etiquetada por un par de enteros o semienteros $\{j_1,j_2\}$ como $\left[j_1\times j_2\right]$, con dimensi\'on $(2j_1+1)(2j_2+1)$. Si $j_1$ o $j_2$ es de esp\'{\i}n semientero son llamados espinoriales, de lo contrario son llamados tensoriales. Las representaciones b\'asicas son caracterizadas por $\left[\ft 12\times 0\right]$, $\left[0\times \ft 12\right]$ y cualquier representaci\'on espinorial y tensorial del grupo de Lorentz puede ser obtenida al `tensorizar' y `simetrizar' estos. Los conjugados complejos son identificados como $\left[j_1 \times j_2 \right]^*=\left[j_2 \times j_1\right]$. La representaci\'on $\left[j_1 \times j_2 \right]$ puede ser llevado a la forma:
$$\left[j_1\times j_2 \right]\rightarrow \left[j_1+j_2,|j_1-j_2|\right]_\pm$$
El producto directo de irreps es determinado por 2 descomposiciones de Glebsch-Gordan.
$$[j_1\times j_2 ]\times [j_1^,\times j_2^,]=\sum_{j,j^,} \left[(j_1+j_1^,-j)\times (j_2+j_2^,-j^,)\right] \ \ \ \mbox{Con} \ \ \ j^,=0,1,..|j_2-j_2^,| \ \ \ j=0,1,..|j_1-j_1^,|$$
} 
\begin {equation}
\D_-\equiv\left[\ft 12,\ft 12\right]_- = \left[\ft 12 \times 0\right] \ \ \ y \ \ \ \D_+\equiv\left[\ft 12, \ft 12\right]_+ = \left[0 \times \ft 12\right]
\end {equation}
Esta transformaci\'on implica:
\begin {equation}
U_{\L}Q_{\a a}U_{\L}^{-1}=Q_{\a' a}\D_+ (\L)_{\a}^{\a'} \ \ \ \ \ \ \ U_{\L}\bar Q_{\dot \b  b}U_{\L}^{-1}=\bar Q_{\dot \b' b}\D_- (\L)_{\dot \b}^{\dot \b'} \ \ 
\end {equation}
Los \'{\i}ndices $\a,\b$ son `rotados' por transformaciones de Lorentz. Considerando la forma m\'as general de anticonmutaci\'on de los $Q's$, estos se transforman como el producto Kronecker: 
\begin {equation}
\D_+\times \D_+ =\left[\ft 12,\ft 12\right]_+\times \left[\ft 12, \ft 12 \right]_+ = \left[ 0\times\ft 12\right] \times \left[ 0\times,\ft 12\right] =\left[ 0\times 0\right]+\left[ 0 \times 1\right]= \left[0\right]+\left[1^2\right]_+
\label{QQ0}
\end {equation}
Por lo tanto: 
\begin {equation}
\{Q_{\a a},Q_{\b b}\}=I \var_{\a \b}Z_{ab}+A_i a_{\a \b}^i Y_{ab} 
\label{QQ1}
\end {equation}
con $A_i={1\over 2}(J_i+iB_i)$, siendo $Z_{ab}$ y $Y_{ab}$ escalares complejos de Lorentz. Como $[0]$ y $[1]$ son las respectivas partes antisim\'etrica /sim\'etrica del cuadrado kronecker de $\D_+$, la estructura espinorial de los dos t\'erminos sobre el lado derecho son antisim\'etricos/sim\'etricos bajo el intercambio $\a \rightarrow \b$, con $Z_{ab}$ antisim\'etrico y $Y_{ab}$ sim\'etricos.\\
\\
Similarmente para los adjuntos $\bar Q_{\dot\b b}$:
\begin {equation}
\{\bar Q_{\dot \a  a},\bar Q_{\dot \b  b}\}=I \var_{\dot\a \dot\b} \bar Z_{ab}+ \bar A_i (\bar a_{\dot\a\dot\b}^{i})\bar Y_{ab} 
\end {equation}
Con $\bar A_i\equiv {1\over 2}(J_i-iB_i)$. Para obtener las relaciones de conmutaci\'on para $A_i$ y $Q_{\a a}$, se debe tener en cuenta que estos se transforman bajo el grupo de Lorentz como:
\begin {equation}
[1^2]_+ \times \D_+ = \left[0\times 1\right ] \times \left [0\times \ft 12\right] =\left[0\times \ft 12\right ]+\left[0\times \ft 32\right]=\D_++\left[\D:1^2\right]_+
\end {equation}
Por lo tanto la relaci\'on de conmutaci\'on toma la forma:
\begin {equation}
[A_i,Q_{\a a}]=Q_{\b b} c_{i\a}^{\b}X_a^b 
\label {14}
\end {equation}
Si $Q_{\a a}$ son los \'unicos generadores fermi\'onicos y no son generadores de esp\'{\i}n $3/2$, se obtiene la identidad:
\begin{equation}
[[A_i,A_j],Q_{\a a}]+[[Q_{\a a},A_i],A_j]+[[A_j,Q_{\a a}],A_i]=0 
\end{equation}
Usando (\ref{14}) se tiene que $[(c_i),(c_j)]=i(c_k)\e_{ij}^k$ y $X^2=X$. Esto significa que $(c_i)$ forma una representaci\'on 2-Dim de un \'algebra $SO(3)$. Una soluci\'on es dada por las matrices de Pauli $\s_i$ con $c_i=(1/2)\s_i$. Por lo tanto:
\begin {equation}
[\left \{ Q_{\a a},Q_{\b b}\right \},A_i]-\left \{[Q_{\b b},A_i],Q_{\a a}\right \}+\left\{[A_i,Q_{\a a}],Q_{\b b}\right\}=0\\ 
\end{equation}
De forma similar a (\ref{14}) se construyen los conmutadores de $\bar A_i$ con $\bar Q_{\dot\a a}$:
\begin {equation}
[\bar A_i,\bar Q_{\dot\a a}]=\bar Q_{\dot\b b}\bar c_{i\dot\a}^{\dot\b}\bar X_a^b 
\end{equation}
Donde $2\bar c_i=-2c_i^*=-\s_i^*=J\s_iJ^{-1}=2Jc_iJ^{-1}$ y $J=-i\s_2$, se puede notar que $J$ tiene un elemento matricial $J_\a^{\dot\a}$ e interviene las dos irreps b\'asicos de esp\'{\i}n.\\
\\
Los conmutadores: $[\bar A_i,Q_{\a a}]$ y $[A_i,\bar Q_{\dot\a a}]$ son cero, ya que no se consideran generadores de esp\'{\i}n $3/2$, donde:
\begin {equation}
[1^2]_-\times \D_+=\left[ 1\times 0\right]\times \left[ 0 \times \ft 12\right]=\left[ 1 \times \ft 12\right]=\left[ \D:1\right]_-\ ,\ \ \ \ \ \ [1^2]_+\times \D_-=\left[ 0\times 1\right]\times \left[ \ft 12 \times 0\right]=\left[ \ft 12 \times 1\right]=\left[ \D:1\right]_+
\end {equation} 
Continuando con los anticonmutadores de $Q_{\a a}$ y los adjuntos:
\begin {equation}
\{Q_{\a a},\bar Q_{\dot\b b}\}=K_\m a_{\a \dot\b}^\m \d_{ab}
\label{20}
\end {equation}
Notando que el lado izquierdo se transforma como: 
\begin{equation}
\D_+\times\D_-= \left[\ft 12,\ft 12\right]_+ \times \left[\ft 12,\ft 12 \right]_-= \left[0\times \ft 12 \right]\times \left[\ft 12 \times 0\right]=\left[\ft 12\times \ft 12\right] =\left[1\right]
\end{equation}
bajo el grupo de Lorentz. Donde $\{Q,\bar Q\}$ es un operador definido positivamente. Tal que el lado derecho de la ecuaci\'on (\ref{20}) debe transformar como un 4-vector de Lorentz. El objeto m\'as general que se puede construir que incluya $K_\m,J_{\m\n}$ y $B_i$ toma la forma $K_\m a_{\a\dot\b}^\m W_{ab}$, con $W_{ab}$ un escalar de Lorentz complejo. Tomando el adjunto del lado derecho de la ecuaci\'on (\ref{20}) y usando $(Q_{\a a})^\dagger=\bar Q_{\dot\a a}$ requiere que $(a_{\a \dot\b}^{\m })^*=a_{\b \dot\a}^{\m}$ y $W_{a b}^{\dagger}=W_{ba}$. Esto significa que podemos siempre escojer una base para los $Q_{\a a}$ con $W_{ab}$ proporcional a $\d_{ab}$.\\
\\
Desarrollando la identidad de Jacobi para $(Q,\bar Q,A)$, se obtiene:
\begin{eqnarray}
[\{Q_{\a a},\bar Q_{\dot\b b}\},A_i]-\{[\bar Q_{\bar {\dot\b} b},A_i],Q_{\a a}\}+\{[A_i,Q_{\a a}],\bar Q_{\dot\b b}\}&=&0\nn\\ 
K_\n(-c_{i\m}^{\n}a_{\a \dot\b}^\m \d_{ab}+a_{\a\dot\g}^{\n}\d_{ac}\bar c_{i\dot\b}^{\dot\g} X_b^c +a_{\g\dot\b}^\n \d_{cb}c_{i\a}^{\g}X_a^c)&=&0\nn\\
K_\n(-c_{i\m}^{\n}a_{\a \dot\b}^\m \d_{ab}+a_{\a\dot\g}^{\n}(Jc_iJ^{-1})_{i\dot\b}^{\dot\g} X_{ab} +a_{\g\dot\b}^\n c_{i\a}^{\g}X_{ba})&=&0
\end{eqnarray}
Si $X_{ab}=\d_{ab}=X_{ba}$ se tiene que:
\begin{equation}
c^\n_{i\m}(a_{\a}J)^\m_\b-(a_{\a}J)_{\g}^\n c_{i\b}^{\g}=(a_{\g}J)^{\n}_{\b}c_{i\a}^{\g}
\end{equation}
En forma matricial $[c_i,a_{\a}J]=a_\b J c_{i \a}^\b$. Con respecto a $SO(3)$ las dos matrices $a_{\a} J$ transforman como un 2-vector. Se puede tomar por lo tanto $a_{\a\dot\b}^\m =2\s_{\a\dot\b}^\m$, donde el factor (2) es simplemente una convenci\'on. Adicionalmente se tiene que los generadores de supersimetr\'{\i}a conmutan con los generadores de traslaci\'on:
\begin {equation}
\left[K_\m , Q_{\a a}\right]=0=\left[K_\m , \bar Q_{\dot\a a}\right] 
\label{26}
\end {equation}
Siendo estos un invariante traslacional. Esto no es obvio dado que la forma m\'as general consistente con la invariancia de Lorentz es determinada de las propiedades de transformaci\'on del lado izquierdo, donde el conmutador se transforma como:
\begin{equation}
[1]\times\D_+= [1]\times\left[ \ft 12,\ft 12 \right]_+ =\left[ \ft 12 \times \ft 12 \right]\times \left[0\times \ft 12 \right]=\left[ \ft 12 \times 0\right]+\left[ \ft 12 \times 1
\right]= \D_-+[\D:1]_+ 
\end{equation}
Por lo tanto:
\begin {equation}
\left[K_\m , Q_{\a a}\right]=\bar Q_{\dot\b b}c_{\m\a}^{\dot\b} V_a^b \ \ \ \ \ \ \  
\left[K_\m , \bar Q_{\dot\a a}\right]=Q_{\b b}\bar c_{\m\dot\a}^\b \bar V_a^b  
\label{28}
\end {equation}
Donde $V_a^b$ y $\bar V_a^b$ son escalares de Lorentz complejos y $\bar c_{\m\dot\a}^\b=-(c_{\m\a}^{\dot\b})^*$. Se ha asumido que estos generadores de simetr\'{\i}a no transforman como: $[\D:1]_+= ({1\over 2},1)$ o $[\D:1]_-= (1,{1\over 2})$. Se puede ver que todos los $V_a^b$ desaparecen, el primer paso es usar la ecuaci\'on (\ref{28}) en la identidad de Jacobi:
\begin {eqnarray}
[[K_{\m},K_{\n}],Q_{\a a}]+[[Q_{\a a},K_\m],K_{\n}]+[[K_\n,Q_{\a a}],K_{\m}]&=&0\nn\\ 
Q_{\g c}(\bar c_{\n\dot\b}^{\g}c_{\m\a}^{\dot\b}-\bar c_{\m\dot\b}^\g c_{\n\a}^{\dot\b})\bar V_b^c V_a^b &=&0\nn\\
Q_{\g c}(\bar c_{\m\b}^{\dot\g}c_{\n\a}^{\dot\b}-\bar c_{\n\b}^{\dot\g} c_{\m\a}^{\dot\b})(\bar V.V)_a^c &=&0
\end {eqnarray}  
Al tomar la contracci\'on con $\e^{\a\b}$ se obtiene que $\bar V V=0$. Se puede obtener m\'as informaci\'on considerando:
\begin{eqnarray}
[\{Q_{\a a},Q_{\b b}\},K_\m]+\{[K_\m,Q_{\a a}],Q_{\b b}\}-\{[Q_{\b b},K_\m],Q_{\a a}\}&=&0 \label{31}\\ 
K_\n c_{i\m}^{\n}a_{\a \b}^i Y_{ab}-K_\n a_{\b \dot\g}^{\n} c_{\m\a}^{\dot\g}\d_{bc}V_a^c-K_\n a_{\a\dot\g}^{\n} c_{\m\b}^{\dot\g}\d_{ac}V_b^c&=&0 \label{32}
\end {eqnarray}
Al tomar contracci\'on con $\e^{\a\b}$, esto se reduce a:
\begin{equation}
K_\n \e^{\a \b}a_{\a \dot\g}^\n c_{\m \b}^{\dot\g}(V_{ba}-V_{ab})=0
\end {equation}
Por lo tanto $V_{ab}$  es sim\'etrico y teniendo en cuenta que $\bar V V=0$ implica que $V_a^b=0$, obteni\'endose la ecuaci\'on (\ref{26}). De la ecuaci\'on (\ref{32}) se establece la parte sim\'etrica de la identidad de Jacobi dada por la ecuaci\'on (\ref{31}). Esto implica que $K_\n f^\n Y_{ab}=0$, lo cual es cierto si $Y_{ab}=0$, entonces en (\ref{QQ1}) obtenemos:
\begin{equation}
\{Q_{\a a},Q_{\b b}\}=I\e_{\a\b}Z_{ab}
\end {equation}
El escalar complejo de Lorentz $Z_{ab}$ es llamado {\it la carga central}, manipulando la identidad de Jacobi se puede mostrar que $Z_{ab}$ conmuta con los generadores $Q_{\a a}\bar Q_{\dot\a a}$:
\begin{eqnarray}
[\{Q_{\a a},\bar Q_{\b b}\},Q_{\g c}]+[\{Q_{\b  b},Q_{\g c}\},Q_{\g a}]+[\{Q_{\g c},Q_{\a a}\},Q_{\b  b}]&=&0\nn\\ 
\e_{\a \b}[Z_{a b},Q_{\g c}]+\e_{\b\g}[Z_{bc},Q_{\a a}]+\e_{\g\a}[Z_{ca},Q_{\b  b}]&=&0\nn\\
\e_{\a \b}Q_{\g d}c_{abc}^d+\e_{\b\g}Q_{\a d}c_{bca}^d+ \e_{\g \a}Q_{\b d}c_{cab}^d&=&0\nn\\
Q_{\g d}(-2c_{abc}^d + c_{bca}^d+c_{cab}^d)&=&0
\end{eqnarray}
Donde $c_{abc}^d=0$, dado que: $c_{bca}^d=c_{abc}^d$ y $c_{cab}^d=c_{abc}^d=-c_{bac}^d$. Esto implica que $[Z_{cd},Q_{\a a}]=0$. Ahora tomando la siguiente identidad de Jacobi: 
$$
[\{Q_{\a a},Q_{\b b}\},Z_{cd}]-\{[Q_{\b  b},Z_{cd}],Q_{\a a}\}+ \{[Z_{cd},Q_{\a a}],Q_{\b  b}\}=0 
$$
$$
[Z_{ab},Z_{cd}]\e_{\a\b}+\{Q_{\b e},Q_{\a a}\}c_{cdb}^e+\{Q_{\a f}, Q_{\b  b}\} c_{cda}^{f}=0 
$$
$$
[Z_{ab},Z_{cd}]\e_{\a\b}+Z_{ea}\e_{\b\a}c_{cdb}^e+ Z_{fb}\e_{\a \b}c_{cda}^f=0 
$$
\begin {equation}
-Z_{ae}c_{cdb}^e+Z_{bf}c_{cda}^f=[Z_{ab},Z_{cd}] 
\end {equation}
se obtiene que: $[Z_{ab},Z_{cd}]=0$. Se puede ver que la carga central genera una sub\'algebra invariante abeliana del \'algebra de Lie compacta (isomorfa a $[U(1)]^p$, con $p=$ n\'umero cargas centrales independientes) generada por el hermitiano $B_i$. Por lo tanto podemos escribir $Z_{ab}=B_ib_{ab}^i$, con la constante de estructura $c_{ia}^b$ definida por:
\begin{equation}
[B_i,Q_{\a a}]=Q_{\a b}c_{ia}^b,\ \ \ \ \ \ \ [B_i,\bar Q_{\dot\a a}]=-\bar Q_{\dot\a b}(c_{ia}^b)^*,
\end{equation}
El coeficiente complejo $b_{ab}^i$ obedece la siguiente relaci\'on:
\begin{equation}
c_{ia}^c b_{bc}^i=b_{ac}^i (c_{ib}^c)^*=0
\end{equation}
Si se considera el conmutador de $J_{\m\n}$ con $Q_{\a a}$ y $\bar Q_{\dot\a a}$, estos pueden ser de la forma:
\begin{equation}
[J_{\m\n},Q_{\a a}]=Q_{\b b}c_{\m\n\a}^{\b}X_a^b,\ \ \ \ \ \ \ [J_{\m\n},\bar Q_{\dot\a a}]=\bar Q_{\dot\b b}\bar c_{\m\n\dot\a}^{\dot\b}\bar X_a^b
\end{equation}
dado que los conmutadores se deben transformar como:
\begin{eqnarray}
([1^2_+]+[1^2_-])\times\D_+&=&(\left[ 0\times 1\right]+\left[ 1\times0\right]) \times (\left[0\times \ft 12 \right])=\left[ 0\times\ft 32\right]+\left[ 0\times\ft 12\right]+ \left[ 1\times\ft 12\right]+\left[ 0\times\ft 12\right]
\\
&=&\D_++\left[\D:1\right]_-+\left[\D:1^2\right]_+
\nn\\
([1^2_+]+[1^2_-])\times\D_-&=&(\left[ 0\times 1\right]+\left[ 1\times0\right]) \times (\left[\ft 12\times 0 \right])=\left[ \ft 32\times 0\right]+\left[ \ft 12\times 0\right]+\left[ \ft 12\times 1\right]+\left[ \ft 12\times 0\right] 
\nn\\
&=&\D_-+\left[\D:1\right]_++\left[\D:1^2\right]_-
\end{eqnarray}
Las constantes de estructura $(c_{\m\n})_\a^\b$ y $(\bar c_{\m\n})_{\dot\a}^{\dot\b}$ son elementos matriciales de la representaci\'on 2-Dim del \'algebra del grupo de Lorentz, relacionadas con $J$ como $\bar c_{\m\n}=Jc_{\m\n}J^{-1}$. Las cuales pueden ser como $c_{\m\n}\sim \s_{\m\n}\equiv {1\over 2i}[\g_\m,\g_\n]$ ($\g_\m\Rightarrow$ las matrices de Dirac), lo que resulta en transformaciones de la forma $\D\equiv \D_++\D_-$. En resumen:
\begin{equation}
\{Q_{\a a},Q_{\b a}\}=I\e_{\a \b}Z_{ab},\ \ \ \ \ \ \ 
\{Q_{\a a},\bar Q_{\dot\b b}\}=K_\m a_{\a \dot\b}^\m \d_{ab} 
\end{equation}
\begin{equation}
[K_\m, Q_{\a a}]=0=[K_\m,\bar Q_{\dot\a a}] \ \ \ \ \ \ \  
[J_{\m\n},Q_{\a a}]=Q_{\b b}c_{\m\n\a}^\b X_a^b\ \ \ \ \ \ \ [J_{\m\n},\bar Q_{\dot\a a}]=\bar Q_{\dot\b b}\bar c_{\dot\m\dot\n\dot\a}^{\dot\b} \bar X_a^b
\label{1000}
\end{equation}
Ahora bien ampliando el grupo de Poincar\'e a un grupo que contenga adem\'as generadores supersim\'etricos $\{Q_{\a a},\bar Q_{\a a}\}$ y otros generadores de simetr\'{\i}a interna. Los cuales cumplen las anteriores relaciones de conmutaci\'on y anticonmutaci\'on, obtenemos un grupo `supersim\'etrico'.\cite{Haase}

\subsubsection{El \'Algebra Supersim\'etrica 4-Dim (N=1)}
Para $N=1$ la carga central $Z_{ab}$ desaparece por antisimetr\'{\i}a, y los coeficientes $c_{i1}^1\equiv c_i$ son reales. La identidad de Jacobi para $[[Q,B],B]$ nos conlleva a ver que la constante de estructura $c_{i\m}^k$ desaparece, tal que el \'algebra de la simetr\'{\i}a interna es abeliana, entonces:
\begin{equation}
[Q_\a, B_i]=Q_{\a}c_i\ \ \ \ \ \ \ [\bar Q_{\dot\a},B_i]=-\bar Q_{\dot\a}c_i \ \ \ \ \ \mbox{re-escalando}\ B_i \ \ \ \Rightarrow \ \ \ [Q_\a, B_i]=Q_{\a}\ \ \ \ \ \ \ [\bar Q_{\dot\a},B_i]=-\bar Q_{\dot\a}
\end{equation}
Claramente solo una combinaci\'on independiente de los generadores abelianos ha de tener conmutadores distintos de cero con $Q_\a$ y $\bar Q_{\dot\a}$, por lo tanto se denota este generador de $U(1)$ por $R$, donde:
\begin{equation}
[Q_\a, R]=Q_{\a}\ \ \ \ \ \ \ [\bar Q_{\dot\a},R]=-\bar Q_{\dot\a}
\end{equation}
Esta \'algebra SUSY $N=1$ en general posee una simetr\'{\i}a interna (global) $U(1)$ conocida como {\it Simetr\'{\i}a R}. Se puede notar que los generadores de SUSY tienen una R-carga $+1$ y $-1$ respectivamente.

\subsection{Casimir SUSY}
Dado que las irreps de SUSY pueden ser caracterizadas por un conjunto de observables que mutuamente conmutan, se necesita mostrar los operadores Casimir, esto es suficiente para el \'algebra SUSY con N generadores.\\
\\
Al construir las representaciones del grupo de Poincar\'e, estas son rotuladas por los Casimir: \begin{itemize}
\item $K.K=K^\m K_\m$, con valores propios $k^2$ $\Rightarrow$ ($k=mc/\hbar$, $m\rightarrow$ la masa de la part\'{\i}cula).
\item $W.W=W^\m W_\m$ con $W_\m$ el vector de Pauli-Lubanski  $\Rightarrow$(operador de esp\'{\i}n generalizado), definido en (\ref{pauli}):
\begin{itemize}
\item Estados Masivos: $W.W$ tiene valores propios $k^2s(s+1)$ con $s=0,1/2,1,...$  
\item Estados No Masivos: $(k=0)$ se tiene que ($K.K=0$)\  y\  ($W.W=0$), cuya soluci\'on en la base de momento es dada como:
\footnote{Para estados no masivos es posible obtener estados en una base standard $k^\m=(k,0,0,k)$, donde:
$$
W^\m=k(\ \ J_{12},\ \ (J_{23}-J_{02}),\ \ (J_{31}+J_{01}),\ \ J_{12}\ \ )\ \ \mbox{ con }\m=0,..3
$$
Con
$$
[W^1,W^2]=0,\ \ [J_3,W^1]=iW^2, \ \ [J_3,W^2]=-iW^1,\ \ \mbox{ con }\ J_3=J_{12}=W_0/k
$$
El \'algebra de estos generadores es $SO(2)\otimes T_2$. Se puede obtener un espectro discreto de estados propios de momento tomando $W_1=W_2=0$, donde los valores propios semienteros $\l$ de $J_3$ introducen los n\'umeros cu\'anticos discretos de helicidad y etiquetan los estados propios de momento. Entonces se obtiene:
$$
W_0|k,\l\rangle=\l k_0|k,\l\rangle=,\ \ \ W_1|k,\l\rangle=W_2|k,\l\rangle=0,\ \ \ W_3|k,\l\rangle=\l k_3|k,\l\rangle
$$ 
De manera compacta.
$$
W_\m|k,\l...\rangle=\l k_\m|k,\l...\rangle \ \ \ \ \ o \ \ \ \ \ W_\m=\l K_\m \ \ \ \ \ \Rightarrow \ \ \ \ \ W.W=0
$$
}  
\begin{equation}
W_\m\equiv\l K_\m \ \ \ \ \ \ \ \l \rightarrow \mbox{ la helicidad.}
\end{equation}
\end {itemize}
\end {itemize}
En general se tiene que:
\begin{equation}
[W_\m,K_\n]=0 \ \ \ \ \ W.K=0 \ \ \ \ \ W_0=(J.K)_0 
\label{1005}
\end{equation}
En el caso de un \'algebra SuperPoincar\'e, el operador $K^\m K_\m$ (cuadrado de la masa) es un Casimir, donde los estados supermultipletes poseen la misma masa. Sin embargo $W.W$ ya no es un Casimir, dado que los supermultipletes contienen estados con diferentes `espines'  rotulados por los valores propios de $W.W$, donde: 
\begin{eqnarray}
[W_\m,Q_{\a a}]&=&{1\over 2}\e^{\m\s\n\r}K_\s [J_{\n\r},Q_{\a a}]
={1\over 2} \e^{\m\s\n\r}K_ \s Q_{\b b}c_{\n\r\a}^{\b} X_a^b
=K_\s Q_{\b b}\tilde {c}_{\b}^{\m\s\a} X_a^b \neq 0
\label{1006}
\end{eqnarray}
Esto muestra que $J^{\m\n}$ no conmuta con $Q$ y $\bar Q$, lo que explica porqu\'e $W.W$ no puede ser un Casimir. Se tiene entonces que los nuevos Casimir para SUSY son $K.K$ y $C:C$. Se introduce entonces un operador vectorial $N_\m$:
\begin{equation}
N_\m\equiv \bar Q_{\dot\a}Q_\b \bar a_\m^{\dot\a \b}
\end{equation}
Donde
\begin{equation}
C:C\equiv {1\over 2}C_{\m\n}C^{\m\n}, \ \ \ \ \ C_{\m\n}\equiv B_\m K_\n -B_\n K_\m, \ \ \ \ \ B_\m \equiv W_\m- {1\over 4}N_\m \equiv W_\m-{1\over 4}\bar Q_{\dot\a}Q_\b \bar a_\m^{\dot\a \b}
\end{equation}
Con $a_\m^{\dot\a \b}\sim a_{\dot\a \b}^\m$. Usando (\ref{1000}) y (\ref{1005}) se puede comprobar de forma inmediata que $B_\m$ cumple:
\begin{equation}
[B_\m,K_\n]=[W_\m-{1\over 4}\bar Q_{\dot\a}Q_\b \bar a_\m^{\dot\a \b},K_\n]=[W_\m,K_\n]-[{1\over 4}\bar Q_{\dot\a}Q_\b \bar a_\m^{\dot\a \b},K_\n]=0
\end{equation}
Usando la identidad $[AB,C]=A\{B,C\}-\{A,C\}B$. Se verifica que:
\begin{equation}
\left[\bar Q_{\dot\b}Q_\g a_\m^{\dot\b\g},Q_\a\right]
=\bar Q_{\dot\b}a_\m^{\dot\b\g}\{Q_\g,Q_\a\}-\{\bar Q_{\dot\b},Q_\a\}Q_\g a_\m^{\dot\b\g}
=-2K_\m Q_\a +4iK_\n Q_\b b_{\m\a}^{\n\b}
\label{1020}
\end{equation}
De los conmutadores (\ref{1006}) y (\ref{1020}) se tiene que:
\begin{eqnarray}
[B_\m,Q_\a]&=&[W_\m,Q_\a]-{1\over 4}[\bar Q_{\dot\n}Q_\b\bar a_\m^{\dot\n \b},Q_\a]={1\over 2}K_\m Q_\a \label{1021}\\
B_\m K^\m &=& {1\over 4} N_\m K^\m
\end{eqnarray}
Esto implica:
\begin{equation}
[C_{\m\n},Q_\a]=[B_{\m},Q_\a]K_\n -[B_{\n},Q_\a] K_\m=0
\end{equation}
Tomando C:C obtenemos:
\begin{equation}
C:C = (B.B)(K.K)-(B.K)^2
\end{equation}
\begin{itemize}
\item Para estados masivos ($m\neq 0$). $(C:C)$ define un n\'umero cu\'antico de supersp\'{\i}n $\x$.
\begin{equation}
[C:C]|m,\x\rangle=[(B.B)-{1 \over K^2}(B.K)^2 ]|m,\x\rangle \equiv m^2\x (\x +1)|m,\x\rangle
\end{equation}
\item Para estados no masivos se sigue que:
\begin{equation}
K_\m a^{\m\dot\a \a}Q_{\a a}=K_\m a^{\m\a\dot\a}\bar Q_{\dot\a}^a=0
\end{equation}
Lo que se reduce a $Q_{1a}=\bar Q_{1a}=0$ en la base standard considerada con anterioridad. 
\end {itemize}
De (\ref{1021}) y (\ref{1006}) se sigue que:
\begin{equation}
\left [W_\m,Q_\a \right ] =-{1\over 2} K_\m Q_\a \label{1100}\ \ \ \ \ 
\left [N_\m,Q_\a \right ] =4 K_\m Q_\a \label{1101}\ \ \ \ \ 
\left [B_\m,Q_\a \right ] ={1\over 2} K_\m Q_\a \label{1102}
\end{equation}
Para una base standard se encuentra que $N_1=N_2=0$ y $N_\m N^\m=0$. Entonces se puede encontrar una representaci\'on en la cual $W_1,W_2$ aniquila los estados (discretos) de n\'umeros cu\'anticos de superhelicidad. Obteniendose por lo tanto la siguiente relaci\'on:
\begin{equation}
B_\m + {1\over 2}R K_\m=-\l K_\m
\end{equation}
Donde
\begin{equation}
\left [R K_\m,Q_\a \right ] =-K_\m Q_\a 
\end{equation}
De igual forma se obtiene que $[R+(1/2K^2)(N.K)]$ conmuta con $Q$ y $\bar Q$. lo que define una carga superquiral en el caso masivo. Para el caso no masivo se sigue de (\ref{1101}) que:
\begin{equation}
\left [(R K_\m + {1\over 4}N_\m),Q \right ] = 0
\end{equation}
Por lo tanto se puede escribir $R K_\m + {1\over 4}N_\m=g_5 K_\m$

\subsection{Clasificaci\'on de Irreps SUSY Sobre Estados Singletes de Part\'{\i}culas}\label{irreps-sp31}
Todas las posibles irreps de SUSY se pueden construir sobre estados f\'{\i}sicos que son clasificados en la base de valores propios de $|k\rangle$ con ($k^2 \geq 0$) y ($k=mc/\hbar$), si ($m\neq 0$) $\Rightarrow$ la masa de la part\'{\i}cula. 
\begin{itemize}
\item Si $K.K|k\rangle=k^2|k\rangle$ \ \ $\Rightarrow$ se considera la base standard $k_\m=(k,0,0,0)$
\item Si $K.K|k\rangle=0$ \ \ \ \ \ \ \ $\Rightarrow$ se considera la base standard $k_\m=(k,0,0,k)$
\end{itemize}
Por lo tanto los estados masivos y no-masivos son tratados por separado. Cabe recordar que para el caso de la simetr\'{\i}a de Poincar\'e, no se considera los tachyones ($k^2$ negativo), dado que $\{Q,\bar Q\}$ es definidamente positivo.

\subsubsection{Estados No-Masivos SUSY N.}
Se considera ahora una representaci\'on no-masiva correspondiente a $K.K=0$. El \'algebra SUSY en una base standard $k_\m=(k,0,0,k)$ se reduce a:
\begin{equation}
\{Q_{\a}^i, Q_{\b}^j\}|k_\m\rangle=\{\bar Q_{\dot \a}^i, \bar Q_{\dot \b}^j\}|k_\m\rangle=0 
\label{4.2}
\end{equation}
y
\begin{equation}
\{Q_{\a}^i,\bar Q_{\dot \b}^j\}|k_\m\rangle=2k\d^{ij}(\s_0 + \s_3)_{\a \dot \b}|k_\m\rangle \ \ \ \ \ \Rightarrow \ \ \ \ \ \{Q_{2}^i \bar Q_{\dot 2}^j + \bar Q_{\dot 2}^j Q_{2}^i \}|k_\m\rangle=0
\label{4.1}
\end{equation}
Por lo tanto $Q_{2}^i,\bar Q_{\dot 2}^i$ mutuamente anticonmutan y actuan como un operador nulo ($Q_2^i=\bar Q_{\dot 2}^i=0$) en la representaci\'on de estados. Se obtiene que los generadores $Q_1^i, \ Q_{\dot 1}^i$ generan un \'algebra de Clifford con $N$ grados de libertad fermi\'onicos:
\begin{equation}
\{Q_1^i, \bar Q_{\dot 1}^j\}=\d^{ij}\ \ \ \ \ \{Q_1^i, Q_1^j\}=\{\bar Q_{\dot 1}^i, \bar Q_{\dot 1}^j\}=0
\label{4.4}
\end{equation}
Reescalando
\begin{equation}
\bar a^i= {1 \over 2\sqrt{k}} Q_{1}^i \ \ \ \ \ a^i=-{1 \over 2\sqrt{k}} \bar Q_{\dot 1}^i
\label{4.5}
\end{equation}
Donde $[W_3,a_\a^i]=-{1\over 2}(\s_3 a^i)_\a$, $[W_3,\bar a^{\dot \a i}]=-{1\over 2}(\s_3 \bar a^i)^{\dot \a}$, se obtiene entonces:
\begin{equation}
W_3 a^i= a^i (W_3- {1 \over 2}) \ \ \ \ \ W_3 \bar a^i= \bar a^i (W_3+ {1 \over 2})
\end{equation}
Una representaci\'on del \'algebra SUSY es caracterizada por un estado base de Clifford, etiquetado por la base de tetramomento $k$ y la helicidad $\l$.
Mostrando que todos los generadores $a^i$ `bajan' la helicidad por $1/2$ y todos los $\bar a^i$ lo `levantan' . Los generadores del \'algebra `menor' relevantes en la f\'{\i}sica son $\{W_3, a^i, \bar a^i\}$ junto con los generadores de simetr\'{\i}a interna. Tomando el estado $|\l_0\rangle$ con el n\'umero m\'aximo de helicidad $\l_0 >0$ en un supermultiplete, este satisface:
\begin{equation}
\bar a^i |\l_0\rangle=0 \ \ \ \ \ i=1,2,...,N
\end{equation}
Se ha suprimido $k$. Entonces se puede construir la siguiente cadena de estados hacia abajo al aplicar los $a^{i,}s$
\begin{equation}
|\l_0\rangle,\ |\l_0-1/2,i_1\rangle...,\ |\l_0-l/2,i_1,...,i_l\rangle,\ |\l_0-N/2,i_1,...,i_N\rangle
\end{equation}
Con
\begin{equation}
\l_{min}= \l_0-N/2 \ \ \ \l_{max}= \l_0 \ \ \ \Rightarrow \d\l = \l_{max}-\l_{min}=N/2
\end{equation}
Donde la multiplicidad de un estado $|\l_0-l/2,i_1,...,i_l\rangle=a^{i_1}...a^{i_l}|\l_0\rangle$ es dada por:
\begin{equation}
{N\choose l} \ \ \ \mbox{con }0\leq l \leq N
\label{multiplicidad-0}
\end{equation}
Este es completamente antisim\'etrico en las etiquetas $(i_1,...,i_l)$, dado que los a's son antisim\'etricos. En general una teor\'{\i}a cu\'antica de Lorentz covariante es invariante CPT, esto implica que cualquier estado con helicidad $(\l)$ puede ser similar a un estado con helicidad $(-\l)$, el cual satisface:  
\begin{equation}
a^i |-\l_0\rangle=0
\end{equation}
Luego, es posible obtener la siguiente cadena de estados.
\begin{equation}
|-\l_0\rangle,\ |-\l_0+1/2,i_1\rangle...,\ |-\l_0+l/2,i_1,...,i_l\rangle,\ |-\l_0+N/2,i_1,...,i_N\rangle
\end{equation}
Donde el estado $|-\l_0+l/2,i_1,...,i_l\rangle=\bar a^{i_1}...\bar a^{i_l}|-\l_0\rangle$ tiene una multiplicidad dada por (\ref{multiplicidad-0}). La consistencia de esto requiere que:
\begin{equation}
-\l_0+N/2   \leq \l_0  \ \ \ \ o \ \ \ \ N\leq 4\l_0
\end{equation}
El supermultiplete CPT auto-conjugado es obtenido cuando $N=4\l_0$, por lo tanto las dos cadenas coinciden. Donde la multiplicidad total en este caso es: 
$$
\sum_{l=0}^N {N\choose l }=(1+1)^N=2^N
$$
con $2^{N-1}$ estados bos\'onicos y $2^{N-1}$ estados fermi\'onicos. Un multiplete de supergravedad con un gravit\'on (y $N$ gravitinos de esp\'{\i}n $3/2$) puede ser construido para $N\leq 8$, donde los supermultipletes con $N=8,\l_0=2$ y $N=4,\l_0=1$ son autoconjugados:
$$
\begin {array} {lcccccccccc} 
\hline 
N=8\ \mbox{(Supergravedad)} & \l_0=2 & & & & & & & & & \mbox{Total}\\
\hline 
\mbox{Helicidad} & -2 & -3/2 & -1 & -1/2 & 0 & 1/2 & 1 & 3/2 & 2 & \\
\hline 
\mbox{Estados}   & 1 & 8 & 28 & 56 & 70 & 56 & 28 & 8 & 1 & =256\\
\hline
\end {array}
$$
$$
\begin {array} {lcccccc} 
\hline 
N=4\ \mbox{(Teor\'{\i}as de Yang-Mills)} & \l_0=1 & & & & &\mbox{Total}\\
\hline 
\mbox{Helicidad} & -1 & -1/2 & 0 & 1/2 & 1 & \\
\hline 
\mbox{Estados}   & 1 & 4 & 6 & 4 & 1 & =16\\
\hline
\end {array}
$$
Para el caso $N=7$, el supermultiplete $\l_0=2$ tiene el mismo contenido de part\'{\i}cula que $N=8, \ \l_0=2$. Lo mismo es cierto para los supermultipletes $N=4, \ \l_0=1$ y $N=3, \ \l_0=1$. Para el caso de esp\'{\i}n $3/2$  se tienen acoplamientos no renormalizables, por lo tanto se requiere que $N \leq 4, \l_0\leq 1$ para teor\'{\i}as renormalizables.\\
\\
Si se parte de una teor\'{\i}a de campos consistente con estados de helicidad menor o igual a 2 (gravitones), entonces es necesario que $N\leq 8$. Para $N\geq 9$ se encuentra una helicidad $5/2$ y superiores, lo cual no es consistente con esto. Adem\'as habr\'{\i}a m\'as de un estado con helicidad 2 (gravitones). Este requerimiento f\'{\i}sico conlleva a un m\'aximo de 8 generadores supersim\'etricos independientes.

\subsubsection{Ejemplo: Estados No-Masivos SUSY N=1.}
Para estados no-masivos ($K.K=0$) desde la base de referencia $k_\m \equiv (k,0,0,k)$, tenemos que:
\begin{equation}
C:C|k_\m\rangle=-2k^2(B_0-B_3)^2|k_\m\rangle=-{1\over 2}k^2 \bar Q_{\dot 2}Q_2\bar Q_{\dot 2} Q_2|k_\m\rangle=0
\label{N1.1}
\end{equation}
Teni\'endo en cuenta (\ref{4.1}), obtenemos:
\begin{equation}
\{Q_{\dot 1},\bar Q_{\dot 1}\}|k_\m\rangle=4k|k_\m\rangle\ \ \ \ \ \ \ \{Q_{\dot 2},\bar Q_{\dot 2}\}|k_\m\rangle=0
\label{N1.2}
\end{equation}
Para definir el estado de vacio $|\O\rangle$ se debe tener en cuenta de las ecuaci\'ones (\ref{N1.1},\ref{N1.2}) que el operador de creaci\'on $\bar Q_{\dot 2}$ construye estados de norma cero
\begin{equation}
\langle \O|Q_2 \bar Q_{\dot 2} |\O\rangle=0
\end{equation}
Esto significa que se puede tomar el conjunto $\bar Q_{\dot 2}$ igual a cero en el sentido de operador, como se mencion\'o antes en (4.1) . Por lo tanto solo queda un par de operadores creaci\'on y destrucci\'on, como en (\ref{4.5}):
\begin{equation}
a={1\over 2\sqrt{k}}Q_1, \ \ \ \ \ \ \ a^{\dagger}={1 \over 2\sqrt{k}} \bar Q_{\dot 1}
\end{equation}
Con $|\O\rangle$ no degenerado y con una helicidad definida $\l_0$. El operador de creaci\'on $a^\dagger$ transforma como $\D_+\equiv (0,{1\over 2})$ bajo el grupo de Lorentz, es decir incrementa la helicidad por ${1\over 2}$. Las irreps no-masivas de SUSY (N=1) contienen cada una (2) estados.
\begin{eqnarray}
|\O\rangle \ \ \ \ \ \ \ & &\mbox { helicidad }\l_0\\
a|\O\rangle \ \ \ \ \ \ \ & &\mbox { helicidad }\l_0 - {1\over 2}
\end{eqnarray}
Sin embargo este no es un estado propio CPT en general, se requiere tener por lo menos (2) irreps no-masivos SUSY para obtener los (4) estados con helicidad $\{-\l_0 ,-(\l_0 -{1\over 2}), \l_0 - {1\over 2}, \l_0\}$. Por ejemplo, para $N=1$ con $4$ estados de helicidad se obtiene:
$$
\begin {array} {lcccccc} 
\hline 
\mbox{Supermultiplete Quiral} & \l_0=1/2 & & \\
\hline 
\mbox{Helicidad} & -1/2 & 0 & 1/2 \\
\hline 
\mbox{Estados}   & 1 & 1+1 & 1 \\
\hline
 & &2(S=0) + (S=1/2)\\
\hline
\end {array}
$$
$$
\begin {array} {lccccc} 
\hline 
\mbox{Supermultiplete Gauge} & \l_0=1 & & & & \\
\hline 
\mbox{Helicidad} & -1 & -1/2 & 1/2 & & 1 \\
\hline 
\mbox{Estados}   & 1 & 1 & 1 & & 1 \\
\hline
 & & (S=1) + (S=1/2)& & & \\
\hline
\end {array}
$$
$$
\begin {array} {lccccc} 
\hline 
\mbox{Multiplete de supergravedad simple} & \l_0=2 & & & & \\
\hline 
\mbox{Helicidad} & -2 & -3/2 & 3/2 & & 2 \\
\hline 
\mbox{Estados}   & 1 & 1 & 1 & & 1 \\
\hline
 & &(S=2) + (S=3/2)& & & \\
\hline
\end {array}
$$
Para $\l_0=3/2$ se obtiene un contenido de part\'{\i}cula con $(S=3/2)+(S=1)$.

\subsubsection{Estados Masivos SUSY $N$.}
Para la representaci\'on masiva ($m\neq 0$), el \'algebra SUSY en el sistema de reposo $k_\m=(k,0,0,0)$), toma la forma:
\begin{equation}
\{Q_\a^i,\bar Q_{\dot\b}^j\}=2k\d_{\a \dot \b} \d_{ij}\ \ \ \ \ 
\{Q_\a^i,Q_\b^j\}=\{\bar Q_{\dot\a}^i,\bar Q_{\dot\b}^j\}=0\ \ \ \mbox{con } i,j=1,...N
\end{equation}
Reescalando se puede tener:
\begin{equation}
a_{\a}^i={Q_\a^i\over \sqrt{2k}} \ \ \ \ \ (a_{\a}^i)^\dagger={\bar Q_{\dot\a}^i\over \sqrt{2k}}
\end{equation}
Los cuales satisfacen el \'algebra de Clifford en 2N dim. Los estados de la representaci\'on pueden ser ordenados en multipletes de esp\'{\i}n de alg\'un estado base (o vacio) $|\O\rangle$ de alg\'un esp\'{\i}n dado, aniquilado por los operadores $a_\a^i$. los otros estados de la representaci\'on son dados como:
\begin{equation}
|a_{\a_1}^{i_1}...a_{\a_N}^{i_N}\rangle=(a_{\a_1}^{i_1})^\dagger...(a_{\a_N}^{i_N})^\dagger|\O\rangle
\end{equation}
El estado base $|\O\rangle$ tiene esp\'{\i}n $s$, el estado de esp\'{\i}n maximal tiene esp\'{\i}n $s+N/2$ y el estado de esp\'{\i}n minimal tiene:
\begin{itemize}
\item Si $s\geq N/2$ \ \ $\Rightarrow$ \ \ esp\'{\i}n \ \ $s-N/2$  
\item Si $s<N/2$ \ \ $\Rightarrow$ \ \ esp\'{\i}n \ \ \ \ $0$  
\end{itemize}
Cuando el estado base $|\O\rangle$ tiene esp\'{\i}n 0, el n\'umero total de estados es igual a $2^{2N}$ con $2^{2N-1}$ estados fermi\'onicos (construidos con un n\'umero impar de operadores $(a_\a^i)^\dagger$) y $2^{2N-1}$ estados bos\'onicos (construidos con un n\'umero impar de 
operadores $(a_\a^i)^\dagger$). El esp\'{\i}n maximal es $N/2$ y el esp\'{\i}n minimal es 0.\\
\\
Como se ilustra en la sgte secci\'on para el caso SUSY N=1, el estado base $|\O\rangle$ tiene sp\'{\i}n j, los estados multiplete tienen esp\'{\i}n $(j,j+1/2,j-1/2,j)$, donde el estado base $|\O\rangle$ tiene esp\'{\i}n 0, el supermultiplete tiene 2 estados de esp\'{\i}n 0 y 1 estado de esp\'{\i}n 1/2. En la siguiente tabla se dan las dimensiones de la representaci\'on masiva con estados base $\O_s$ (de esp\'{\i}n $s$) para N=1,2,3,4. 
$$
\begin {array} {lcccc} 
\hline
            &          &N=1           &          &              \\  
\hline 
\mbox{Esp\'{\i}n} & \O_0 & \O_{1/2} & \O_1 & \O_{3/2} \\
\hline 
0           & 2        & 1            &          &              \\
\hline 
1/2         & 1        & 2            &    1     &              \\
\hline
1           &          & 1            &    2     &      1       \\
\hline
3/2         &          &              &    1     &      2       \\
\hline
2           &          &              &          &      1       \\
\hline
\end {array}
\ \ \    
\begin {array} {lccc} 
\hline
            &          & N=2          &         \\
\hline 
\mbox{Esp\'{\i}n} & \O_0 & \O_{1/2} & \O_1 \\
\hline 
0           & 5        & 4            &    1     \\
\hline 
1/2         & 4        & 6            &    4     \\
\hline
1           & 1        & 4            &    6     \\
\hline
3/2         &          & 1            &    4     \\
\hline
2           &          &              &    1     \\
\hline
\end {array}
\ \ \  
\begin {array} {lcc} 
\hline 
            & N=3      &              \\
\hline 
\mbox{Esp\'{\i}n} & \O_0 & \O_{1/2} \\
\hline 
0           & 14       & 14           \\
\hline 
1/2         & 14       & 20           \\
\hline
1           & 6        & 15           \\
\hline
3/2         & 1        & 6            \\
\hline
2           &          & 1            \\
\hline
\end {array}
\ \ \  
\begin {array} {lc} 
\hline 
            & N=4      \\
\hline 
\mbox{Esp\'{\i}n} & \O_0 \\
\hline 
0           & 42       \\
\hline 
1/2         & 48       \\
\hline
1           & 27       \\
\hline
3/2         & 8        \\
\hline
2           & 1        \\
\hline
\end {array}
$$

\subsubsection{Ejemplo: Estados Masivos SUSY N=1.}
Se analiza estos estados desde la base $k_\s \equiv (k,0,0,0)$, donde:
\begin{equation}
C_{\m\n}|k_\s\rangle=B_i k|k_\s\rangle\d_{\m}^i\d_\n^0-B_i k|k_\s\rangle\d_\m^0\d_\n^i
\end{equation}
y $B_i=W_i-\bar Q\times Q:a_i^{-1}$
\begin{equation}
B_i|k_\s\rangle=(J_i k-\bar Q \times Q:a_i^{-1})|k_\s\rangle\equiv kS_i|k_\s\rangle
\end{equation}
Por lo tanto se puede escribir:
\begin{equation}
C:C|k_\s\rangle=k^2 B.B|k_\s \rangle=k^4 S.S|k_\s\rangle, \ \ \ \ \ \ \ S_i\equiv J_i-{1\over k}\bar Q\times Q:a_i^{-1}
\end{equation}
Con $S_i$ el operador espinorial, definido aqu\'{\i} como un operador de rotaci\'on en la base standard. teni\'endose que $\bar Q\times Q:a_i^{-1}$ obedece el \'algebra $SO(3)$.  
\begin{eqnarray}
[\bar Q \times Q:a_i^{-1},\bar Q \times Q:a_j^{-1}]&=&(\bar Q_{\dot\a}Q_{\b}\bar Q_{\dot\g}Q_{\d}-\bar Q_{\dot\g}Q_{\d}\bar Q_{\dot\a}Q_{\b})(a^{-1})_{i}^{\dot\a\b}(a^{-1})_{j}^{\dot\g\d}\\
&=&K_\m(\bar Q_{\dot\a}Q_\d a_{\dot\b\g}^\m+\bar Q_{\dot\g}Q_\b a_{\d\dot\a}^\m)(a^{-1})_{i}^{\dot\a\b}(a^{-1})_{j}^{\dot\g\d}
\end{eqnarray}
Donde
\begin{equation}
[\bar Q \times Q:a_i^{-1},\bar Q \times Q:a_j^{-1}]|k_\s\rangle=k(\bar Q_{\dot\a}Q_{\d}a_{\b\dot\g}^0+\bar Q_{\dot\g}Q_\b a_{\d\dot\a}^0)(a^{-1})_{i}^{\dot\a\b}(a^{-1})_{j}^{\dot\g\d}|k_\s\rangle
\end{equation}
y
\begin{equation}
[S_i,S_j]=iS_k \e_{ij}^k
\end{equation}
$S.S$ tiene valores propios $s(s+1)$, con $s$ entero o semientero. Los conmutadores de $S_i$ con $Q$ y $\bar Q$ son proporcionales a los $K_i s$
\begin{equation}
[S_i,Q_{\a}]=[J_i,Q_\a]-{1\over k}[\bar Q\times Q:a_i^{-1},Q_{\a}]=Q_{\b}c_{i\a}^\b- {1\over k}(K_i Q_{\a}+K_\n Q_{\b}b_{\a i}^{\n\b})
\end{equation}
Usando $J_i=A_i+\bar A_i$, `el cual es vacio' en la base.
Los generadores $Q_{\a},\bar Q_{\dot\a}$ se comportan como 2 pares de operadores de creaci\'on y destrucci\'on en irreps masivas SUSY $N=1$ con $k$ y $s$ fijos.
\begin{equation}
\{ Q_\a,\bar Q_{\dot\b}\}=Ik a_{\a\dot\b}^0=2Ik\d_{\a\dot\b}
\end{equation}
Dado un estado definido por $|k,s\rangle$ se puede ahora definir un nuevo estado.
\begin{equation}
|\O\rangle\equiv Q_1 Q_2 |k,s\rangle, \ \ \ \ \ \ \ Q_1|\O\rangle=0=Q_2|\O\rangle
\end{equation}
$|\O\rangle$ es el estado de vacio de Clifford con respecto a los operadores de aniquilaci\'on fermi\'onicos $Q_1,Q_2$. Se puede notar que $|\O\rangle$ es degenerado $2s+1$ donde $s_3$ toma los valores $-s,...,+s$. Actuando sobre $|\O\rangle$, Dado que $S_i$ se reduce a $J_i$, se obtiene que $|\O\rangle$ es ahora un estado propio de rotaci\'on.
\begin{equation}
|\O\rangle=|k,j,m_j\rangle
\end{equation}
Por lo tanto todos las irreps en SUSY pueden ser caracterizadas por masa y esp\'{\i}n. Es conveniente definir convencionalmente los operadores de creaci\'on y destrucci\'on.
\begin{equation}
a_\a \equiv {1\over \sqrt {2k}}Q_\a, \ \ \ \ \ \ \ a_\a^{\dagger} \equiv {1\over \sqrt {2k}}\bar Q_\a
\end{equation}
Para un $|\O\rangle$ la irreps masiva SUSY es dada como:
\begin{equation}
|\O\rangle,\ \ \ \ \ a_1^{\dagger}|\O\rangle,\ \ \ \ \ a_2^{\dagger}|\O\rangle,\ \ \ \ \ {1\over 2}a_1^{\dagger}a_2^{\dagger}|\O\rangle=-{1\over 2}a_2^{\dagger}a_1^{\dagger}|\O\rangle
\end{equation}
Donde el total de estados en la irreps masiva son $4(2j+1)$\\
\\
Calculando el esp\'{\i}n de estos estados con el uso de los conmutadores:
\begin{equation}
[S_3,a_1^{\dagger}]=-{1\over 2}a_1^{\dagger}\ \ \ \ \ \ \ [S_3,a_2^{\dagger}]={1\over 2}a_2^{\dagger}
\end{equation}
Por lo tanto para $|\O\rangle=|k,j,m_j\rangle$ se encuentran los estados con esp\'{\i}n $m_s=m_j,m_j-{1\over 2},m_j+{1\over 2},m_j$.\\
\\
Como un ejemplo se considera ahora $j=0$ o la irreps fundamental masiva (N=1). Donde $|\O\rangle$ tiene esp\'{\i}n cero, por lo tanto tenemos un total de (4) estados en la irreps, con esp\'{\i}n $m_s=0,-{1\over 2},{1\over 2}$ y $0$ respectivamente. El operador paridad intercambia $a_1^{\dagger}$ con $a_2^{\dagger}$, entonces uno de los estados de esp\'{\i}n cero es un pseudoescalar. Para estos cuatro estados corresponde un estado masivo de Weyl fermi\'onico, un escalar real y un pseudoescalar real. 

\newpage

\section  {Supersimetr\'{\i}a Desde Grupos $SO(n,m)$}\label{susy-ads}
El estudio de T.C.C. en espacios de Sitter $SO(d,1)$ y Anti-de Sitter $SO(d-1,2)$ ha tenido mucho empuje en los \'ultimos a\~nos y muestra ser un ambiente propicio para acercarce a un esquema de unificaci\'on. Dado que:
\begin{itemize}
\item En teor\'{\i}as de supergravedad gauge: si los estados base son invariantes bajo $N$ supersimetr\'{\i}as independientes, puede ser adicionalmente invariante $SO(N)$. Donde el grupo de simetr\'{\i}as total que incluye simetr\'{\i}as fermi\'onicas, es caracterizado por la extensi\'on graduada de $Sp(4)\times SO(N)$ (siendo $Sp(4)\sim SO(3,2)$ en el \'algebra) denotado por $OSp(N|4)$ (ver \cite{Ferrara-PVN,Freedman}). Dado que las representaciones de \'algebras supersim\'etricas con $N>8$ contienen estados de helicidad mayor que dos, no se construyen. Por lo tanto el grupo de simetr\'{\i}a mas grande permisible es dado por $OSp(8|4)$. Teniendo en cuenta que los campos en un supermultiplete no necesariamente deben tener la misma masa a diferencia del supergrupo $\ ^SP(N|3,1)$. 
\item Muy recientemente Maldacena \cite{Mald} encontr\'o una relaci\'on dual (correspondencia $AdS$/$CFT$) entre: teor\'{\i}as de campos superconformales ($CFT$), definidas en espacios planos d-Dimensionales que habitan sobre la frontera de un espacio-tiempo $AdS$ y la teor\'{\i}a-M (teor\'{\i}a de cuerdas) compactificada en un espacio $AdS$ d+1-Dimensional, la cual corresponde a una teor\'{\i}a de supergravedad gauge. Por ejemplo es posible establecer una correspondencia $AdS_7/CFT_6$ entre una teor\'{\i}a-M compactificada en $S^4$ y una teor\'{\i}a de campos comformales en seis dimensiones. El supergrupo de la teor\'{\i}a-M asociado con $N=4$ es $OSp(8^*|4)$ y es definido en un espacio $AdS_7\times S^4$ con un algebra bos\'onica $SO(6,2)\times USp(4)$ (ver \cite{Gunaydin}).
\end{itemize}
Cuando se construyen teor\'{\i}as de supergravedad gauge, estas permiten campos vectoriales gauge no-abelianos $SO(N)$. Donde el lagrangiano asociado contiene un t\'ermino cosmol\'ogico $\epsilon$ proporcional a $g^2$ y el t\'ermino masivo del gravitino $\bar \Psi_\mu \gamma^{\mu\nu}\Psi^\nu$ proporcional a $g$, con $g$ la constante de acoplamiento de los campos gauge no-abelianos. En el l\'{\i}mite $g\to 0$ esta teor\'{\i}a se reduce a una 
teor\'{\i}a de supergravedad ordinaria (SuperPoincar\'e $\Rightarrow ^SP(N|3,1)$). 

\subsection{El Grupo $SO(d-1,2)$}
Empezando con el grupo $SO(n,m)$, el cual se puede representar como el conjunto de las matrices $\{T\}_{p\times p}$ con $n+m=p$ que preservan $g$ (con $g$ una m\'etrica sobre $R^p$) y signatura $(n,m)$, es decir:
\begin{equation}
g(Tv,Tu)=g(v,u)  \ \ \ \ \ \mbox{con } \ v,u \in R^p \ \ \ \mbox{y } \ \ det(T)=1
\end{equation}
Con $\ft 12 (p-1)p$ generadores de grupo asociados.\\ 
\\
Tomando $n=d,\ m=1$ $(n=d-1,m=2)$, se puede hacer extensiones de las propiedades del grupo de Sitter $SO(4,1)$ (Anti-de Sitter $SO(3,2)$) al grupo $SO(d,1)$ $(SO(d-1,2))$ con $d\geq 4$. Este corresponde al grupo de simetr\'{\i}as maximal en un espacio $dS$ ($AdS$), con ${1\over 2}d(d+1)$ isometrias. El cual es descrito por una hipersuperficie embuida en un espacio $(d+1)$-dimensional definida como:\footnote{El espacio $dS$ ($AdS$) es un espacio homogeneo, lo cual significa que cualquier 2 puntos en \'el pueden ser relacionados por una isometr\'{\i}a. Para el caso $AdS$, este tiene la topolog\'{\i}a de $S^1 \mbox{ [time] }\times {\bf R}^{d-1}$. Al Desenrrollar $S^1$, se encuentra el espacio de cubertura universal denotado por $CadS$, El cual tiene la topolog\'{\i}a de ${\bf R}^{d}$.} 
\begin{equation}
(y^1)^2+(y^2)^2+...+(y^{d-1}) \pm (y^d)^2 - (y^{d+1})^2 = \h_{ij} \,y^i y^j = -g^{-2} 
\ \ \ \ \ \ \ \mbox{con }g\simeq 1/R
\end{equation}
Siendo $y^i$ ($i= 1,2,...,d-1,d,d+1$) las coordenadas y $R$ el radio $dS$($AdS$). Esta hipersuperficie es invariante bajo transformaciones lineales que preserven la  m\'etrica $\h_{ij}  =diag(++...+\pm,-)$, las cuales constituyen el grupo $SO(d,1)$ ($SO(d-1,2)$) con ${1\over 2}d(d+1)$ generadores denotados por $J_{ij}$ y un \'algebra definida como:
\begin{equation}
\left[ J_{ij} ,J_{kl} \right] = i(J_{ik}  \h_ {jl}-J_{il}  \h_ {jk}+J_{jl}  \h_ {ik}-J_{jk}  \h_ {il})
\label{ads/algebra}
\end {equation}
Para $SO(d-1,2)$ (caso de nuestro inter\'es), los generadores se pueden representar de forma espinorial como una combinaci\'on de matrices gamma: 
\begin{equation}
J_{ij}\ \Rightarrow \ \ \ \left\{ 
\begin{array}{lll}
\ft 12 \G_{ij}& \mbox{si} & i,j=1,\ldots,d-1,d \ \ \ \G_{ij}\sim [\G_i,\G_j]\\
\ft 12\G_j  &\mbox{si} & i=d+1\,, j= 1,\ldots,d-1,d \end{array}\right.
\end{equation}
Las cuales satisfacen la propiedad de Clifford $\{\G^i\,,\, \G^j\} = 2\, \h^{ij}\, {\bf 1}$, con $\h^{ij} = {\rm diag}\,(++...+-)$.

\subsection{Super\'algebras Anti-de Sitter} \label{superadS}
En general cuando se considera un super\'algebra Anti-de Sitter significa construir un \'algebra combinada de simetr\'{\i}as bos\'onicas y fermi\'onicas, donde su estructura cambia dr\'asticamente si $d\geq 7$ (Ver \cite{Nahm}). Para $d\leq7$ el sub\'algebra bos\'onica correspondiente a las simetr\'{\i}as del espacio-tiempo coincide con el \'algebra Anti-de Sitter. Cuando introducimos $N$ generadores de supersimetr\'{\i}a, cada uno transforma como un espinor bajo el grupo Anti-de Sitter, donde el sub\'algebra bos\'onica ya no es restringida a un \'algebra Anti-de Sitter, se necesitan generadores bos\'onicos extras $(N)$ que transformen como un tensor antisim\'etrico de rango superior bajo el grupo de Lorentz. Estos $N$ generadores transforman bajo un grupo compacto y aparecen en los anticonmutadores $\{Q,\bar Q\}$. Algo similar ocurre cuando se construye un \'algebra SuperPoincar\'e (N-extensi\'on) asociada con un espacio de Minkowski plano, donde los generadores bos\'onicos corresponden al grupo Poincar\'e, aumentados con los generadores de un grupo compacto asociado a las rotaciones de las supercargas. Los cuales son considerados nulos para una super\'algebra con cargas centrales (ver ecuaciones (\ref{QQ0}) y (\ref{QQ1})).\\
\\
Tomando la clasificaci\'on de Nahm \cite{Nahm}. Los requerimientos para un \'algebra superconformal en $d$ o un super-\'algebra Anti-de Sitter en $d+1$ son:
\begin{itemize}
\item $SO(d-1,2)$ \ o \ $SO(d,1)$  debe aparecer como un subgrupo factorizado de la parte bos\'onica del super\'algebra. Para Nahm, este requirimiento est\'a motivado por el teorema de Coleman--Mandula, pero se puede exijir igualmente que el \'algebra bosonica sea el \'algebra de isometrias de un espacio que tenga al espacio $AdS$ como una factorizaci\'on.
\item  Los generadores fermi\'onicos deben colocarse en una representaci\'on espinorial del grupo.
\end{itemize}
Obteniendose la siguiente clasificaci\'on para las \'algebras bos\'onicas simples con $4\leq d\leq 7$ que contengan $SO(3,2)$ o $SO(4,1)$:
\begin{equation}
\begin{array}{lll}
SO(3,2) \oplus SO(N)    &\mbox{con}\ N=1,2,...  & \qquad SO(4,1) \oplus U(1) \\
SO(4,2) \oplus U(N)     &\mbox{con}\ N\neq 4    & \qquad SO(4,2) \oplus SU(4)\\
SO(5,2) \oplus SU(2)    &                       & \qquad SO(6,1) \oplus SU(2)\\
SO(6,2) \oplus SU(N,H)  &\mbox{con}\ N=1,2,...  & 
\end{array}
\end{equation}
$H$ indica los cuaterniones. Sobre las \'algebras se establece el siguiente isomorfismo:
\begin{eqnarray}
SO(3,2)&=&Sp(4) \ \ \ \ \ \ \ \ \ \ SO(4,1)=USp(2,2) \nn\\ 
SO(4,2)&=&SU(2,2)\ \ \ \ \ \ \ SO(6,2)=SO^*(8) 
\label{isomBosAlg}
\end{eqnarray}
Dado que el prop\'osito de este trabajo es encontrar grupos $SO(n,m)$ que permitan generar un grupo $P(3,1)$ mediante una contracci\'on ya sea del grupo de Sitter o Anti-de Sitter y a la vez su extensi\'on supersim\'etrica permita acoplar grupos gauge de simetr\'{\i}a interna. Usando la anterior clasificaci\'on: el grupo $SO(n,m)$ minimal que cumple estas condiciones es dado por $SO(3,2)$. Donde su extensi\'on supersim\'etrica es dada por $OSp(N|4)$. El cual clasifica las representaciones de part\'{\i}culas en teor\'{\i}as de supergravedad y es la extensi\'on graduada de $SO(3,2)\times SO(N)$. En lo que sigue se considerar\'a este grupo, donde $SO(3,2)$ tiene 10 generadores herm\'{\i}ticos $J_{ij}$ y $N(N-1)/2$ generadores $Y_{ab}$ de $SO(N)$ con $a,b=1,...,N$, con un \'algebra asociada:
\begin{eqnarray}
\{Q_\a^a, \bar Q_\b^b\} &=& i(\d^{ab} c_{\a\b}^{ij}J_{ij}+\d_{\a\b} Y^{ab})\nn\\
\left [J_{ij}, Q_\a^a\right]&=&-ic_{ij\a}^\b Q_{\b}^a \nn\\
\left [Y^{ab}, Y^{cd}\right]&=&-i(\d^{ac}Y^{bd}-\d^{bc}Y^{ad}+\d^{bd} Y^{ac}-\d^{ad}Y^{bc})\nn\\
\left [Y^{ab}, Q_\a^c\right ] &=& i(\d^{ac} Q_{\a}^{b}-\d^{bc} Q_{\a}^a) 
\label{ads-superalgebra} 
\end{eqnarray}
Junto con (\ref{ads/algebra}), estas relaciones completan el super\'algebra Anti-de Sitter. Como se muestra el \'algebra Anti-de Sitter cambia su forma cuando consideramos $N$ generadores de supersimetr\'{\i}a, que rotan bajo la acci\'on de un grupo compacto. Surgen unas cargas bos\'onicas extras asociadas con tensores de Lorentz de mayor rango, que aparecen en el lado derecho de los anticommutadores $\{Q,\bar Q\}$. A diferencia de un espacio de Minkowski (plano), en un espacio $AdS$ el grupo de simetr\'{\i}a interna es $SO(N)$ en vez del grupo abeliano $[U(1)]^p$ generado por las cargas centrales ($p=$ n\'umero total de cargas centrales independientes). (ver \cite{Ferrara-PVN,Freedman,Ferrara})\\ 
\\
Las matrices $c^{ij}$ en (\ref{ads-superalgebra}) son definidas como:
\begin{equation}
c^{ij} = \left\{ 
\begin{array}{lll}
-\ft 12 \g^i & \mbox{para } & j= 5\ \ \ \ i=1,2,3,4\\[3mm]
\ft 14 \left[\g^i,\g^j\right]  &\mbox{para} & i,j=1,2,3,4 \end{array}\right.
\end{equation}
Aqu\'{\i} se usa la siguiente representaci\'on para las $\g$-matrices:
\begin{equation}
\g^0=\left (
\begin {array}{cc} 
1 & 0 \\
0 & -1 
\end{array}
\right )
\ \ \ \ \ 
\g^i=\left (
\begin {array}{cc} 
0 & \s^i \\
-\s^i & 0 
\end{array}
\right )
\end{equation}
Adem\'as se usa la parametrizaci\'on espinorial de Majorana como:
\begin{equation}
Q_\a^a=\left (
\begin {array}{c} 
a_\a^a \\
\e_{\a\b} \bar a_\b^a 
\end{array}
\right )
\ \ \ \ \ 
\e =\left (
\begin {array}{cc} 
0  & 1 \\
-1 & 0 
\end{array}
\right )
\label{spinormajorana}
\end{equation}
Siendo $\bar a_\a^a$ es el adjunto herm\'{\i}tico de $a_\a^a$. De (\ref{ads-superalgebra}) y (\ref{spinormajorana}) se obtiene que:
\begin{eqnarray}
\left [J_{45}, a_\a^a \right ] &=& -{1\over 2} a_{\a}^{a} \ \ \ \ \ \ \ \ \ \ \ \ \ \ \ \
\left [J_{45}, \bar a_{\a}^a \right ] = {1\over 2} \bar a_{\a}^a \nn\\
\left [J_{i}, a_\a^a \right ] &=& -{1\over 2} (\s_{a})_{\a}^{\b} a_{\b}^a\ \ \ \ \ \ \ \ \ \    
\left [J_{i}, \bar a_{\a}^a \right ] = {1\over 2} \bar a_{\b}^{a}(\s_{a})_{\a}^{\b}\ \ \ \ \ \ \ J_{ij}\equiv \e_{ij}^k J_k
\label{comajorana} 
\end{eqnarray}
Donde $\bar a_{\a}^a$ y $a_{\a}^a$ suben y bajan la energ\'{\i}a respectivamente por $1/2$, con $\bar a_{\a}^a$ un operador espinorial standard irreducible con respecto a las rotaciones espaciales. Usando (\ref{spinormajorana}) se puede escribir expl\'{\i}citamente (\ref{ads-superalgebra}) como:
\begin{eqnarray}
\{a_{\a}^a, a_\b^b \} &=& \d^{ab}(J_{45}+\s^iJ_i)_{\a\b} + i \d_{\a\b}Y^{ab} \nn\\
\e_{\a\g}\{\bar a_{\g}^a, \bar a_\b^b \}&=&\d^{ab}\s_{\a\b}^i J_i^+ \ \ \ \ \ \ \ 
\{a_{\a}^a, a_\g^b \} \e_{\g\b} = \d^{ab}\s_{\a\b}^i J_i^+
\label{newalgebra} 
\end{eqnarray}
Se puede ver que los operadores compactos en el sub\'algebra bos\'onica $SO(3,2)\times SO(N)$ pueden ser escritos como anticonmutadores de operadores de creaci\'on y aniquilaci\'on mientras que los operadores no-compactos siempre involucran ya sean 2 operadores de creaci\'on o 2 operadores de destrucci\'on. Esta exposici\'on es cierta para otras \'algebras no-compactas y super\'algebras cuando se expresan estos generadores en t\'erminos de productos de operadores: creaci\'on, aniquilaci\'on fermi\'onicos y bos\'onicos. (ver \cite{GunaNieuWarn})\\
\\
Como se mostr\'o en la secci\'on (\ref{superpoincare}), el espacio plano es consistente con supersimetr\'{\i}a. Esto tambi\'en es cierto para un espacio $AdS$, con un mismo n\'umero de isometrias pero ahora corresponden al grupo Anti-de Sitter. Los cuales est\'an relacionados al tomar el l\'{\i}mite $R\rightarrow \infty$ (o $g\to 0$, que corresponde a un espacio plano). Donde el algebra Anti-de Sitter es contraida en el \'algebra de Poincar\'e (ver secci\'on (\ref{sittertopoincare})). Mecanismo que tambi\'en puede ser aplicado a las super\'algebras reescribiendo los generadores como:
\begin{eqnarray}
Q_{\a}^a &\to& \varphi_{\a}^a\equiv {1\over \sqrt{R}} Q_{\a}^a \ \ \ \ \ \hbox{ con }\ \lim_{R\to \infty} \varphi_{\a}^a\equiv Q_{\a}^a\nn\\
J_{5\n} &\to& \P_{\n}\equiv {1\over R} J_{5\n} \ \ \ \ \ \ \ \hbox{ con }\ \lim_{R\to \infty} \P_{\n}\equiv K_\n \nn\\
J_{\m\n} &\to& J_{\m\n}
\end{eqnarray}
Por lo tanto el super\'algebra $OSp(N|4)$ toma la forma:
\begin{eqnarray}
\left[J_{\m\n},J_{\r\s}\right]&=&i(J_{\m\r}\h_{\n\s}-J_{\m\s}\h_{\n \r}+J_{\n\s}\h_{\m\r}-J_{\n\r}\h_{\m\s}) \label{contr-osp1}
\\ 
\left[\P_{\m},J_{\r\s}\right]&=&i(\P_{\r}\h_{\m\s}-\P_{\s}\h_{\m\r}) 
\ \ \ \ \ 
\left[\P_{\m},\P_{\n}\right]={i\h_{z}\over R^2_z}J_{\m\n}
\\
\left[Y^{ab},Y^{cd}\right]&=&-i(\d^{ac}Y^{bd}-\d^{bc}Y^{ad}+\d^{bd}Y^{ac}-\d^{ad}Y^{bc}) \label{contr-osp2}
\\
\left[Y^{ab},\varphi_\a^c\right]&=&i(\d^{ac}\varphi_{\a}^{b}-\d^{bc}\varphi_{\a}^a) \ \ \ \ \ \left[J_{\m\n},\varphi_\a^a\right]=-ic_{\m\n\a}^\b \varphi_{\b}^a \ \ \ \ \ \left[\P_{\n},\varphi_\a^a\right]=-\ft 1R (ic_{5\n\a}^\b \varphi_{\b}^a) 
\\ 
\{\varphi_\a^a,\bar\varphi_\b^b\}&=&i(\d^{ab}c_{\a\b}^{\m}\P_{\m}+\ft 1R (\d^{ab}c_{\a\b}^{\m\n}J_{\m\n}+\d_{\a\b}Y^{ab}))
\label{ads-superalgebra-mod} 
\end{eqnarray}
Tomando $R \to \infty$, obtenemos un \'algebra $\ ^SP(N|3,1)\times SO(N)$:
\begin{eqnarray}
\left[K_{\m},J_{\r\s}\right]&=&i(K_{\r}\h_{\m\s}-K_{\s}\h_{\m\r}) \ \ \ \ \ 
\left[K_{\m},K_{\n}\right]=0 \nn\\
\left[Y^{ab},Q_\a^c\right]&=&i(\d^{ac}Q_{\a}^{b}-\d^{bc}Q_{\a}^a) \ \ \ \ \ \left[J_{ij},Q_\a^a\right]=-ic_{ij\a}^\b Q_{\b}^a \ \ \ \ \ \left[K_{\n},\varphi_\a^a\right]=0
\nn\\ 
\{Q_\a^a,\bar Q_\b^b\}&=&i\d^{ab}c_{\a\b}^{\m}K_{\m}
\label{ads-superalgebra-mod-1} 
\end{eqnarray}
Junto con (\ref{contr-osp1}) y (\ref{contr-osp2}). En lo que sigue ver \cite{Heidenreich,GunaWarn,GunaNieuWarn} como complemento.

\subsection{Representaciones Unitarias Irreducibles de $OSp(N|4)$} \label{adSreps}
Para estudiar las irreps unitarias de $OSp(N|4)$ se emplea un m\'etodo an\'alogo al usado en la secci\'on (\ref{condicion-unitaria}). Entonces se asume primero la existencia de una representaci\'on espacial sobre la cual todos los operadores de $OSp(N|4)$ puedan actuar. Esto requiere que exista un estado de baja energ\'{\i}a y junto con la \'ultima relaci\'on en (\ref{newalgebra}) se obtenga la condici\'on:
\begin{equation}
a_{\a}^a|(E_0,s,...)E_0sm...\rangle=0
\label{condicion} 
\end{equation}
Lo cual implica automaticamente (\ref{ecua-propia}). Aqu\'{\i} se necesitan m\'as etiquetas que en (\ref{t-ecua-propia}) para designar los estados y su contenido $SO(N)$ de isosp\'in, donde los puntos indican las etiquetas de la representaci\'on de $SO(N)$. De (\ref{newalgebra}) se sigue que todas las combinaciones pares de operadores de simetr\'{\i}a $\bar a_\a^a$ pueden ser expresadas como elementos pares del \'algebra $OSp(N|4)$. Para construir la representaci\'on espacial del estado de vacio, se considera primero las combinaciones sim\'etricas:
\begin{equation}
\begin{array}{lllll}
\bar B = \bar B_0 \cup \bar B_1 \cup ... \cup \bar B_{2N}&\Rightarrow& 
\bar B_0 &=& \{ 1 \}, \ \ \ \ \ \bar B_1 = \{\bar a_{\a}^{a} \} \\
&&\bar B_2 &=& \{ \left [ \bar a_{\a}^a , \bar a_{\b}^b \right ] \} \ \ \ \ \ 
\bar B_3 = \{ \S \pm \bar a_\a^a \bar a_\b^b \bar a_\g^c \} \ \ \ \ \ \ \ etc. 
\end{array}
\label{antisimetrico} 
\end{equation}
$\S$ indica las sumas hasta segundo orden. Se tiene que $\bar B_n$ contiene ${2N \choose n}$ operadores y el conjunto $\bar B$ contiene $2^{2N}$ operadores. La representaci\'on espacial es definida como el generador de todos los vectores de la forma:
\begin{equation}
(M_1^+)^{n_1}(M_2^+)^{n_2}(M_3^+)^{n_3} \bar B |(E_0,s,...)E_0 s m ...\rangle
\label{generado} 
\end{equation}
Esta relaci\'on  es importante, ya que se necesita para estudiar la acci\'on de $\bar B$ sobre el vacio, dados los resultados previos en la representaci\'on de $SO(3,2)$. Todas las representaciones unitarias de $OSp(N|4)$ son del tipo:
\begin{equation}
D(E_0^{(1)},s^{(1)}) \oplus ... \oplus D(E_0^{(\t)},s^{(\t)}) \ \ \ \hbox{ con } \t < \infty
\label{unitaria} 
\end{equation}
Aqu\'{\i} se han suprimido las etiquetas de isosp\'{\i}n por simplicidad. En un espacio de Minkowski plano se conoce que todos los campos pertenecen a un supermultiplete que est\'a sujeto a unas ecuaciones de campo con la misma masa. Esto se debe a que los operadores de momentum conmutan con las cargas supersim\'etricas, siendo $K.K$ un operador Casimir. Una caracter\'{\i}stica que diferencia la representaciones de supersimetr\'{\i}a Anti-de Sitter de supersimetr\'{\i}a en el grupo Poincar\'e es que, en general, $E_0^{(1)} \neq E_0^{(2)} \neq ... \neq E_0^{(\t)}$. Teniendose que las part\'{\i}culas en un supermultiplete dado no necesariamente poseen una masa igual si se identifica $E_0 \rightarrow m$.\\
\\
Para $N=1$ el sub\'algebra bos\'onica de (\ref{ads-superalgebra}) es caracterizada por $SO(3,2)$, donde los nuevos operadores que se consideran son $\bar B_1=\{\bar a_\a , \a=1,2\}$ y $\bar B_{2}=\e^{\a\b} \bar a_\a \bar a_\b $. Teniendo en cuenta la ecuaci\'on (\ref{comajorana}), el operador $\bar a_\a$ es un operador $\D=1/2$, donde $\bar a_1$ y $\bar a_2$ suben y bajan respectivamente la componente z de esp\'{\i}n en un medio de la unidad. Al actuar $\bar a_1$ sobre el estado $|(E_0,s)E_0,\ s,\ m\rangle$ obtenemos:    
\begin{eqnarray}
\bar a_1|(E_0,s)E_0,\ s,\ m\rangle&=&R_+ \langle sm\ft 12 \ft 12|s+\ft 12,m+\ft 12\rangle|(E_0,s)E_0+\ft 12,\ s+\ft 12,\ m+\ft 12\rangle+\nn\\
&&R_- \langle sm\ft 12 \ft 12|s-\ft 12,m+\ft 12\rangle|(E_0,s)E_0+\ft 12,\ s-\ft 12,\ m+\ft 12\rangle
\label{a1-barra-propios}
\end{eqnarray}
Teniendo en cuenta los coeficientes de Clebsh-Gordan:
\begin{equation}
\langle sm\ft 12 \ft 12|s+\ft 12,m+\ft 12\rangle = \left ({s+m+1 \over 2s+1 }\right )^{\ft 12} \ \ \ \ \ \ \ \ \ \ 
\langle sm\ft 12 \ft 12|s-\ft 12,m+\ft 12\rangle = -\left ({s-m \over 2s+1 }\right )^{\ft 12} 
\end{equation}
Usando (\ref{newalgebra}) junto con (\ref{condicion}) para $N=1$. se obtiene que:
\begin{equation}
|R_+|^2=E_0+s\ \ \ \ \ \ \ |R_-|^2=E_0-s-1
\end{equation}
Partiendo de la representaci\'on $D(E_0,s)$ de $SO(3,2)$ en (\ref{a1-barra-propios}), aqu\'{\i} se han producido dos (2) nuevas representaciones $D(E_0+\ft 12,s+\ft 12)$ y $D(E_0+\ft 12,s-\ft 12)$. La segunda est\'a ausente si $s=0$ o $E_0=s+1$ para $s\geq \ft 12$.\\
\\
Se puede considerar la acci\'on del operador $\left [\bar a_1,\bar a_2 \right ]= \e_{\a\b}\bar a_\a \bar a_\b$ sobre el vacio, el cual baja la energ\'{\i}a por una unidad y deja igual los n\'umeros cu\'anticos de momento angular $j$ y $m$ del vacio. El nuevo estado puede ser obtenido como una superposici\'on  de estados de vacio $|(E_0+1,s)E_0+1,s,m\rangle$ de la representaci\'on $D(E_0+1,s)$ de $SO(3,2)$  y de estados de la representaci\'on $D(E_0,s)$ (ver ec. (\ref{a1-barra-propios})). Para $m=s$, obtenemos que:
\begin{eqnarray}
\left [\bar a_1\,\bar a_2\right ]|(E_0,s)E_0,\ s,\ s\rangle&=&R_0 |(E_0+1,s)E_0+1,\ s,\ s\rangle+
\a M_{1+2i}^+ |(E_0,s)E_0,\ s,\ s-1\rangle+\nn\\
&&\b J_{3}^+ |(E_0,s)E_0,\ s,\ s\rangle
\end{eqnarray}
Como en la secci\'on (\ref{representacion-unitaria-so32}) todos los estados se asumen ortogonales. Para calcular los coeficientes $R_0$, $\a$ y $\b$ se debe tener en cuenta que el estado $|(E_0+1,s)E_0+1,s,s\rangle$ debe ser aniquilado por el operador $M_a^-$ es cual es un estado de vacio de la representaci\'on $D(E_0+1,s)$. Usando (\ref{conm-pm}) y 
\begin{equation}
\left [M_a^-,\bar a_\a\right ]=(\e \s_a)_{\a}^{\b}a_\b
\end{equation}
Se obtiene:
\begin{eqnarray}
\a&=&-{\sqrt{s}\over E_0-1} \ \ \ \ \ \b=-{s\over E_0-1}\ \ \ \hbox{ para }\ s\neq0 \nn\\
\a&=&\b=0 \ \ \ \ \ \ \ \ \ \ \ \ \ \ \ \ \ \ \ \ \ \ \ \ \ \ \ \ \ \ \hbox{ para }\ s=0
\end{eqnarray}
Donde
\begin{equation}
||\left [\bar a_1\,\bar a_2\right ]|(E_0,s)E_0,\ s,\ s\rangle||^2=2E_0(2E_0-1)-4s(s+1)
\end{equation}
Comparando la norma en ambos lados, se obtiene:
\begin{equation}
|R_0|^2= {2\over E_0-1}(2E_0-1)(E_0+s)(E_0-s-1)
\end{equation}
De aqu\'{\i} se puede notar que la representaci\'on $D(E_0+1,s)$ est\'a ausente si se toma $s=0$ y $E_0=\ft 12$ o si $E_0=s+1$ para $s\geq \ft 12$. En resumen, se pueden tener las siguientes representaciones unitarias para $OSp(1|4)$:  
\begin{itemize}
\item Representaci\'on de Wess-Zumino ($E_0>\ft 12$)
\begin{equation}
D(E_0,0)\oplus D(E_0+\ft 12,\ft 12)\oplus D(E_0+1,0)
\end{equation}
\item Representaci\'on `masiva' de esp\'{\i}n superior ($E_0>s+1,s\geq \ft 12$ con $s=\ft 12,1,\ft 32,...$)
\begin{equation}
D(E_0,s)\oplus D(E_0+\ft 12,s+\ft 12)\oplus D(E_0+\ft 12,s-\ft 12)\oplus D(E_0+1,s)
\end{equation}
\item Representaci\'on `no-masiva' de esp\'{\i}n superior ($E_0=s+1,s \geq \ft 12$ con $s=\ft 12,1,\ft 32,...$)
\begin{equation}
D(s+1,s)\oplus D(s+\ft 32,s+\ft 12)
\end{equation}
\item Representaci\'on s\'{\i}ngleton de Dirac
\begin{equation}
D(\ft 12,0)\oplus D(1,\ft 12)
\label{Dirac-representacion}
\end{equation}
\end{itemize}
La construcci\'on de irreps unitarias de $OSp(N|4)$ para $N\geq 2$ es similar y se facilita dado que cualquier irreps de $OSp(N|4)$ se puede descomponer en t\'erminos de los multipletes base de $OSp(1|4)$. Si bien las $N-$extensiones de multipletes de orden superior consiste de multipletes de $OSp(1|4)$. Se debe tener cuidado ya que surgen nuevos rasgos en el an\'alisis con $N>1$. Por otro lado la estructura de $OSp(N|4)$ admite nuevos tipos de multipletes que no tienen an\'alogos en SuperPoincar\'e. Aunque pueden tener una estructura similar en SuperPoincar\'e con cargas central, el grupo $SO(N)$ se mantiene aqu\'{\i}.

\subsection {Aplicaci\'on al grupo $OSp(8|4)$}
Al tomar el grupo $OSp(8|4)$ este corresponde al grupo de invariancia de los estados base en supergravedad $11-Dim.$ sobre $s^7$, donde las flutuaciones del estado base corresponden a irreps de $OSp(8|4)$, las cuales tienen un esp\'{\i}n maximal $s=2$. Los campos gauge $2,\ft 32,1$ que pertenecen a un multiplete gravitacional son precisamente los campos relacionados a las traslaciones, cargas espinoriales y las cargas centrales respectivamente. En particular los campos vectoriales etiquetan un grupo abeliano (El centro) y las simetr\'{\i}as internas (generalmente no-abelianas) transforman como cargas espinoriales. Ahora se procede a clasificar las representaciones supersim\'etricas $N=8$ con un esp\'{\i}n maximal $s=2$. Como antes (cap. \ref{superpoincare}) se hacen las cuentas de estados de helicidad para $N=8$, con un n\'umero total de estados: 
$$
\sum_{p=0}^{N=8} {8\choose p }=(1+1)^8=256 \ \ \ \ \ \Rightarrow \ \ \ \ \ 
\begin {array} {cccccccccc} 
\hline 
         s       & -2 & -\ft 32 & -1 & -1/2 & 0 & 1/2 & 1 & \ft 32 & 2 \\
\hline 
\mbox{Estados}   &  1 &  8      & 28 &  56  & 70& 56  & 28&     8  & 1 \\
\hline
\end {array}
$$
Cada t\'ermino es la contribuci\'on del n\'umero de \'{\i}ndices en un tensor antisim\'etrico de rango-$p$, clasificados como:
\begin {itemize} 
\item N\'umero total de fermiones $\Rightarrow 128=8+56+56+8$
\item N\'umero total de bosones \ \ $\Rightarrow 128=1+28+70+28+1$
\end {itemize}
Como ya se vio en la secci\'on (\ref{superpoincare}) el n\'umero total de bosones es igual al n\'umero total de fermiones. 
\begin{equation}
\begin{array}{clclcl}
(1) & \hbox{Gravit\'on} & (28) & \hbox{Campos de esp\'{\i}n}-1    & (35) & \hbox{Escalares} \\                     
(8) & \hbox{Gravitinos} & (56) & \hbox{Campos de esp\'{\i}n}-\ft 12 & (35) & \hbox{Pseudoescalares} \\
\end{array}
\end{equation}
Dado el grupo $OSp(8|4)$ este permite las siguientes descomposiciones:
\begin{equation}
OSp(8|4)\supset OSp(1|4)\times SO(7)\ \ \ \ \ \ \ OSp(8|4)\supset Sp(4)\times SO(8)
\label{descomp}
\end{equation}
Lo cual permite las correspondientes reducciones:
\begin{equation}
OSp(8|4)\downarrow OSp(1|4)\times SO(7)\ \ \ \ \ \ \ OSp(8|4)\downarrow Sp(4)\times SO(8)
\label{descomp1}
\end{equation}
En resumen, partiendo del supergrupo $OSp(8|4)$, se puede obtener el resultado deseado por varias rutas:
$$
\begin{array}{c}
\ \ \ \ \ \ \ \ \ \ \ \ \ \ \ \ \ \ \ \ \bf{OSp}(8|4) 
\\
  \begin{array}{cc}
  \downarrow 
  &
  \downarrow  
  \\
  OSp(1|4)\times SO(7)
  &
  Sp(4)\times SO(8)\\ 
  &\sim SO(3,2)\times SO(8)  
  \\
    \begin{array}{ll}
      \begin{array}{c}
      \hookrightarrow \\       
      \ ^SP(1|3,1) \times SO(7)\\ 
      \downarrow \\       
      \bf{P}(3,1)\times \bf{SO}(7)
      \end{array}
    & 
      \begin{array}{c}  
      \downarrow \\   
      Sp(4)\times SO(7)\\
      \sim SO(3,2)\times SO(7)\\
      \hookrightarrow \\          
      \bf{P}(3,1)\times \bf{SO}(7)
      \end{array} 
    \end{array} 
  &
    \begin{array}{c} 
    \hookrightarrow \\    
    \bf{P}(3,1)\times \bf{SO}(8)
    \end{array}  
  \end{array}
\end{array}
$$
El anterior diagrama se puede explicar como sigue:
\begin{itemize}
\item Suponiendo la reducci\'on $OSp(8|4)\downarrow OSp(1|4)\times SO(7)$
\begin{itemize}
\item Se usa la contracci\'on $OSp(1|4)\to \ ^SP(1|3,1)$, expuesto en la secci\'on (\ref{superadS}) y luego se aplica otra reducci\'on $\ ^SP(1|3,1) \downarrow P(3,1)$ para obtener el grupo producto directo $P(3,1)\times SO(7)$.  
\item Se supone otra reducci\'on $OSp(1|4)\downarrow Sp(4)$ (con $Sp(4)\sim SO(3,2)$ en el \'algebra) y usando la contracci\'on $SO(3,2)\to P(3,1)$, se obtiene $P(3,1)\times SO(7)$, lo cual est\'a expuesto en la secci\'on (\ref{sittertopoincare}). 
\end{itemize} 
\item Suponiendo la reducci\'on $OSp(8|4)\downarrow Sp(4)\times SO(8)$ y usando la contracci\'on $SO(3,2) \to P(3,1)$, se obtiene $P(3,1)\times SO(8)$, expuesto en la secci\'on (\ref{sittertopoincare}). 
\end{itemize}
Lo cual completa nuestro acercamiento. Cual es la ruta f\'{\i}sica?. Ya que por reducci\'on no es posible encontrar los grupos del Modelo Standard aqu\'{\i}, dado que el grupo minimal en teor\'{\i}as de gran unificaci\'on (GUT) es $SO(10)\supset SU(5)\supset SU(3)\times SU(2)\times U(1)$. En que forma es posible?. Estas son preguntas a resolver.

\end{document}